\documentclass[review]{elsarticle}

\usepackage{hyperref}

\journal{Journal of Systems and Software}

\newtheorem{theorem}{Theorem}
\newtheorem{lemma}{Lemma}
\newdefinition{definition}{Definition}
\newproof{proof}{Proof}

\usepackage{etoolbox}
\usepackage{subcaption}

\usepackage{mathtools}
\usepackage{cspsymb}
\usepackage{eucal}

\usepackage{thmtools}
\usepackage{stmaryrd}
\usepackage{pgf}
\usepackage{tikz}
\usepackage{listings}
\usepackage{relsize}

\tikzset{filled/.style={fill=black!0, draw=black!100, thick}, outline/.style={draw=black!100, thick}}
\usetikzlibrary{arrows}

\usepackage[lined,boxed]{algorithm2e}
\LinesNumbered

\SetAlFnt{\Large}
\SetAlCapFnt{\LARGE}
\SetAlCapNameFnt{\LARGE}
\SetAlCapSty{}

\DeclareCaptionFont{subfontHuge}{\Huge}
\DeclareCaptionFont{subfontSmall}{\small}

\usetikzlibrary{arrows,automata}

\renewcommand{\cat}{\hspace{0.1em}\mbox{\textasciicircum}\hspace{0.1em}}
\newcommand{\dcomp}{~\asymp~}

\newcommand{\mcal}[1]{\mathcal{#1}}

\DeclareMathSymbol{\backprime} {\mathord}{AMSa}{"38}
\newcommand{\ssub}{~\backprime~}

\newcommand{\intercomp}{~[\interleave]~}
\newcommand{\commcomp}[2]{[#1\leftrightarrow#2]}
\newcommand{\feedcomp}[2]{[#1~\hookrightarrow~#2]}
\newcommand{\refcomp}[2]{[#1~\bar{\hookrightarrow}~#2]}

\newcommand{\comp}[2]{#1~~  \textbf{com} ~~  #2}

\newcommand{\refcvg}[2]{#1 ~~\mathtt{io \un cvg}~~#2}
\newcommand{\refecvg}[2]{#1~~\mathtt{io \un ecvg}~~#2}
\newcommand{\cvg}[2]{#1~~\mathtt{cvg}~~#2}
\newcommand{\ecvg}[2]{#1~~\mathtt{ecvg}~~#2}
\newcommand{\defcong}[3]{#1~~\mathbf{def}\mbox{-}\mathbf{cong}({#3)}~~#2}
\newcommand{\inpcong}[3]{#1~~\mathbf{inp}\mbox{-}\mathbf{cong}({#3)}~~#2}

\newcommand{\frefbric}{\refinedby_{\mcal{B}}}

\newcommand{\subcvg}{\leftarrowtriangle_{cvg}}
\newcommand{\subecvg}{\leftarrowtriangle_{ecvg}}

\newcommand{\ctdef}[5]{#1 : \langle \mcal{#2}, \mcal{#3}, \mcal{#4}, \mcal{#5}\rangle}

\newcommand{\ce}[1]{\mcal{#1}} 
\newcommand{\cc}[2]{\mcal{#1}_{#2}}  

\newcommand{\sm}[1]{\mathcal{S} \llbracket #1 \rrbracket} 

\newcommand{\un}{\underline{\hspace{3pt}}}

\newcommand{\BRIC}{\mcal{BRIC}}
\newcommand{\bricB}{\mcal{B}}
\newcommand{\bricR}{\mcal{R}}
\newcommand{\bricI}{\mcal{I}}
\newcommand{\bricC}{\mcal{C}}

\newcommand*\circled[2]{\tikz[baseline=(char.base)]{
            \node[shape=circle,draw,inner sep=#2pt] (char) {#1};}}

\begin{document}

\begin{frontmatter}

\title{A refinement checking based strategy for component-based systems evolution
}

\author[ifba]{Jos\'e Dihego\corref{cor1}}
\ead{jose.dihego@ifba.edu.br; josedihego@gmail.com}

\author[ufpe]{Augusto Sampaio}
\ead{acas@cin.ufpe.br}

\author[ufrn]{Marcel Oliveira}
\ead{marcel@dimap.ufrn.br}

\cortext[cor1]{Corresponding author}



\address[ifba]{Coordena\c{c}\~ao de Inform\'atica, IFBA, Feira de Santana-BA, Brazil}

\address[ufpe]{Centro de Inform\'atica, Universidade Federal de Pernambuco, Recife-PE, Brazil}

\address[ufrn]{Departamento de Inform\'atica e Matem\'atica Aplicada, UFRN, Natal-RN, Brazil}

\begin{abstract}
We propose inheritance and refinement relations for a CSP-based component model ($\BRIC$), which supports a constructive design based on composition rules that preserve classical concurrency properties such as deadlock freedom. The proposed relations allow extension of functionality, whilst preserving behavioural properties. A notion of extensibility is defined on top of a behavioural relation called \textit{convergence}, which distinguishes inputs from outputs and the context where they are communicated, allowing extensions to reuse existing events with different purposes. We mechanise the strategy  for extensibility verification using the FDR4 tool, and illustrate our results with an autonomous healthcare robot case study.
\end{abstract}


\end{frontmatter}


\section{Introduction}\label{sec:Introduction}


\noindent
Designing and reasoning about computational systems that interoperate in heterogeneous environments, and are expected to cope with a variety of application domains, is a challenge of increasing importance. Modelling and analysis techniques become manageable, and possibly scalable, when systems can be understood and developed by smaller (less complex) units. This is the core idea of component-based model driven development (CB-MDD) \cite{Szyperski98}, by which a system is defined as a set of components and their connections. This has become an essential infrastructure to the emergence of some other important development paradigms like Service Oriented Computing~\cite{Papazoglou2003} and Systems of Systems~\cite{Russell71}.

There are several component models for MDD such as, for instance, those presented in \cite{Jifeng2006,Meng2006,Hennicker2008,Arbab2004}, and a variety of approaches to analyse and refine component-based systems as, for example, the ones reported in\cite{Chen2009,Buchi1998,KurkiSuonio1999}. Nevertheless, formal support to component evolution, possibly involving an increment of interface via the addition of new functionality, has not been properly addressed. Component inheritance arises as a natural aspect to be provided by a CB-MDD approach, as a way to support evolution. It has been a successful feature present in object oriented languages from the beginning \cite{Liskov94, Wegner88, America90}; however, differently from object-orientation, our focus is not on defining subtyping, but extension relations that preserve some notions of conformance, a safe way to evolve component systems considering structure and behaviour. 

To achieve this goal in a controlled manner, component inheritance must obey the substitutability principle \cite{Liskov1987,Wegner88}: an instance of the subcomponent should be usable wherever an instance of the supercomponent was expected, without a component, playing the role of a client, being able to observe any difference. 

Some works have proposed inheritance relations for behavioural specifications \cite{Liskov94, America90,Bowman99,Puntigam96, Wehrheim03, DihegoPedroAugusto2013}. The first four define behaviour in terms of pre- and postconditions, and structure by method signatures (covariance and contravariance), but do not address reactive behaviour. Although the approaches in \cite{Wehrheim03, DihegoPedroAugusto2013} consider active objects, they do not focus on structure. Furthermore, none of the mentioned works  differentiate the nature of input and output events nor the context in which they are communicated. This differentiation is relevant for a large class of specifications (such as Enterprise JavaBeans (EJB) \cite{Rubinger2010}, client-server protocols and Model-View-Controller (MVC) components \cite{Krasner1988}) where components control outputs, while the environment controls inputs. Finally, these works do not consider the impact of these relations over behavioural properties, such as preservation of deadlock freedom. Therefore, although these approaches deal with substitutability, the proposed inheritance relations do not guarantee deadlock free evolutions, in some cases because it is not an explicit concern \cite{Liskov94} or just because it admits, implicitly, the introduction of deadlock through weaker inheritance relations \cite{Wehrheim03}.

In the current work, we propose component extension relations that address how components behave, how they are structured (channels and interfaces), distinguish the nature of inputs and outputs and, moreover, guarantee safe evolution by means of classical properties preservation.

We develop a fully formal and mechanised approach to component based model driven development. Particularly, our work is in the context of $\BRIC$ \cite{Ramos2009Systematic}, which formalises the core CB-MDD concepts \cite{Szyperski98} (components, interfaces, channels and behaviour) and, moreover, supports compositions, where deadlock freedom is ensured by construction. This approach covers not only tree topologies, but also cyclic ones.

In previous work \cite{DihegoAugustoMarcel2015}, we defined $\BRIC$ component refinement and inheritance notions based on the concept of behavioural convergence. Here, we significantly extend our previous work  with the following contributions.
\begin{itemize}
\item Detailed proofs of lemmas  that relate the proposed notions of inheritance, as well as the proofs of theorems that relate inheritance with refinement, and the fact that inheritance preserves deadlock freedom.
\item A strategy, based on refinement checking, using the FDR4 tool \cite{RobinsonABR14}, to mechanically verify whether two given component models are related by the proposed inheritance notions.  
\item An elaborate case study, an autonomous healthcare robot, used to illustrate the overall approach. 

\item A more detailed account of related work. 
\end{itemize}


It is important to emphasise that the focus of this paper is on an infrastructure we have devised to formally and mechanically check model evolution based on a notion of convergence. Nevertheless, an approach to design model extensions that ensure convergence, in a constructive way, is out of the scope of this paper. This important, complementary contribution, is considered as one of our major topics for future work, and is discussed in more detail in the concluding section.   

We structure the paper as follows. Section \ref{sec:BRIC component model} introduces the $\BRIC$ component model. Section \ref{sec:BRIC Inheritance} presents a congruent semantics for  $\BRIC$, a refinement notion and two inheritance relations, based on a  concept called behavioural \textit{convergence}, which allows extensions but preserves conformance. A strategy to mechanically verify conformance with respect to the proposed relations is the subject of Section \ref{sec:Checking convergence via refinement}. Our results are illustrated by a case study of an autonomous healthcare robot in Section \ref{sec:Case Study}. We conclude with related work in Section \ref{sec:Related work} and summarise our contributions and future work in Section \ref{sec:Conclusions}.

\section{The $\BRIC$ component model}\label{sec:BRIC component model}

In the $\BRIC$ component model, one specifies components, connectors and their behaviour in the Communicating Sequential Processes (CSP) language \cite{Roscoe1998Theory}; $\BRIC$ provides a set of rules to assemble components. The behavioural properties (particularly, deadlock freedom) of compositions using the $\BRIC$ rules are guaranteed by construction, verified by local analyses \cite{Ramos11}, which include cyclic networks \cite{Pedro14}. More recently, some additional rules were proposed to  ensure livelock freedom~\cite{Filho2016, Madiel2018} and to avoid nondeterminism \cite{Otoni2017}.

\subsection{CSP}

A process algebra like CSP can be used to describe systems composed of interacting components, which are independent self-contained processes with interfaces used to interact with the environment. Such formalisms provide mechanisms to specify and reason about interaction between components. Furthermore, phenomena that are exclusive to the concurrent world, that arise from the combination of components rather than from individual components, like deadlock and livelock, can be more easily understood and controlled using such formalisms. Tool support is another reason for the success of CSP in industrial applications. For instance, FDR4 \cite{RobinsonABR14} provides an automatic analysis of model refinement and of properties like deadlock, livelock and determinism.  Each of these classical properties is very complex  to verify in a behavioural model. In this context, we contribute to $\BRIC$ by developing a strategy to evolve component specifications without introducing deadlock. More recently, some additional rules were proposed~\cite{Filho2016} to also ensure livelock freedom.

Here we use the machine-readable version of CSP (\CSPM~\cite{Roscoe1998Theory}). The two basic CSP processes are \verb"STOP" (deadlock) and \verb"SKIP" (successful termination). The prefixing \verb"c -> P" is initially able to perform only the event \verb"c"; afterwards it behaves like process \verb"P". A boolean guard may be associated with a process: given a predicate \verb"g", if it holds, the process \verb"g & c?x -> A" inputs a value through channel \verb"c" and assigns it to the variable \verb"x", and then behaves like \verb"A", which has the variable \verb"x" in scope; the process deadlocks otherwise. It can also be defined as  \verb"if g then c?x -> A else STOP". Multiple inputs and outputs are also possible. For instance, \verb"c?x?y!z" inputs two values that are assigned to \verb"x" and \verb"y" and outputs the value resulting from the evaluation of expression \verb"z".

The external choice \verb"P1 [] P2" initially offers events of both processes. The engagement of the process in an event resolves the choice in favor of the process that performs it. On the other hand, the environment has no control over the internal choice \verb"P1 |~| P2". The sequence operator \verb"P1;P2" combines processes \verb"P1" and \verb"P2" in sequence. The synchronised parallel composition \verb"P1 [| cs |] P2" synchronises \verb"P1" and \verb"P2" on the channels in the set \verb"cs"; events that are not in \verb"cs" occur independently. Processes composed in interleaving, as in \verb"P1 ||| P2", run independently. The event hiding operator \verb"P \ cs" encapsulates (internalises) in P the events that are in the channel set \verb"cs", which become no longer visible to the environment.

In this work we use two denotational models of CSP: traces ($\tmodel$) and stable failures, or just failures ($\fmodel$). A trace of a process \verb"P" is  a sequence of events that it can perform and we define $\tmodel(\verb"P")$ to be the set of all its finite traces.  For example $\tmodel(\verb"e1 -> e2 -> STOP") = \{\emptyseq, \langle \verb"e1"\rangle, \langle \verb"e1", \verb"e2" \rangle\}$. The set $\fmodel(\verb"P")$ consists of all stable failures $(s,X)$, where $s$ is a trace of \verb"P" ($ s \in \tmodel(\verb"P")$) and $X$ is a set of events \verb"P" can refuse in some stable state after $s$. A stable state is one in which a process can only engage in a visible event (registered in the process trace). The invisible event is represented in CSP as $\tau$;  it can happen, for example, in the internal choice  \verb"P |~| Q", which will be implemented as a process that can take an invisible $\tau$ event to each of \verb"P" and \verb"Q". We summarise this discussion in \ref{appendix:CSP}.

\subsection{$\BRIC$}

A component is defined as a contract (Definition \ref{def:componentcontract}) that specifies its behaviour, communication points (or channels) and their types.

\begin{definition}[Component contract] \label{def:componentcontract}  A component contract $Ctr{:}\langle \bricB, \bricR, \bricI, \bricC \rangle$  comprises an observational behaviour $\bricB$ specified as a CSP process, a set of communication channels $\bricC$, a set of interfaces $\bricI$, and a \textbf{total} function $\bricR: \bricC \mapsto \bricI$ between channels and interfaces of the contract, such that $\bricB$ is an I/O process.
\end{definition}

We require the CSP process $\bricB$ to be an I/O process, which is a non-divergent processes with infinite traces. Moreover, it offers to the environment the choice over its inputs (external choice) but reserves the right to choose among its outputs (internal choice). It represents a wide range of specifications, including the server-client protocol, where the client sends requests (inputs) to the server, which decides the outputs to be returned to the client.

\begin{definition}[I/O process] \label{def:ioprocess}We say that a CSP process $P$ is an  I/O process if it satisfies the following five conditions, which are formally presented in~\cite{Ramos11,Ramos2009Systematic}: \\
(1)  \label{def_inf:io-channels} \textbf{I/O channels}:~Every channel in $P$ has its events partitioned into inputs and outputs.\\
(2) \label{def_inf:inf-traces} \textbf{infinite traces}:~$P$ has an infinite set of traces~(but finite state-space).\\
(3) \label{def_inf:div-free} \textbf{divergence-freedom}:~$P$ is divergence-free.\\
(4) \label{def_inf:input-det} \textbf{input determinism}: If a set of input events in $P$ are offered to the environment, none of them are refused.\\
(5)\label{def_inf:strong-output-decisive} \textbf{strong output decisive}: All choices~(if any) among output events on a given channel in $P$ are internal; the process, however, must offer at least one output on that channel. 
\end{definition}
 
Contracts can be composed using any of the four rules available in the model: interleaving, communication, feedback, or reflexive composition. Each of these rules impose associated side conditions, which must be satisfied by the contracts and channels involved in the composition in order to guarantee deadlock freedom by construction.

\begin{figure*}[t] 
\resizebox{.23\textwidth}{!}{
 \begin{minipage}{.35\textwidth}
  \begin{tikzpicture}
\draw  (-8,3.5) rectangle (-4.5,0);
\draw  (-2.5,2.5) rectangle (4,0.5);
\draw  (6,3) rectangle (9.5116,0.1048);
\draw  (-7.5,3) rectangle (-5,2);
\draw  (-7.5,1.5) rectangle (-5,0.5);
\draw  (3.5,2) rectangle (1.5,1);
\draw  (0,2) rectangle (-2,1);
\draw  (6.5,2) node (v2) {} rectangle (9,1);

\draw [semithick] arc (-59.9993:-140:0.5);
\node at (-6.3,2.5) {$\bf P$};
\node at (-6.3,1) {$ \bf Q$};
\node at (-1,1.5) {$ \bf P$};
\node at (2.5,1.5) {$ \bf Q$};
\node at (7.8,1.5) {$ \bf P$};
\draw (-4,2.7) arc (100:270:.2);
\draw (-4,1.2) arc (100:270:.2);
\draw  (-5,2.5) rectangle (-4.15,2.5);
\draw  (-5,1) rectangle (-4.15,1);
\draw  (-7.5,2.5) rectangle (-8.5,2.5);
\draw  (-7.5,1) rectangle (-8.5,1);
\draw  (-8.7,2.5) ellipse (0.2 and 0.2);
\draw  (-8.7,1) ellipse (0.2 and 0.2);
\node at (-6.5,.2) {$ \bf Interleave$};
\draw (0.5,1.8) node (v1) {} arc (100:270:0.3);
\draw  (0,1.5) rectangle (0.25,1.5);
\draw  (0.6,1.5) ellipse (0.2 and 0.2);
\draw  (.8,1.5) rectangle (1.5,1.5);
\draw  (3.5,1.5) rectangle (4.5,1.5);
\draw  (-2,1.5) rectangle (-3,1.5);
\draw  (-3.2,1.5) ellipse (0.2 and 0.2);
\draw (4.65,1.7) arc (100:270:0.2);
\draw  (9,1.2) rectangle (9.8,1.2);
\draw  (6.5,1.2) rectangle (5.7,1.2);
\draw  (5.5,1.2) ellipse (0.2 and 0.2);
\draw (9.95,1.4) arc (100:270:0.2);
\draw  (8.9,2.5) node (v4) {} rectangle (8.9,2);
\draw  (6.6,2.5) node (v3) {} rectangle (6.6,2);
\draw  (v3) rectangle (7.6,2.5);
\draw  (v4) rectangle (8,2.5) node (v6) {};
\draw  (7.75,2.5) node (v5) {} ellipse (0.15 and 0.15);
\draw (7.8,2.7) arc (90:-90:0.2);
\node at (1,0.727) {$ \bf Communication$};
\node at (8,0.7681) {$ \bf Feedback /$ };
\node at (8,0.4257) {$ \bf Reflexive$};
\end{tikzpicture}
  \end{minipage} 
}
\caption{Composition rules}
\label{f:CompositionRules}
\end{figure*}
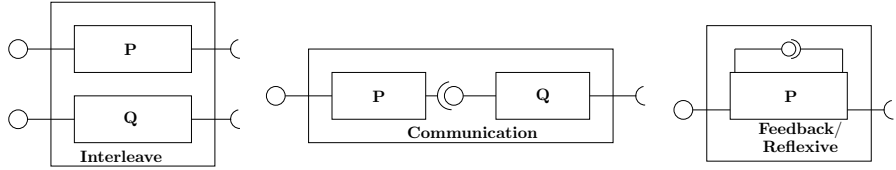

The rules provide asynchronous pairwise compositions, mediated by buffers, and focus on the preservation of deadlock freedom in the resulting component. Using the rules, developers may connect channels of two components, or even of the same component. The four rules are illustrated in Figure \ref{f:CompositionRules}. Two of  them, feedback and reflexive, are merged (see the rightmost scheme in the figure) as both entail connection of channels of a same component; more explanation is presented in the sequel for the composition rule and the detailed formalisation in \ref{appendix:BRIC} for the rest of the rules.

The interleave composition rule is the simplest form of composition. It aggregates two independent entities such that, after composition, these entities still do not communicate between themselves. They communicate directly with the environment as before, with no interference from each other.

The communication composition (Definition \ref{def:communicationCompositionMainText}) states the most common way for linking channels \cite{Ramos2009Systematic} of two different entities. It is given in terms  of asynchronous binary composition of channels from $P$ and $Q$ (Definition \ref{def:Asynchronous binary compositionMainText}).

\begin{definition}[Communication composition] \label{def:communicationCompositionMainText}
Let $P$ and $Q$ be two component contracts, and $ic$ and $oc$ two communication channels. The communication composition of $P$  and $Q$  (namely $P \commcomp{ic}{oc} Q$) via $ic$ and $oc$ is defined as follows:
\begin{align*}
    P \commcomp{ic}{oc} Q  =  P {}_{\langle ic\rangle}\dcomp {}_{\langle oc \rangle} Q
\end{align*}%
\end{definition}

This rule assumes the components behaviours on channels $ic$ and $oc$ are I/O confluent, strong compatible and satisfy the finite output property (FOP). These properties are detailed in \cite{Roscoe1987pursuit,Roscoe2006Confluence,Ramos11}: I/O confluence means that choosing between inputs (deterministically) or outputs (non-deterministically) does not prevents other inputs/outputs offered alongside from being communicated afterwards; two processes are strong compatible if all outputs produced by one are consumed by the other, and vice versa; finally, FOP guarantees that a process cannot output forever, so eventually it inputs after a finite sequence of outputs. The resulting component $P {}_{\langle ic\rangle}\dcomp {}_{\langle oc \rangle} Q$ is the binary composition of $P$ and $Q$ on channels $ic$ and $oc$ (Definition \ref{def:Asynchronous binary compositionMainText}).

The asynchronous binary composition hooks two components, say $P$ and $Q$, with disjoint communication points, by their respective channels $c$ and $z$. Instead of communicating directly, their communications are buffered.

\begin{definition}[Asynchronous binary composition] \label{def:Asynchronous binary compositionMainText} 
Let $P$ and $Q$ be two distinct component contracts, and $c \in \cc{C}{P}$ and $z \in \cc{C}{Q}$ two channels, such that $\cc{C}{P}$ and $\cc{C}{Q}$ are disjoint. Then, the asynchronous binary composition of $P$ and $Q$, $P  {}_{\langle c \rangle}\asymp_{\langle z \rangle} Q$, is given by:
\begin{align*}
P  {}_{\langle c \rangle}\asymp_{\langle z\rangle} Q & = 
\langle \cc{B}{P} 
     	\parallel[\eset{c}] BUFF^{n}_{IO}(R_{IO}^{~c \rightarrow z}, R_{IO}^{~z \rightarrow c}) \parallel[\eset{z}] 
         \cc{B}{Q}, 
         \ce{R}', \ce{I}', \ce{C}' \rangle
\end{align*}

\text{where } $\ce{C}' = (\cc{C}{P} \cup \cc{C}{Q}) \backprime \set{c,z}$, $\ce{R}' = \ce{C}' \lhd (\cc{R}{P} \cup \cc{R}{Q})$,\\
\hspace*{1.6cm} $\ce{I}' = \ran \ce{R}'$ and $R_{IO}^{~a \rightarrow b} = \set{a.out.x \mapsto b.in.x}$.
\end{definition}

In this composition, the channels $c$ and $z$ are combined such that output events from one channel are consumed by input events of the other, and vice versa. This correspondence is made by two mapping relations, $R_{IO}^{~c \rightarrow z}$ and $R_{IO}^{~z \rightarrow c}$, which are used to input/output from the buffer  $BUFF^{n}_{IO}$. The resulting component behaviour is that of $P$ synchronised with the buffer $BUFF^{n}_{IO}(R_{IO}^{~c \rightarrow z}, R_{IO}^{~z \rightarrow c})$ on $c$ and with $Q$ on $z$. The interface, $\ce{C}'$, of the resulting component, $P  {}_{\langle c \rangle}\asymp_{\langle z\rangle} Q$, contains channels of both $P$ and $Q$ except for the hooked channels $c$ and $z$ ($(\cc{C}{P} \cup \cc{C}{Q}) \backprime \set{c,z}$). Therefore, only channels in $\ce{C}'$ appear in the resulting relation $\ce{R}'$ ($S \lhd R$ restricts the domain of $R$ to $S$) and in the resulting interface $\ce{I}'$ ($\ran \ce{R}'$).

There are more complex systems that present cycles of dependencies in the topology of their structure. The next two compositions, depicted together in Figure \ref{f:CompositionRules},  allow the link of two channels of a same entity without introducing deadlock (it is proved by Theorem \ref{thm:Deadlock-free Component Systems} from \cite{Ramos2009Systematic,Ramos11}). The feedback composition can only be used to assemble channels that are decoupled: $ch_1$ and $ch_2$ are decoupled if the interleaving of the projected component behaviour on $ch_1$ and on $ch_2$ is equivalent to 
the projected  behaviour of the component on the set formed of these two channels. This means that there is no interference between these two channels (see \ref{appendix:BRIC} for a formal account). Reflexive composition is more general than the feedback rule, as it does not require channels to be decoupled; however, it is also more costly regarding verification, since, in general, it requires a global analysis to ensure deadlock freedom. These rules ensure that systems developed in $\BRIC$ are deadlock-free. Theorem \ref{thm:Deadlock-free Component Systems} is proved in \cite{Ramos2009Systematic}.

\begin{theorem}[Deadlock-free Component Systems] \label{thm:Deadlock-free Component Systems}
Consider $P$ a deadlock-free component and $c_1$ and $c_2$ channels. Any system S in normal form, as defined below, built from deadlock-free components, is deadlock-free.

\vspace*{-1cm}

\begin{align*}
    S  ::= P \ \ \ | \ \ \ S \intercomp S \ \ \ \textnormal{(interleave)} \ \ \ | \ \ \ S \commcomp{c_1}{c_2} S \ \ \ \textnormal{(communication)} \\ 
 | \ \ \ S \feedcomp{c_1}{c_2} \ \ \ \textnormal{(feedback)}\ \ \ | \ \ \ S \refcomp{c_1}{c_2} \ \ \ \textnormal{(reflexive)}
\end{align*}%
\end{theorem}

\section{$\BRIC$ extensibility}\label{sec:BRIC Inheritance}

In this section we present the main contributions of this work: the development of inheritance relations for behavioural specifications that distinguish inputs from outputs, in the context of rigorous trustworthy component development, where structural aspects are also considered. Our relations rely on a behavioural relation called \textit{convergence}. It captures the idea that components can evolve  by accepting new inputs or establishing a communication session after these inputs but the components are able to \textit{converge} to the behaviour exhibited by their abstractions. This is a concept that cannot be captured only by the hiding operator as in the case of other inheritance relations \cite{Wehrheim03}. The reason is that, when hiding an event, it is removed from the traces a behaviour exhibits but, in our inheritance relations, an event can have different meanings based on the context it appears and, in some of these contexts we want to hide them, in others we do not. This is exemplified by our motivating example.

Additionally, we propose a denotational semantics and a refinement notion for $\BRIC$ components, which, alongside the proposed inheritance notions, make it a fully formal approach to CB-MDD. In the sequel we present a motivating example that illustrates the purpose and intuition behind this new behavioural relation.

\subsection{Motivating example}\label{subsec:Motivating example}

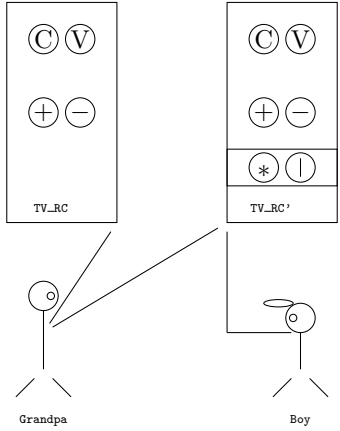
\begin{figure*}[t] 
\centering
\resizebox{0.4\textwidth}{!}{
\begin{tikzpicture}
\draw  (-4,4.5) rectangle (-2.5,1.5);
\draw  (-7,4.5) rectangle (-5.5,1.5) node (v9) {};
\draw  (-6.5,4) ellipse (0.2 and 0.2);
\draw  (-6,4) ellipse (0.2 and 0.2);
\draw  (-6.5,3) ellipse (0.2 and 0.2);
\draw  (-6,3) ellipse (0.2 and 0.2);
\draw  (-4,2.5) rectangle (-2.5,2);
\draw  (-3.5,4) ellipse (0.2 and 0.2);
\draw  (-3,4) ellipse (0.2 and 0.2);
\draw  (-3.5,3) ellipse (0.2 and 0.2);
\draw  (-3,3) ellipse (0.2 and 0.2);
\draw  (-3.5,2.25) ellipse (0.2 and 0.2);
\draw  (-3,2.25) ellipse (0.2 and 0.2);
\draw  (-6.5,0.5) node (v7) {} ellipse (0.2 and 0.2);
\draw  (-6.5,0.3) rectangle (-6.5,-0.5) node (v1) {};
\node (v2) at (-7,-1) {};
\node (v3) at (-6,-1) {};
\draw  (v1) edge (v2);
\draw  (v1) edge (v3);
\draw  (-3,0.2) ellipse (0.2 and 0.2);
\draw  (-3,0) node (v11) {} rectangle (-3,-0.5) node (v4) {};
\node (v5) at (-3.5,-1) {};
\node (v6) at (-2.5,-1) {};
\draw  (v4) edge (v5);
\draw  (v4) edge (v6);
\draw  (-3.3,0.4) ellipse (0.2 and 0.05);
\draw  (-3.1,.2) ellipse (0.05 and 0.05);
\draw  (-6.4,.5) ellipse (0.05 and 0.05);
\node (v8) at (-6.5,0) {};
\draw  (v8) edge (v9);
\node (v10) at (-4,1.5) {};
\draw  (v8) edge (v10);
\node (v12) at (-4,0) {};
\draw  (v11) edge (-4,0);
\draw  (-4,0) edge (v10);
\node at (-6.5,-1.2) {\tiny\ttfamily Grandpa};
\node at (-3,-1.2) {\tiny\ttfamily Boy};
\node at (-6.5,4) {C};
\node at (-6,4) {V};
\node at (-6.5,3) {$+$};
\node at (-6.4,1.7) {{\tiny\ttfamily TV\_RC}};
\node at (-6,3) {$-$};
\node at (-3.5,4) {C};
\node at (-3,4) {V};
\node at (-3.5,3) {$+$};
\node at (-3,3) {$-$};
\node at (-3.5,2.2) {$*$};
\node at (-3.4,1.7) {{\tiny\ttfamily TV\_RC'}};
\node (v13) at (-3,2.5) {};
\node (v14) at (-3,2) {};
\draw  (v13) edge (v14);
\end{tikzpicture}
}
\caption{TV remote control extension}
\label{fig:TV remote control extension}
\end{figure*}

Consider a TV remote control that offers the options for controlling the TV audio volume and switching between channels. It is represented by the device \verb"TV_RC" in Figure \ref{fig:TV remote control extension}; the user presses the button \circled{C}{.5}, then he can go forward or backward on the channel list by pressing \circled{$+$}{.5} or \circled{$-$}{.5}, respectively. The same applies when \verb"Grandpa" wants to increase/decrease the TV volume by pressing \circled{V}{.5}. The user \verb"Boy" wants to do more by having the option to adjust the TV brightness and contrast by pressing the buttons \circled{$*$}{1.5} and \circled{$|$}{.2}, respectively. The user \verb"Boy" is aware of a lid on the device \verb"TV_RC'" (Figure \ref{fig:TV remote control extension}), by which he can access these new functionalities, which are unknown by \verb"Grandpa".

\begin{figure*}[t]
\resizebox{1\textwidth}{!}{
\begin{tabular}{r}
\begin{minipage}{1\textwidth}
\centerline{\includegraphics[scale=.5]{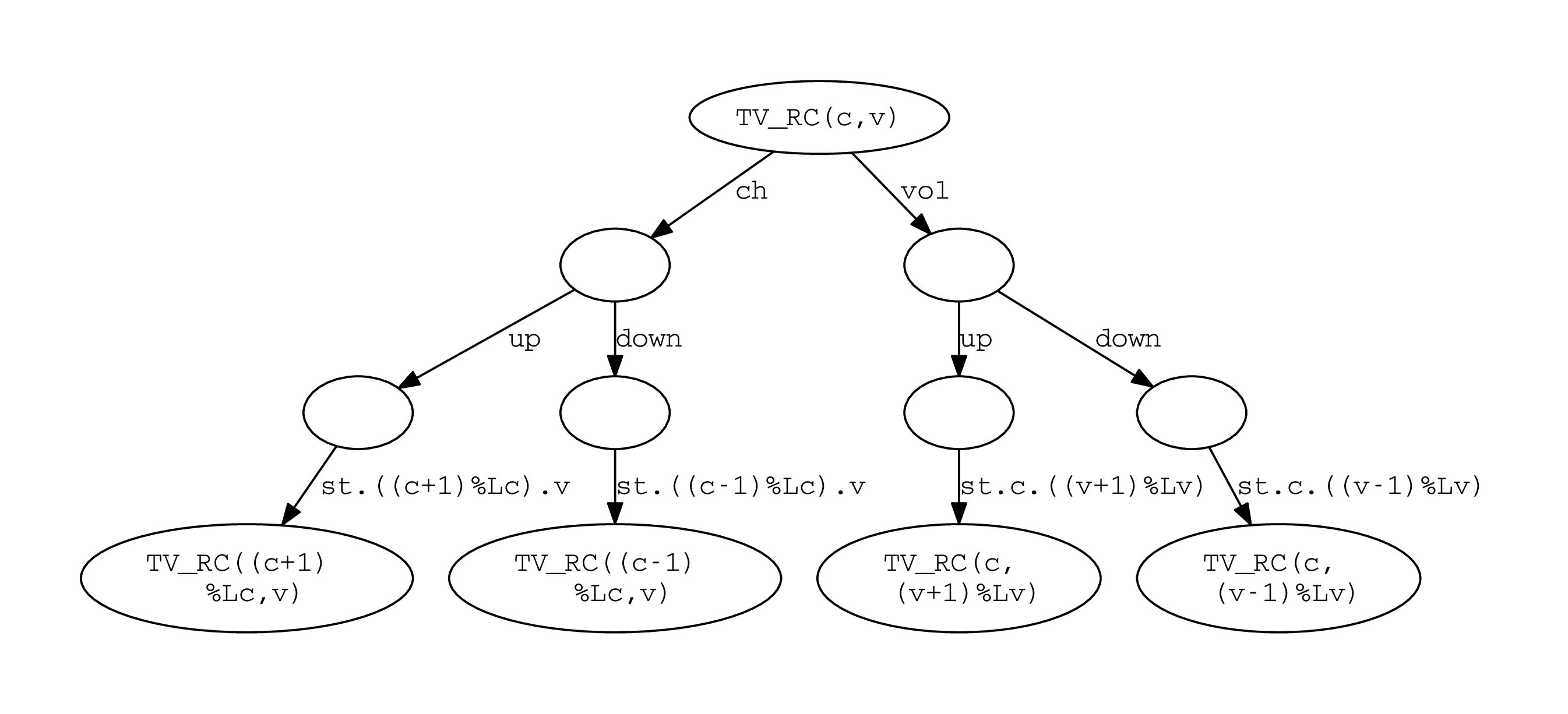}}
\end{minipage}
\end{tabular}
}
\caption{Labelled transition system of the \texttt{TV\_RC(c,v)} process}
\label{fig:RCTV}
\end{figure*}

\begin{figure*}[t]
\resizebox{1\textwidth}{!}{
\begin{tabular}{r}
\begin{minipage}{1\textwidth}
\centerline{\includegraphics[scale=.4]{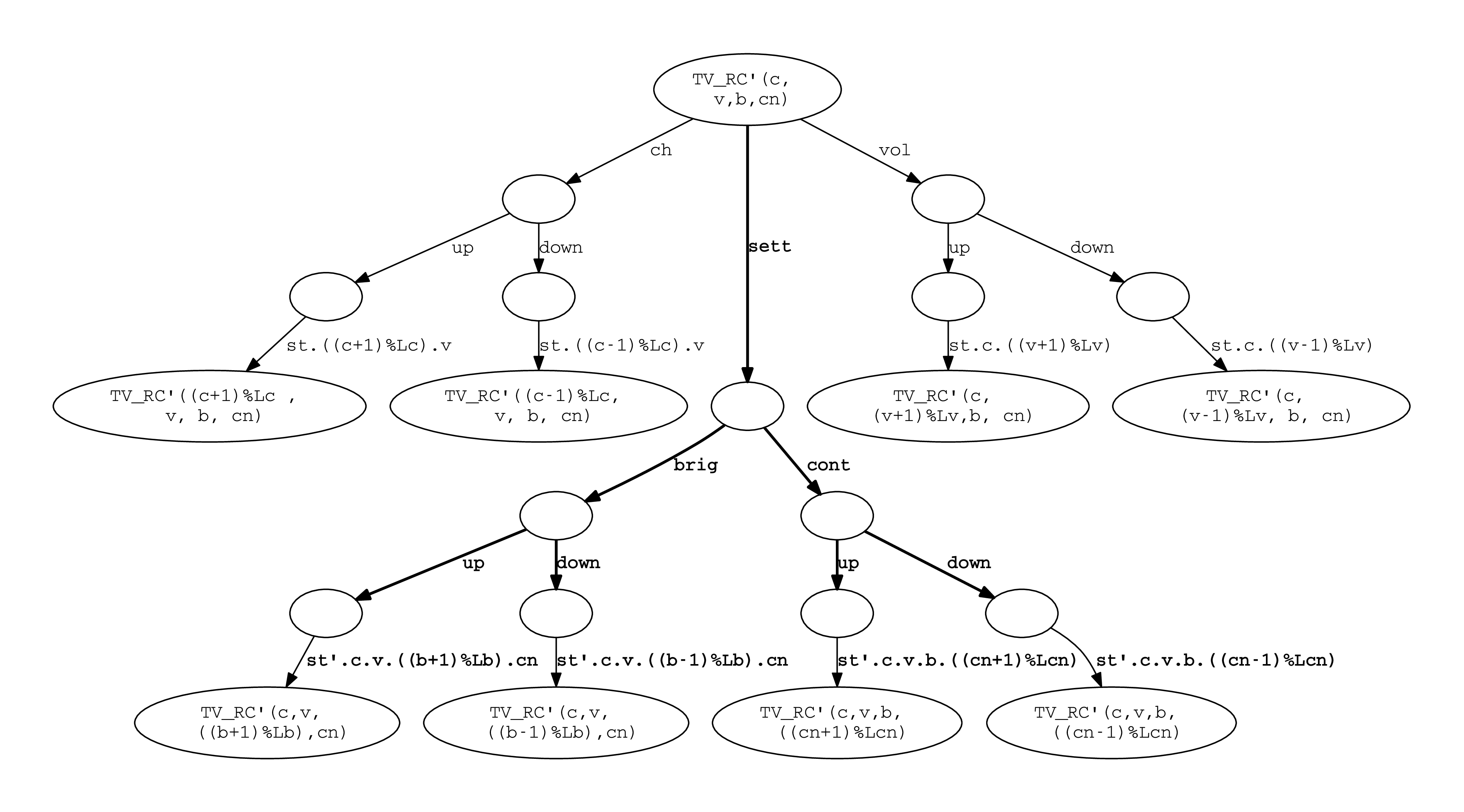}}
\end{minipage}
\end{tabular}
}
\caption{Labelled transition system of the \texttt{TV\_RC'(c,v)} process}
\label{fig:RCTV2}
\end{figure*}

We specify the behaviours of \verb"TV_RC" and \verb"TV_RC'" in the LTSs (Labelled Transition System) depicted in Figures \ref{fig:RCTV} and \ref{fig:RCTV2}, respectively. The process \verb"TV_RC(c,v)" represents a state where channel \verb"c" and volume \verb"v" are the last values sent to the TV. The user provides the input event \verb"vol" to increase (\verb"up") or decrease (\verb"down") the TV volume or \verb"ch" to navigate forward  (\verb"up") or backward (\verb"down")  on TV channels. Status output events (\verb"st") acknowledge the user of its commands. Updates in the volume or channel are captured by modular arithmetic; for instance, \verb"(c+1)%Lc" is addition modulo a maximum allowed value \verb"Lc" for channels. Modular arithmetic avoids results outside the range between zero and the maximum value \verb"Lc".

The process \verb"TV_RC'(c,v,b,cn)" in Figure \ref{fig:RCTV2} extends \verb"TV_RC(c,v)" by adding the new input event \verb"sett", which occurs when the user opens the remote control lid; it gives access to the TV settings menu, in which the user can adjust brightness  (\verb"brig") or contrast (\verb"cont"), whose values are registered by the state variables \verb"b" and \verb"cn", respectively. After providing \verb"brig" or \verb"cont" as input, it can use the regular input events \verb"up" and \verb"down" to make the image adjustments, which are echoed by output events through the channel \verb"st'".  

If we consider \verb"TV_RC'(c,v,b,cn)" as a valid extension to \verb"TV_RC(c,v)" (in the sense that it preserves convergence) we must decide which relation captures this kind of extension. Such a relation must allow new-in-context input events (those that are not necessarily new in the process alphabet, but that are not among the events offered by the process in a particular context), as \verb"sett", followed by a finite number of new-in-context input/output events, as \verb"brig", \verb"up", and \verb"st'". Furthermore, the extension must converge (offering what was expected before the new-in-context event happened) to the original behaviour. For instance, this is what \verb"TV_RC'(c,v,((b+1)%Lb),cn)" does after the trace $\trace{\texttt{sett}, \texttt{brig}, \texttt{up}}$, see Figure \ref{fig:RCTV2}: it offers \verb"ch" and \verb"vol", with the variables \verb"c" and \verb"v" unchanged by what happened after \verb"sett". In summary, \verb"Grandpa" can share the new TV remote control with \verb"Boy", without being stuck with, or even perceiving, the new features.

One can recognise the CSP failures refinement as a candidate to capture the relationship between \verb"TV_RC" and \verb"TV_RC'", provided the new events are hidden in the extension, but a closer investigation shows this is not the case. This happens because the events \verb"up" and \verb"down" are also used in new contexts in \verb"TV_RC'" to adjust brightness and contrast. This kind of relationship in which we use existing events in a different context cannot be captured by failures refinement. 

In \cite{Wehrheim03} four subtyping relations are defined for behaviour specifications, which allow functionality extension. The first issue with these candidate relations is that they do not differentiate inputs from outputs (as further discussed in the next section) and, moreover, extensions can only be defined in terms of: new events, which can be concealed (they may not be communicated but are not made internal~\cite{Wehrheim03}), hidden (made internal), explained (in terms of existing events) or restricted (completely forbids them). Except for hiding, the other extensions are not directly supported by CSP, and none of them can capture the intended relationship between \verb"TV_RC" and \verb"TV_RC'".

The ioco (input-output conformance) relation \cite{Petrenko04} allows extensions to admit new inputs (more functionalities) and to restrict outputs (more deterministic), but it does not obligate extensions, after a new input, to adhere (converge) to the original behaviour. Taking our example into account, ioco would admit \verb"TV_RC'" to engage in \verb"sett" and then behave as \verb"STOP" (or anything else, including a divergent process). Therefore, the user \verb"Boy" would not be able to navigate on the TV channel list, if he tried to adjust the image settings. Although ioco is a relation adopted in the context of conformance testing, whereas we are concerned with model evolution, we considered it here because it also allows extension of functionality, but in a more restrictive manner than we need.  

This discussion highlights the fact that the current behaviour relations cannot cope with functionality extension in a scenario where we need to add new events, to use existing events in different contexts and to distinguish input from output events. 

Before formalising convergence we need to say that \verb"TV_RC'" process, and inheritance behaviours in general, can be achieved by a variety of mechanisms, including design patters. Nevertheless, this work does not address the mechanisms to achieve convergent behaviours. In fact, our focus is on the definition of convergence and how it can be mechanically verified to ensure deadlock freedom in the evolution of behaviour component specifications.


\subsection{Convergence}\label{subsec:convergence}

Inheritance in the object-oriented paradigm is a well known concept with a comprehensive literature \cite{Liskov94}. More recently, efforts have been made to extend this concept to process algebras as CSP. Notably, in  \cite{Wehrheim03} the author proposes four types of behavioural inheritance relations for Labelled Transition Systems (LTS). Although very promising, as already discussed, these relations do not consider specifications that distinguish inputs from outputs, as required in $\BRIC$, in which a component behaviour is modelled as a CSP I/O process. Since I/O processes must satisfy behaviour restrictions such as  input determinism and  strong output decisiveness, we need  relations that can capture these restrictions. We base our approach on a concept of \textit{convergence}: a convergent process is allowed to do the same as or more inputs than its parent process, but is restricted to do the same or less outputs in convergent points. A convergent point represents a state reachable by both the original and the convergent process when doing two convergent sequences of events; these sequences differ only because the convergent process is allowed to do extra inputs (inputs not allowed by the original process) in converging points. First, we formalise the concept of convergent traces as follows.

In Definition \ref{def:I/O convergent traces} and others that follow, $\Sigma$ stands for the alphabet of all possible events, $ \Sigma^{*}$ is the set of possible sequences of events from $\Sigma$, the input events are contained in $\Sigma$ ($inputs \subseteq \Sigma$) and $ in(T,t)$ is a function that yields the set of input events that can be communicated by the I/O process $T$ after some trace $t \in \tmodel(T)$; therefore $in$ has type $ I/OProcess \cross \Sigma^{*} \to \mathcal{P} (inputs)$, where $\mathcal{P}$ stands for the powerset. Additionally $t_1  \leq t_2$ means that $t_1$ is a prefix of $t_2$. 

A trace $t'$ (of a process $T'$) is I/O convergent to a trace $t$ (of a process $T$) if they are equal or if $t = {t_1} \cat t_3$ and $t' = {t_1} \cat \trace{ne} \cat t'_3$ such that $ne \in in(T',t_1)$ but $ne \notin in(T,t_1)$ and $t$ is I/O convergent to $t_1 \cat t'_3$. This recursive definition means that $t'$ and $t$ might differ because $T'$ can do new-in-context inputs (inputs not allowed by $T$) where $T$ cannot, but, in spite of that, the trace $t'$ of $T'$ has always a counterpart $t$ of $T$. In other words, a trace $t'$ is I/O convergent to a trace $t$ if they differ only by certain inputs allowed by $T'$ but not by $T$. Also, because the CSP hiding operator renames visible events (events that can appear in traces) to the invisible event $\tau$, it is not an alternative to define convergence, where events have different meanings depending on the context they are communicated.
\\
\\
\begin{definition}[I/O convergent traces]\label{def:I/O convergent traces}
Consider an I/O process $T$. Let  $t$  and $t'$ be two traces, such that $t \in \traces(T)$. We say that $t'$ is an I/O convergent trace of $t$ ($\cvg{t'}{t}$) if, and only if:
\vspace*{-.2cm}
{\small \begin{align*}
\begin{array}{l}
(t' = t) \lor
	\left( 
	\begin{array}{l}
 	(\# t' > \# t) ~~\land~~ \exists t_1, t_3 :  \Sigma^{*}, \exists ne : \Sigma  ~|~ \\
 		\quad \left( 
		\begin{array}{c}
 		t' = t_1 \cat \trace{ne}  \cat t_3 ~~\land~~ t_1 \leq t \land \\
 		ne \in inputs \land ne \notin in(T,t_1 ) \land\\
 		\cvg{t_1 \cat t_3}{t}
 		\end{array}
		\right)
	\end{array}
	\right)
\end{array}	
\end{align*}
}
\end{definition}

Based on the definition of convergent traces, we are now able to define behavioural convergence.

\begin{definition}[I/O convergent behaviour] \label{def:I/O convergent behaviour}
Consider two I/O process $T$ and $T'$.  $T'$ is an I/O convergent behaviour of $T$ ($ \refcvg{T'}{T}$) if, and only if:
\vspace*{-.2cm}
{\small \begin{align*}
\forall (t',X) \in \fmodel(T'), \exists(t,Y) \in \fmodel(T)  \spot 
	   \left( 
		\begin{array}{c}
 		\cvg{t'}{t} \land \\
 		Y \cap inputs \supseteq X \cap inputs \land \\
 		Y \cap outputs \subseteq X \cap outputs\\
 		\end{array}
		\right)
\end{align*}}
\end{definition}

An I/O process $T'$ is convergent to $T$ if, for any trace $t' \in \tmodel(T')$, there exists a trace $t \in \tmodel(T)$, such that $\cvg{t'}{t}$, and $T'$ after $t'$ can offer more or equal inputs ($Y \cap inputs \supseteq X \cap inputs$) but is restricted to offer less or equal outputs ($Y \cap outputs \subseteq X \cap outputs$) when compared with $T$ after $t$. 

\begin{figure*}[t]  

\resizebox{.92\textwidth}{!}{
\begin{tabular}{lc}
\begin{minipage}{1.4\textwidth}
\begin{subfigure}{\textwidth}
  \includegraphics[scale=0.9]{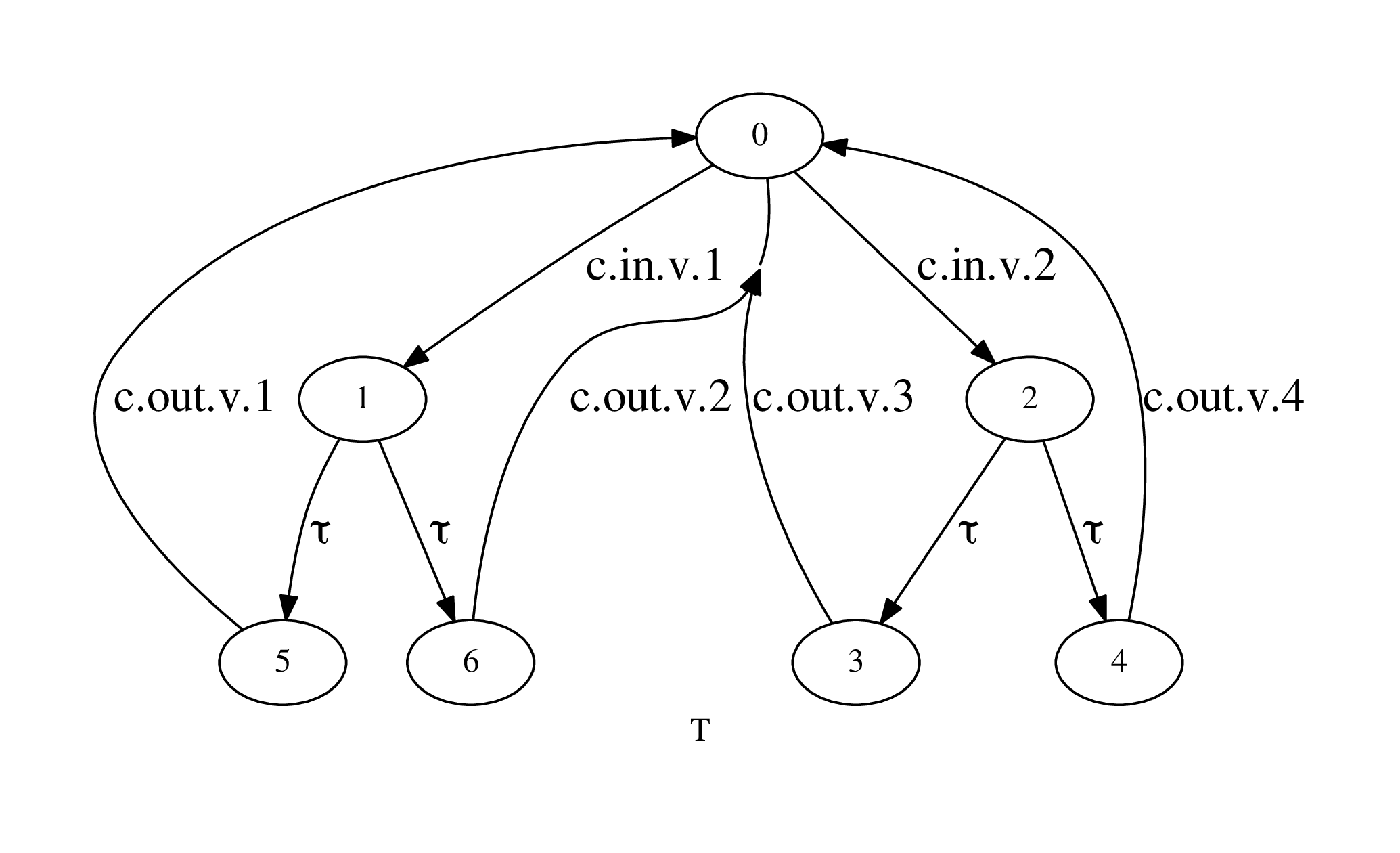}
  \vspace*{-.5cm}
  \caption{{\Huge Original process}}
  \label{fig:original_t}
  \end{subfigure}
\end{minipage}
&
 \begin{minipage}{1.4\textwidth}
  \begin{subfigure}{\textwidth}
 \includegraphics[scale=0.9]{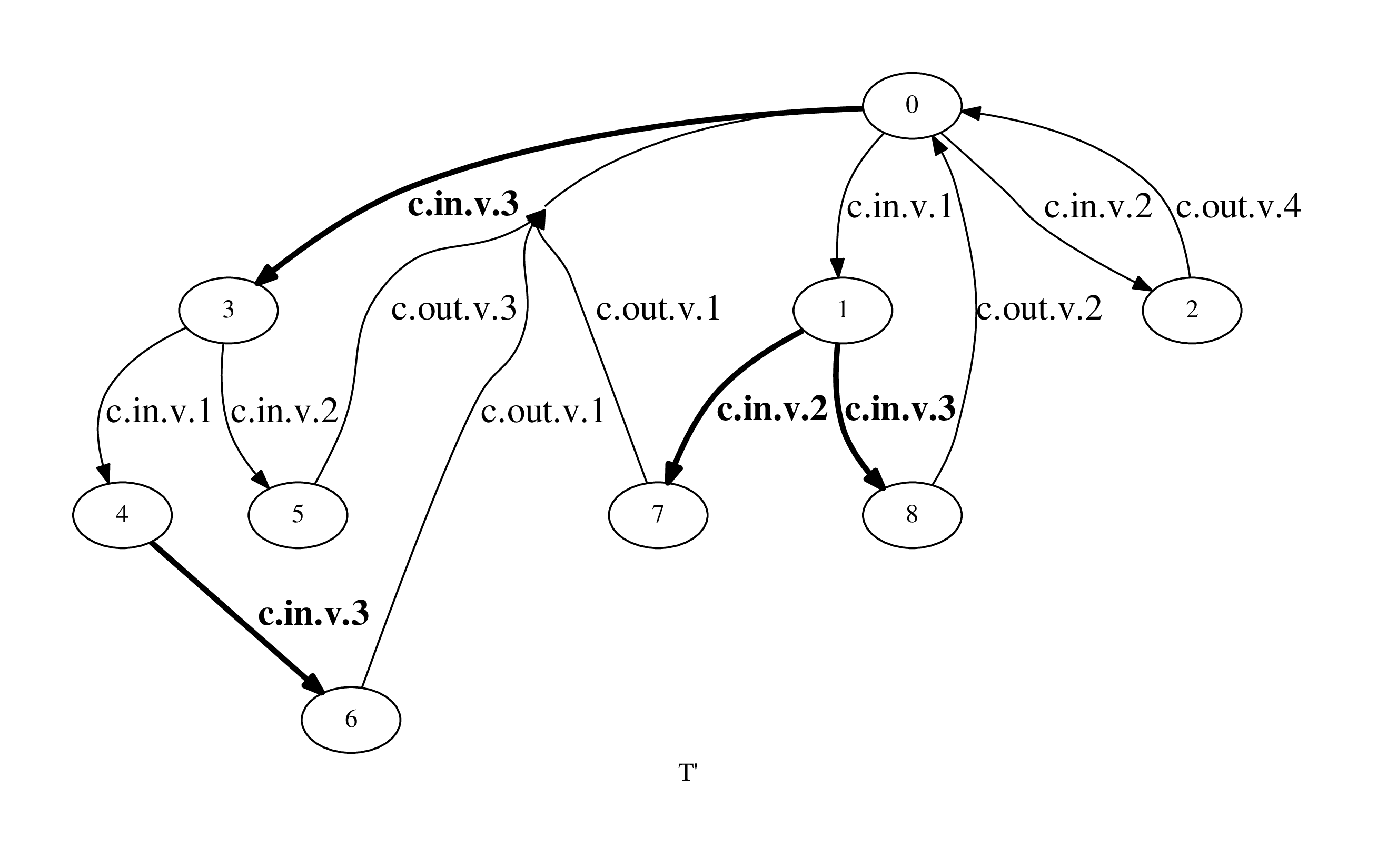}
  \vspace*{-.5cm}
  \caption{{\Huge Convergent process}}
  \label{fig:cvg_t}
 \end{subfigure}
\end{minipage}
\\
 \begin{minipage}{1.2\textwidth}
  \begin{subfigure}{\textwidth}
 \includegraphics[scale=0.9]{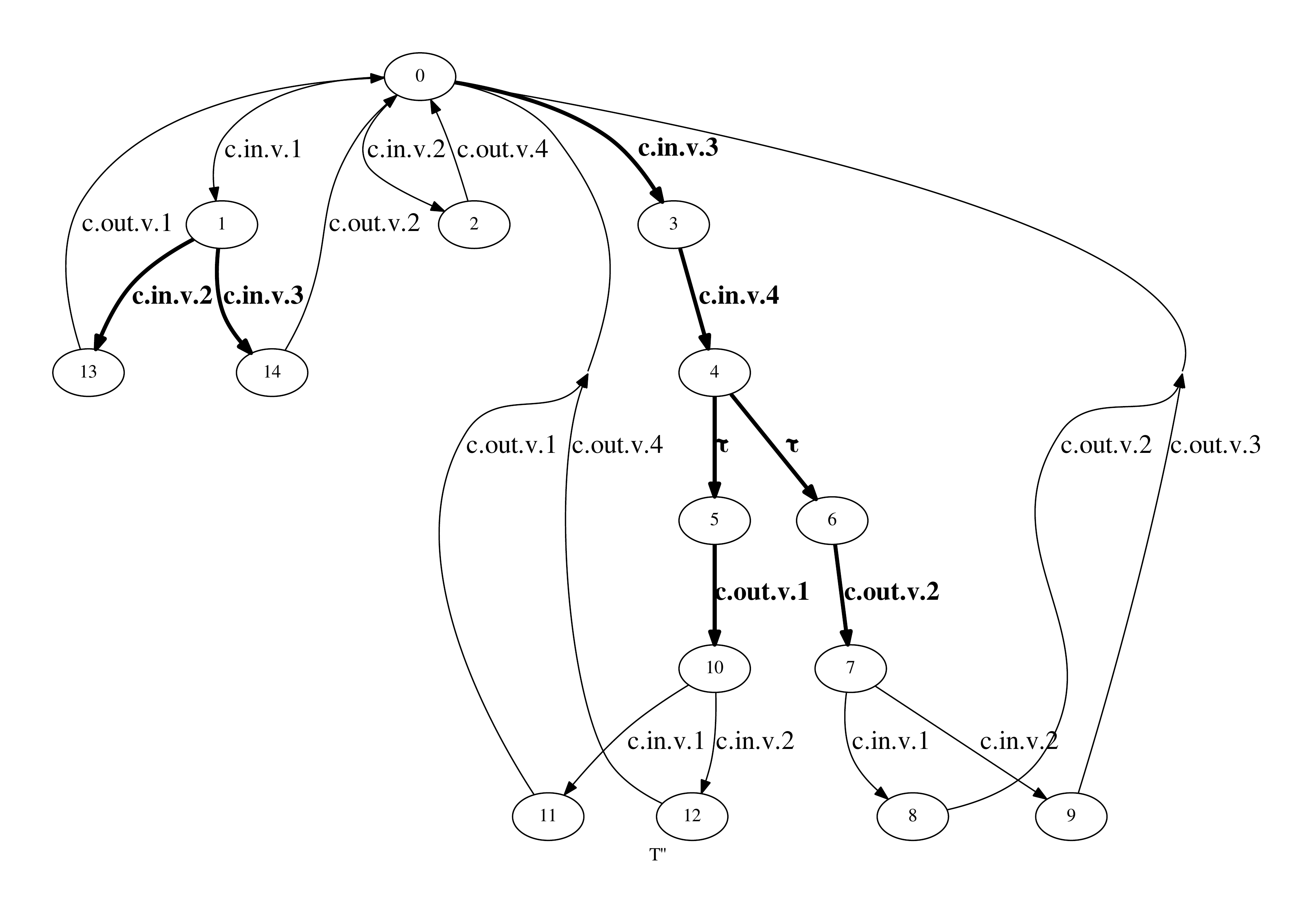}
   \vspace*{-.7cm}
  \caption{{\Huge Extended convergent process}}
  \label{fig:ecvg_t}
 \end{subfigure}
\end{minipage}
\end{tabular}
}
\caption{I/O convergent behaviours}
\end{figure*}

Convergence is a more restrictive relation than those based solely on covariation of inputs and contravariation of outputs (such as ioco \cite{Petrenko04}), because it requires, after a new-in-context event, the process to converge to its parent, which allows extensions but ensures substitutability as we discuss later. Let us consider the I/O processes $T$ (Listing \ref{lst:cvg:T}) and $T'$ (Listing \ref{lst:cvg:Tcvg}), whose LTSs are depicted in Figures \ref{fig:original_t} and \ref{fig:cvg_t}, respectively. Recall that  the suffixes \texttt{in} and \texttt{out} are used to mark input and output events, respectively.

\begin{lstlisting}[frame=single, mathescape=true, caption={I/O process T} , label={lst:cvg:T}]
T = c.in.v.1 -> (c.out.v.1 -> T |~| c.out.v.2 -> T)
    []
    c.in.v.2 -> (c.out.v.3 -> T |~| c.out.v.4 -> T)
\end{lstlisting}

\begin{lstlisting}[frame=single, escapeinside={(*}{*)},caption={I/O process T'(version 1)}, label={lst:cvg:Tcvg}]
T' = c.in.v.1 -> ((*\bf{c.in.v.2}*) -> c.out.v.1 -> T' 
                 [] 
                 (*\bf{c.in.v.3}*) -> c.out.v.2 -> T')
    []
    c.in.v.2 ->  c.out.v.4 -> T'
    []  	
    (*\bf{c.in.v.3}*) -> (c.in.v.1 ->  (*\bf{c.in.v.3}*) -> c.out.v.1 -> T' 
    	         [] c.in.v.2 -> c.out.v.3  -> T')
\end{lstlisting}

\begin{lstlisting}[frame=single, escapeinside={(*}{*)}, caption={I/O process T'(version 2)}, label={lst:cvg:Tecvg}]
T' = c.in.v.1 -> ( (*\bf{c.in.v.2}*) -> c.out.v.1 -> T' 
                 [] (*\bf{c.in.v.3}*) -> c.out.v.2 -> T')
     []
     c.in.v.2 ->  c.out.v.4 -> T' 
     []
     (*\bf{c.in.v.3}*) -> (*\bf{c.in.v.4}*) -> ((*\bf{c.out.v.1}*) -> 
                               (c.in.v.1 -> c.out.v.1 -> T'  
                               [] c.in.v.2 -> c.out.v.4 -> T')
                             |~| 
                             (*\bf{c.out.v.2}*) -> 
                               (c.in.v.1 -> c.out.v.2 -> T' 
                               [] c.in.v.2 -> c.out.v.3 -> T')) 
\end{lstlisting}

Based on Definition \ref{def:I/O convergent behaviour}, we have that $\refcvg{T'}{T}$. To explain why this is the case, let $(t',X)$ and $(t,Y)$ be failures of $T'$ and $T$, respectively. Then, by a non-exhaustive analysis, where $\eset{c}$ stands for all events that can be communicated through the channel $c$: 

\begin{itemize}
\item let $(t',X) = (\seq{c.in.v.3, c.in.v.1, c.in.v.3}, \eset{c}\ssub \set{c.out.v.1})$, which means that after trace $t'$, $T'$ can only communicate $c.out.v.1$, rejecting everything else (i.e., $\eset{c}\ssub \set{c.out.v.1}$). Considering that $(t,Y)$ $= (\seq{c.in.v.1}, \eset{c}\ssub $ $\set{c.out.v.1})$ is a failure of $T$, we have $\cvg{t'}{t}$ as $c.in.v.3$ is a new-in-context input of $T$ in both states 0 and 3 of its LTS (see Figure \ref{fig:original_t}); also $Y \cap inputs = X \cap inputs = \eset{c.in}$ and $Y \cap outputs = X \cap outputs = \eset{c.out} \ssub \set{c.out.v.1}$;

\item let $(t',X) = (\seq{c.in.v.1}, \eset{c}\ssub \set{c.in.v.2, c.in.v.3})$, which means that after trace $t'$, $T'$ can only communicate $c.in.v.2$ or $c.in.v.3$  rejecting everything else. Considering that $(t,Y) = ( \langle c.in.$ $v.1 \rangle, \eset{c}\ssub \set{c.out.v.2})$ is a failure of $T$, we have $\cvg{t'}{t}$ as equal traces are also convergent by definition; also $X \cap inputs = \eset{c.in}\ssub$ $\{c.in.v.2, c.in.v.3\}$, $X \cap outputs = \eset{c.out}$, $Y \cap inputs = \eset{c.in}$ and $Y \cap outputs = \eset{c.out} \ssub$ $\set{c.out.v.2}$.
\end{itemize}

A convergent I/O process can engage in more inputs so that, when converging, it can take more deterministic decisions on what to output. Nevertheless, it can be useful to offer other events after a new input and before converging to its original behaviour according to the relation. This extension to convergence allows convergent processes to add more implementation details. We define this relation in the traces and failures behavioural models in Definitions \ref{def:I/O extended convergent traces} and \ref{def:I/O extended convergent behaviour}, respectively.
\\
\\
\begin{definition}[I/O extended convergent traces]\label{def:I/O extended convergent traces}
Consider two I/O processes $T$ and $T'$. Let  $t$  and $t'$  be two of their traces, respectively. We say that $t'$ is an I/O extended convergent trace of $t$ ($\ecvg{t'}{t}$) if and only if:
\vspace*{-.2cm}
{\small \begin{align*}
\begin{array}{l}
	(t' = t) \lor \left( 
	\begin{array}{l}
 	(\# t' > \# t) ~~\land~~ \exists t_1, t_2, t_3 :  \Sigma^{*}, \exists ne \in \Sigma  ~|~ \\
 		\quad \left( 
		\begin{array}{c}
 		t' = t_1 \cat \trace{ne} \cat t_2  \cat t_3 \land t_1 \leq t \land \\
 		ne \in inputs \land ne \notin in(T,t_1 ) \land\\
 		set(t_2) \cap (in(T, t_1 ) \cup out(T, t_1 )) = \emptyset \land\\
 		\ecvg{t_1 \cat t_3}{t}
 		\end{array}
		\right)
	\end{array}
	\right)
\end{array}	
\end{align*}
}
\end{definition}

A trace $t'$ is I/O extended convergent to $t$ if they are the same or if it is possible to equate them by concealing each event $ne$ that is offered by $T'$ but not for $T$ after a common subtrace, say $t_1$, of $t$ and $t'$ ($ne \notin in(T,t_1 )$, but $ne \in in(T',t_1 )$); furthermore, since we allow more events after a new input $ne$, we also conceal them ($set(t_2) \cap (in(T, t_1 ) \cup out(T, t_1 )) = \emptyset$).   

\begin{definition}[I/O extended convergent behaviour]\label{def:I/O extended convergent behaviour}
Consider two I/O processes $T$ and $T'$. We say that $T'$ is an I/O extended convergent behaviour of $T$ ($\refecvg{T'}{T}$), if and only if:
\vspace*{-.2cm}
{\small \begin{align*}
\forall (t',X) \in \fmodel(T'),   \exists(t,Y) \in \fmodel(T)  \spot
 \ecvg{t'}{t} \land \left( 
	 \begin{array}{c}  
	  \left( 
		\begin{array}{c}
 		Y \cap inputs \supseteq X \cap inputs \land \\
 		Y \cap outputs \subseteq X \cap outputs\\
 		\end{array}
 	   \right) \\
 	   	\begin{array}{l}
        \lor ( \Sigma \backprime Y \subseteq X )
         \end{array}
   \end{array}
 	   \right)      
\end{align*}}

\end{definition}

Definition \ref{def:I/O extended convergent behaviour} is very similar to Definition \ref{def:I/O convergent behaviour}, but allows the extended convergent process $T'$ to accept any event not expected by $T$ ($\Sigma \backprime Y \subseteq X$) in an extended convergent point of their execution ($\ecvg{t'}{t}$), provided a new input $ne$ (see Definition \ref{def:I/O extended convergent traces}) has happened, marking the start of the extended convergent behaviour of $T'$. 

We illustrate this definition with an example. Let us consider the processes $T$ (Listing \ref{lst:cvg:T}) and $T''$ (Listing \ref{lst:cvg:Tecvg}), whose LTSs are depicted in Figures \ref{fig:original_t} and \ref{fig:ecvg_t}, respectively. By Definition \ref{def:I/O extended convergent behaviour}, we have that $\refecvg{T''}{T}$. To gather some evidence that this is the case, we analyse a couple of failures of these processes. Assume that $(t',X) \in \fmodel(T'')$ and $(t,Y) \in \fmodel(T)$, then:

\begin{itemize}
\item  considering $(t', X)= (\trace{c.in.v.3,c.in.v.4}, \eset{c} \ssub \set{c.out.v.1})$, we can find $(t, Y) = (\emptyseq, $ $\eset{c} \ssub \{c.in. \allowbreak v.1, c.in.v.2\})$, such that $\ecvg{t'}{t}$, $\Sigma \backprime Y = \set{c.in.v.1, \allowbreak c.in.v.2}$ and, therefore,  $\Sigma \backprime Y \subseteq X$;

\item if $(t', X)= (\trace{c.in.v.1,c.in.v.3}, \eset{c} \ssub \set{c.out.v.2})$, we can find the failure $(t, Y) = (\trace{c.in.v.1}$ $, \eset{c}$ $\ssub \set{c.out.v.2})$, such that $\ecvg{t'}{t}$, $X=Y$ and, therefore, Definition \ref{def:I/O extended convergent behaviour} holds;

\item finally, assuming $(t', X)= (\trace{c.in.v.2}, \eset{c} \ssub \set{c.out.v.4})$, we can find the failure $(t, Y)= (\trace{c.in.v.2}, \eset{c} \ssub \set{c.out. \allowbreak v.4})$ and Definition \ref{def:I/O extended convergent behaviour} trivially holds.
\end{itemize}

As one might expect, extended convergence is a generalisation of convergence, which comes from Lemmas \ref{lem:cvg_implies_ecvg_on_prefixing}, \ref{lem:cvg subseteq ecvg} and \ref{lem:io_cvg subseteq io_ecvg} proved in \ref{appendix:Proofs}. Consider two traces with a common prefix; the first lemma ensures that if one of these traces is convergent to the other, starting from that common prefix trace, it is also extended convergent.

\begin{lemma}[$cvg$ implies $ecvg$ on trace prefixing]\label{lem:cvg_implies_ecvg_on_prefixing} Consider two I/O proc-- esses $T$ and $T'$ such that, $t_1 \cat t_3 \in \tmodel(T)$ and $t' \in \tmodel(T')$. If $\cvg{t'}{t_1 \cat t_3}$, where $t_1 \leq t'$, then $\ecvg{t'}{t_1 \cat t_3}$.
\end{lemma}

Lemma \ref{lem:cvg subseteq ecvg} formalises extended convergence as a generalisation of convergence; therefore, if two traces are convergent they are also extended convergent. Lemma \ref{lem:io_cvg subseteq io_ecvg} proves the same for I/O processes.

\begin{lemma}[$\mathtt{cvg} \subseteq \mathtt{ecvg}$] \label{lem:cvg subseteq ecvg} Consider two I/O processes $T$ and $T'$, and $t$ and $t'$ such that $t \in \tmodel(T)$ and $t' \in \tmodel(T')$. If $\cvg{t'}{t}$ then $\ecvg{t'}{t}$.

\end{lemma}

\begin{lemma}[$\mathtt{io \un cvg} \subseteq \mathtt{io \un ecvg}$] \label{lem:io_cvg subseteq io_ecvg} Consider two I/O processes $T$ and $T'$. If we have $\refcvg{T'}{T} $ then $\refecvg{T'}{T}$.
%
\end{lemma}

\subsection{Extensibility}\label{subsec:Extensibility, refinement and substitutability}

Our definition of inheritance deals with component structural and be\-hav\-ior\-al aspects. Structurally, it guarantees that the inheriting component preserves at least its parent's channels and their types: if $T'$ extends $T$ we have that $\cc{R}{T} \subseteq \cc{R}{T'}$. Regarding behaviour, they are related by convergence. Additionally, it guarantees, for substitutability purposes, that the inherited component $T'$ refines the protocols exhibited by common channels (default channel congruence, as in Definition \ref{def:Default channel congruence}) or that additional inputs (new in context, see Definitions \ref{def:I/O convergent behaviour} and \ref{def:I/O extended convergent behaviour}) over common channels are not exercised by any possible client of its parent $T$ (input channel congruence, Definition \ref{def:Input channel congruence}).

The $\BRIC$ component model restricts how components can be assembled to avoid deadlock. Channel congruence aims at paving a safe way to extend a specification by using convergence without introducing deadlock; it does not reduce possible inputs, but disciplines the way in which existing inputs can be used in convergent extensions. The simplest, but restrictive, form of achieving this is by guaranteeing that the protocol over a channel must be refined, in the failures classical sense (Definition \ref{def:Default channel congruence}).

\begin{definition}[Default channel congruence] \label{def:Default channel congruence} An  I/O process $T_{dc}$ has a default congruent channel, say $c$, to another I/O process $T$ ($\defcong{T_{dc}}{T}{c}$), when there is a failures refinement relation between their projections over $c$:

\vspace*{-.8cm}

\begin{align*}
& \fmodel(T_{dc} \hide (\Sigma ~\backprime~ \set{c}))  \subseteq \fmodel(T \hide (\Sigma ~\backprime~ \set{c})), \\& \quad \quad \text{ which is equivalent to } \fmodel(T_{dc} \project \set{c})  \subseteq \fmodel(T \project \set{c})
\end{align*}
\end{definition}

A more flexible way is given by Definition \ref{def:Input channel congruence}: I/O processes $T_{ic}$ and $T$ are input congruent on an I/O channel $c$ if, after both have done the same trace $t$, either of the following holds:

\begin{enumerate}[i.]
\item if $T$ cannot engage in any input, then $T_{ic}$ cannot input on $c$: it avoids $T$'s clients from deadlocking when interacting with $T_{ic}$, since these clients do not communicate (after the trace $t$) outputs on $c$ to $T$, otherwise they deadlock; therefore $T_{ic}$, as $T$, cannot expect inputs on $c$ after the trace $t$;

\item  if $T$ is not able to input on $c$, $T_{ic}$ can only do it for events outside $T$'s alphabet: it avoids $T$'s clients to engage in a possible unexpected communication over $c$, which can (but not necessarily) lead to deadlock. It is worth saying that in places where $T$ can input over $c$, $T_{ic}$ can also input new-in-context events over $c$; it is possible because they are offered in external choice, so $T$'s clients will not be able to communicate these new-in-context inputs offered by $T_{ic}$. This happens because $T$'s clients are not ready to engage in these new-in-context inputs; otherwise, this would mean that their composition with $T$ deadlocks, because $T$ is not able to offer such new-in-context inputs.
\end{enumerate}

An I/O process $T_{ic}$ has an input congruent channel $c$ to an I/O process $T$ ($\inpcong{T_{ic}}{T}{c}$), if after a common trace $t$, such that $(t,X) \in \fmodel(T_{ic})$ and $(t,Y) \in \fmodel(T)$, the following holds: $T$ refuses to input ($inputs \subseteq Y$) and $T_{ic}$ refuses to input over $c$ ($\eset{c.in} \subset X$) or $T$ refuses to input over $c$ ($\eset{c.in} \subseteq Y$) and $T_{ic}$ refuses the events on $c$ that can be communicated by $T$ ($\eset{c.in} \ssub (\alpha T_{ic} \ssub \alpha T) \subset X$), where $\alpha T_{ic}$ and $\alpha T$ stand for the alphabets of $T_{ic}$ and $T$, respectively. The formal definitions is as follows.
 
\begin{definition}[Input channel congruence]  \label{def:Input channel congruence}  Given two I/O processes $T_{ic}$ and $T$, and a channel $c$, we say that $\inpcong{T_{ic}}{T}{c}$ if, and only if:
\vspace*{-.6cm}

\allowdisplaybreaks
\begin{align*}
& \forall (t,X) \in \fmodel(T_{ic}) \spot \exists (t,Y) \in \fmodel(T) \implies
 \left( \begin{array}{c}
   inputs \subseteq Y \  \implies \   \eset{c.in} \subset X\\
   \lor\\
  \eset{c.in} \subseteq Y \   \implies \   \eset{c.in} \ssub (\alpha T_{ic} \ssub \alpha T) \subset X\\
\end{array} \right) \\
& \textbf{provided}\\
& (\exists (t,X') \in  \fmodel(T_{ic}) \implies X' \subseteq X) ~~\land~~ (\exists (t,Y') \in  \fmodel(T) \implies Y' \subseteq Y)
\end{align*} 
\end{definition}

\noindent In the definition above, the proviso guarantees that $(t,X)$ and  $(t,Y)$ are maximal failures.

The following is the most important definition of this work, the component inheritance relation, which allows behaviour extension and, moreover, guarantees substitutability.

\begin{definition}[$\BRIC$ inheritance]\label{def:BRIC inheritance}
Consider $T$ and $T'$ two $\BRIC$ components, such that $\cc{R}{T} \subseteq \cc{R}{T'}$. We say that $T'$ inherits from $T$:
\allowdisplaybreaks
\begin{align*}
& \spot \text{by convergence:  } \ T \subcvg T' \iff \refcvg{\cc{B}{T'}}{\cc{B}{T}}\\
& \spot \text{by extended convergence:  } \  T \subecvg T' \iff \refecvg{\cc{B}{T'}}{\cc{B}{T}}\\
& \textbf{provided} \\
& \forall c : \cc{C}{T} \spot (\defcong{\cc{B}{T'}}{\cc{B}{T}}{c}) \lor  (\inpcong{\cc{B}{T'}}{\cc{B}{T}}{c})
\end{align*}
\end{definition}

The provided clause guarantees that, when interacting with $T'$, a component originally designed to interact with $T$ will not engage in $T'$ extensions trigged by inputs also used by $T$, which in $T'$ have a different meaning.

\subsection{Semantics and refinement}

In this section, we contribute with a denotational semantics and a refinement relation for $\BRIC$. Furthermore, we show that the refinement and inheritance relations form a hierarchy. We also prove that our relations preserve deadlock freedom and moreover, that they respect the substitutability principle. We start by defining a function $\sm{\cdot}$ from a $\BRIC$ component to an underpinning mathematical model. Consider the component $\ctdef{T}{B}{R}{I}{C}$, the semantics of $T$ is given by:

\vspace*{-1cm}

\begin{align*}
\sm{T} & = (\failures(\mathcal{B}), \{ (c,\failures(\mathcal{B} \project \set{c.i})) ~|~ c \in \mathcal{C} \land i \in \ce{I} \land (c,i) \in \mathcal{R}\})
\end{align*}

This semantics captures the relevant properties of a $\BRIC$ component: its overall behaviour (given by the failures of the I/O process $\mathcal{B}$ that defines the component behaviour) and those exhibited through its channels (a set of pairs where each channel $c$ maps into the failures of the overall behaviour projected to $c$); such projected behaviours are crucial in composition rules. Whenever the type of a particular channel $c$ is known we can simplify its semantics to $\sm{T} = (\failures(\mathcal{B}), \{ (c,\failures(\mathcal{B} \project c.i)) ~|~ c \in \mathcal{C} \})$. It is important to note that we are not presenting  $\sm{\cdot}$ in a compositional manner, by induction on the process structure; rather, the more concise and simpler presentation is sufficient in this work. With a component semantics we can define a refinement notion (monotonic with respect to the $\BRIC$ composition rules \cite{COMPASS}).

\begin{definition}[$\BRIC$ refinement]\label{def:BRIC refinement based on failures}
Consider two components $T$ and $T'$.  If we consider $\sm{T} = (f,fp)$ and $\sm{T'} = (f',fp')$, with $f$ and $f'$ standing for the overall behaviour of $T$ and $T'$, and $fp$ and and $fp'$ for the projected behaviours on their channels, respectively, then we say that $T'$ refines $T$ ($T \frefbric T'$) if, and only if:

\vspace*{-1cm}

 \begin{align*}
(f' \subseteq f) \land (\dom fp = \dom fp') \land (\forall c \in \dom fp \spot fp'(c) \subseteq fp(c))
\end{align*}
\end{definition}

\vspace*{-.1cm}

This definition ensures that $T$ and $T'$ have the same interaction points. Moreover, it guarantees that the component behaviour of $T'$ refines that of $T$, which is equivalent to: 

\vspace*{-.8cm}

\begin{align*}
(\cc{B}{T} \frefinedby  \cc{B}{T'}) \land (\cc{C}{T} = \cc{C}{T'}) \land (\forall c: \cc{C}{T} \spot \cc{R}{T}(c) \subseteq \cc{R}{T'}(c))
\end{align*}

Theorem \ref{thm:Hierarchical relations} (proved in \ref{appendix:Proofs}) states that the refinement and inheritance relations form a hierarchy. Component refinement is the strongest relation between $\BRIC$ components; it implies inheritance. We can see refinement as the strongest form of inheritance. As expected, inheritance by convergence is a more strict form of inheritance than extended convergence, as their names suggest.

\begin{theorem}[Hierarchy] \label{thm:Hierarchical relations} The relations $\frefbric$, $\subcvg$ and $\subecvg$ form a hierarchy: $\frefbric ~~~\subseteq~~~ \subcvg ~~~\subseteq~~~ \subecvg$.
\end{theorem}

This concludes an important result that relates refinement and the notions of inheritance based on (extended) convergence. 


\subsubsection{Substitutability}
We prove, in Lemma \ref{lem:BRIC inheritance preserves deadlock freedom}, that $\BRIC$ inheritance preserves deadlock freedom. A more interesting result, proved by Theorem \ref{thm:Substitutability for BRIC inheritance}, guarantees that a component $T'$ can replace $T$, in any context produced by the $\BRIC$ composition rules, without introducing deadlock, provided $T'$ inherits by convergence from $T$. These results are proved in \ref{appendix:Proofs}.

\begin{lemma}[Inheritance preserves deadlock freedom]\label{lem:BRIC inheritance preserves deadlock freedom} Consider $T$ and $T'$ two $\BRIC$ components, such that $T$ is deadlock free. If $T \subecvg T'$ then $T'$ is deadlock free.

\end{lemma}

\begin{theorem}[Substitutability] \label{thm:Substitutability for BRIC inheritance}
Let $T$ and $T'$ be two components such that $T \subecvg T'$. Consider $S[T]$ a deadlock free component contract that includes, as part of its behaviour, a deadlock free component contract $T$; then $S[T']$, which stands for $S$ with $T$ replaced with $T'$, is deadlock free.
\end{theorem}

Note that this result also holds for the other two relations ($\frefbric$ and $\subcvg$), as a consequence of the hierarchy established by Theorem \ref{thm:Hierarchical relations}.

\section{Checking convergence} \label{sec:Checking convergence via refinement}

Behaviour convergence and the relations built on top of it are the backbone of this work; therefore, we must have an automated strategy to check whether two I/O processes are related by convergence. We start addressing this issue by choosing FDR4 (Failures-Divergence Refinement) \cite{RobinsonABR14} as the model-checker to carry out the analysis; it seems a natural choice given the widespread use of FDR4 both in academy and industry, which makes it a \textit{de facto} standard tool for analysing CSP specifications. Its method of establishing whether a property holds is to check for the refinement  between CSP specifications, internally represented by labelled transition systems.  

To check conformance of I/O Processes, say $P$ and $P'$, our strategy is to construct, for each relation, a verification strategy of the form:

\vspace*{-.8cm}

\begin{align*}
& \refcvg{P'}{P} \Longleftrightarrow GLB\_CVG(P) \frefinedby P'\\
& \refecvg{P'}{P} \Longleftrightarrow GLB\_ECVG(P) \frefinedby P'
\end{align*}

\vspace*{-.3cm}

To explain how these parametrised $GLB\_$ processes can be constructed, we need to present some additional background on failures refinement, convergence and I/O process alternative representations. Consider $P$ an I/O process, then $cvg^+ P$ stands for a set of I/O processes such that: $\forall P' \in cvg^+ P  \ \bullet \ \refcvg{P'}{P}$; it contains every I/O process convergent to $P$, including $P$ itself. The set $cvg^+ P$ is infinite, which makes its use prohibitive for any implementation that aims to traverse it. A finite subset of $cvg^+ P$ is given by $cvg^{+n} P$, which stands for the $P$ convergent processes whose depth differ from that of $P$ by at most $n $. An I/O process depth is given by the longest trace after which the process returns (for the first time) to its initial state or, by considering its LTS, the maximum number of transitions (labelled with visible events) from the initial state to itself. An I/O process depth can be equivalently expressed in two ways: (a) in terms of traces and failures-semantics or (b) based on its LTS representation, where $ P \Trans[t] P$ means that the I/O process $P$ returns to its initial state after the trace $t$:

\vspace*{-.7cm}

\begin{align*}
& \text{(a) } depth(P) =  max \set{\# t \ | \ t \in \tmodel(P) \land P \fequiv P / t \land (\nexists s < t \ | \  P \fequiv P / s) } \\
& \text{(b) } depth(P) =  max \set{\#t  \ | \ P \Trans[t] P \land (\nexists s < t \ | \ P \Trans[s] P)}
\end{align*}

\vspace*{-.1cm}

For example, the depths of the processes in Figures \ref{fig:original_t}, \ref{fig:cvg_t} and \ref{fig:ecvg_t} are, respectively, 2, 4 and 5. The core of our strategy is to build a CSP process $GLB$, such that it belongs to $cvg^{+n} P$ and every member $Q$ of $cvg^{+n} P$ refines it, $GLB \frefinedby Q$. Furthermore, if there is any other process, say $R$, which satisfies this property, then $R \frefinedby GLB$. It means that the process $GLB$ is the \textit{Greatest Lower Bound} \cite{Enderton77,Roscoe1998Theory} of the set $cvg^{+n} P$ under the CSP failures refinement relation ($\frefinedby$). Our intention is to construct $GLB\_CVG(P)$ to be failures equivalent to $GLB$, $GLB\_CVG(P) \fequiv GLB$. Therefore, to verify if a process $P'$ is convergent to $P$, i.e., if $P'$ belongs to $cvg^{+n} P$, one needs only to verify if $GLB\_CVG(P) \frefinedby P'$. The same reasoning applies to the extended convergence relation. The next section details how we construct $GLB\_CVG(P)$ and $GLB\_ECVG(P)$ for an I/O process $P$.

%
%
%

\subsection{Building $GLB\_{CVG}$}

Let $P$ and $P'$ be I/O processes that differ in depth by $n$. To test whether $\refcvg{P'}{P}$, we must build from $P$ a new process $GLB\_CVG(P)$, which must be able to do \textit{at most} $n$ new-in-context inputs in every state of $P$. Such a process is, by Definition \ref{def:I/O convergent behaviour}, convergent to $P$. As we stated before, any process convergent to $P$ (which differs in depth by at most $n$) must refine in failures $GLB\_CVG(P)$. An important practical question to build $GLB\_CVG(P)$ is to compute the new-in-context inputs for any state of $P$. Given a state of $P$, the easiest way of computing the new-in-context inputs available is to know which inputs can be accepted in this state; the result will be its complement. The problem is that CSP does not have a native mechanism for \textit{backtracking} a process execution: we cannot synchronise on an event and then go back to the state before this communication. A complicating factor is that we are dealing with new-in-context events, not new-in-alphabet events; if this were the case, alphabetised parallelism and hiding could be sufficient to compare the behaviours of the two components modulo the new events, as already demonstrated in \cite{Wehrheim02checkFDR}. 

To circumvent this problem we define an alternative representation for I/O processes, with a finite LTS. We serialise an I/O process as a sequence of tuples of the form $(ev, a\_ev, l)$, for a particular state, where the event $ev$ is possible from this state and if it happens the next state can accept the events in the sequence $a\_ev$, which is at the level $l+1$. The initial state's level is zero; each subsequent state has the level of its immediate predecessor increased by one. 


Let us consider a practical example on how an I/O process can be serialised. Consider the process $P$ in Figure \ref{fig:original_t}. We define two special events \verb"start" and \verb"end"; these are used only as marking events, and will not be part of the $GLB\_{CVG}$ behaviour, but will play a role in its construction: \verb"start" indicates that a process is ready to engage by offering its initial events; \verb"end" marks the state where it is ready to come back to its initial state. Line 1 of \verb"P_serial" (see Listing \ref{lst:Serialisation}) indicates that $P$ can, initially (state level zero) accept \verb"c.in.v.1" or \verb"c.in.v.2";  if \verb"c.in.v.1" happens (line 2, state level 1) then it can output (non-deterministically) \verb"c.out.v.1" or \verb"c.out.v.2"; if it outputs \verb"c.out.v.1" (line 3, state level 2) then it can only go back to its initial state (\verb"end"). Note that we follow a nested-structural pattern, which allows us to backtrack an I/O process by traversing its serial representation in a recursive manner.

\vspace*{-.2cm}

\begin{center}
\begin{minipage}{.75\textwidth}
\begin{lstlisting}[frame=single,framexleftmargin=15pt, numbers=left,caption={Serialisation},label={lst:Serialisation}]
P_serial = <(start, <c.in.v.1, c.in.v.2>, 0),
            (c.in.v.1, <c.out.v.1, c.out.v.2>, 1),
            (c.out.v.1, <end>, 2),
            (end, <>, 3),
            (c.out.v.2, <end>, 2),
            (end,<>, 3),
            (c.in.v.2, <c.out.v.3, c.out.v.4>, 1),
            (c.out.v.3, <end>, 2),
            (end, <>, 3),
            (c.out.v.4, <end>,2),
            (end, <>, 3)> 
\end{lstlisting}
\end{minipage}
\end{center}

\vspace*{-.1cm}

\begin{figure*}[t]
\resizebox{1\textwidth}{!}{
\begin{tabular}{r}
\begin{minipage}{1\textwidth}
\centerline{\includegraphics[scale=1]{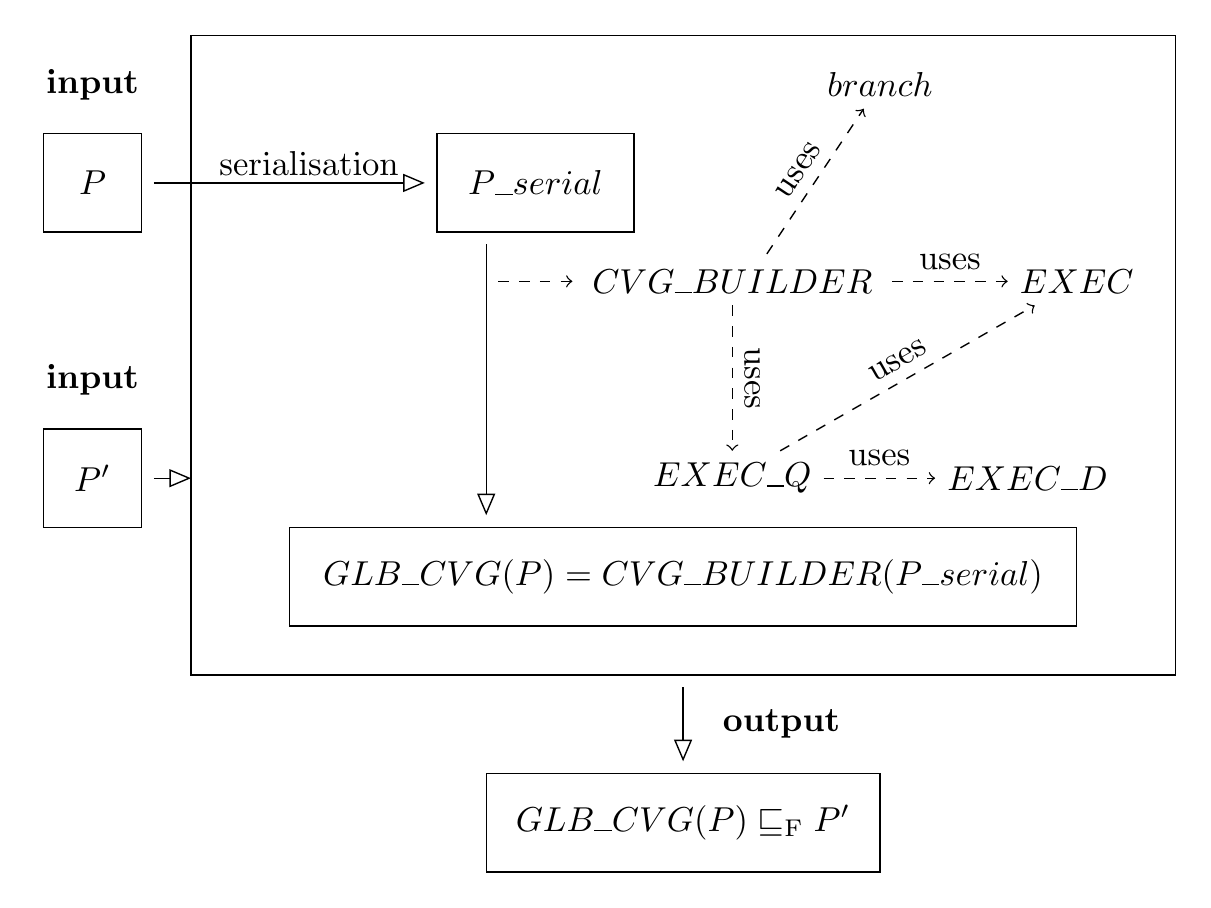}}
\end{minipage}
\end{tabular}
}
\caption{Conformance checking strategy}
\label{fig:validation_implementation}
\end{figure*}

A question that arises is how to deal with parallelism in such a representation. We take advantage of the fact that any parallel process has a unique sequential representation in terms of the operators $\then$, $\intchoice$ and $\extchoice$ \cite{Roscoe1998Theory}. This is the background we need to present our strategy (Figure \ref{fig:validation_implementation}). Given an I/O process $P$ we serialise it as $P\_serial$, which is passed to the process $CVG\_BUILDER$; $CVG\_BUILDER(P\_serial)$ is our strategy to build precisely $GLB\_CVG(P)$, by coordinating the processes $EXEC$ and $EXEC\_Q$. $CVG\_BUILDER$ traverses the $P\_serial$ tuples testing whether it has: (a) found an output event, which makes it offer, prior to this output, all inputs in internal choice iteratively, for $n$ times (behaving as $EXEC$), then recursing on the next branch of $P\_serial$; (b) found an input event, in which case it offers in internal choice iteratively, for $n$ times (by using $EXEC\_Q$) the complementary inputs, while ensuring this complement do not take precedence over expected inputs; (c) reached  $P\_serial$ end, then recursing to its start. Processes $EXEC$ and $EXEC\_Q$ rely on the $branch$ function to traverse all behavioural paths of $P$ (subsequences of $P\_serial$). The construction of $GLB\_CVG(P)$ ensures it offers, at any point, at least $n$ new-in-context inputs, therefore, a process $P'$ convergent to $P$ (differing in depth by at most $n$) must refine $GLB\_CVG(P)$. 

We detail these processes in the sequel, using the CSP syntax (\ref{appendix:CSP}). The process $EXEC(evs,n)$, in Listing \ref{lst:EXEC}, offers the internal choice between the events $evs$ iteratively, up to $n$ times. The process $EXEC\_D$ (Listing \ref{lst:EXEC_D}) is quite similar to $EXEC$ but differs as it offers $evs$ in external choice.

\vspace*{-.7cm}

\setcounter{algocf}{4}

\begin{center}
\resizebox{1\textwidth}{!}{
\begin{tabular}{ll}
\begin{minipage}{1\textwidth}
\begin{algorithm*}[H]
EXEC(evs,n) = \\
 \eIf {$n > 0$} 
  { $\Intchoice$ x : evs $\at$ ((x $\then$ EXEC(evs, n-1)) $\intchoice$ \\
  \Indp \Indp  \Indp \Indp \Indp  \Indp  EXEC(evs, n-1))}
 {SKIP} 
 \caption{EXEC}
 \label{lst:EXEC}
\end{algorithm*}
\end{minipage}
&
\begin{minipage}{1\textwidth}
\begin{algorithm*}[H]
EXEC$\_$D(evs,n) = \\
 \eIf {$n > 0$} 
 {$\Extchoice$ x : evs $\at$ ((x $\then$ EXEC$\_$D(evs, n-1)) $\extchoice$  \\
  \Indp \Indp  \Indp \Indp \Indp  \Indp  EXEC$\_$D(evs, n-1))}
 {SKIP}
 \caption{EXEC$\_$D}
 \label{lst:EXEC_D}
\end{algorithm*}
\end{minipage}
\end{tabular}
}
\end{center}

The process $EXEC\_Q$ (Listing \ref{lst:EXEC_Q}) combines $EXEC$ and $EXEC\_D$ in parallel. Let $evs1$ and $evs2$ be sets of events such that $evs2 \subseteq evs1$, then $EXEC$ offers $evs1 \ssub evs2$ in internal choice and $EXEC\_D$ offers $evs1$ in external choice, iteratively, up to $n$ times, synchronising, in each step, on the set $evs1 \ssub evs2$.

\vspace*{-.8cm}

\begin{center}
\resizebox{1\textwidth}{!}{
\begin{tabular}{ll}
\begin{minipage}{1\textwidth}
\LinesNumberedHidden
\begin{algorithm*}[H]
EXEC$\_$Q(evs1, evs2, n) = \\
\Indp EXEC(diff(evs1, evs2),n)\\
$~~~~~\parallel[diff(evs1,evs2)]$ \\
EXEC$\_$D(evs1,n)\\            
\caption{process EXEC$\_$Q}
\label{lst:EXEC_Q}
\end{algorithm*}

\end{minipage}
&
\begin{minipage}{1\textwidth}
\LinesNumberedHidden
\begin{algorithm*}[H]
channel start, end  \\
e1((e,$\_$,$\_$)) = e \\
e2(($\_$,e,$\_$)) = e \\
e3(($\_$,$\_$,e)) = e \\
subset(s1,s2) = empty(diff(s1,s2))
\caption{Helper functions}
\label{lst:functions_and_channels}
\end{algorithm*}
\end{minipage}
\end{tabular}
}
\end{center}

Given a 3-tuple of an I/O process serialised representation, the functions $e1$, $e2$ and $e3$ (Listing \ref{lst:functions_and_channels}) yield the first, the second and the third element of the tuple, respectively. We declare the aforementioned channels $start$ and $end$ and define the function $subset(s1,s2)$ to check whether $s1$ is a subset of $s2$.

The $branch$ function (Listing \ref{lst:branch}) yields a local view of a serialized I/O process: giving some event $key$ at a particular level $l$, it provides the tuples (a branch) reached from $key$ (at level $l$) until the end mark (the event $end$). For example, considering \verb"P_serial" (Listing \ref{lst:Serialisation}), the $branch$ of the event \verb"c.in.v.1", at the level one, and of the event \verb"c.out.v.2", at the level two, gives, respectively, the local serialised views at Listings \ref{lst:branchLevel1} e  \ref{lst:branchLevel2}. 

\vspace*{.2cm}

\setcounter{lstlisting}{8}
 
\begin{minipage}{.5\textwidth}
\begin{lstlisting}[frame=single,numbers=left,framexleftmargin=12pt, caption={Branch at level one}, label={lst:branchLevel1}]
<(c.in.v.1, <c.out.v.1, c.out.v.2>, 1), 
 (c.out.v.1, <end>, 2),
 (end, <>, 3), 
 (c.out.v.2, <end>, 2), 
 (end, <>, 3)>
\end{lstlisting}
\end{minipage}
\hfill
\begin{minipage}{.35\textwidth}
\begin{lstlisting}[frame=single,numbers=left, framexleftmargin=12pt, caption={Branch at level two}, label={lst:branchLevel2}]
<(c.out.v.2, <end>, 2), 
 (end, <>, 3)>
\end{lstlisting}
\end{minipage}

The $branch$ function (Listing \ref{lst:branch}) works by finding the tuples, in a sequence $s$, whose events are offered together, by looking for one of them, say $key$, at a specific level $l$. If $s$ is empty (line 2) the search is finished and the result remains unchanged (line 3). Otherwise, we check if the current tuple has the event at the level we are looking for (line 5); if it is the case, we append this tuple to the result and call $branch$ recursively, passing to it the remainder of the sequence and setting the parameter $b$ (marking the first occurrence of a tuple with $key$) to $true$ (line 6). 

If we have found the key event ($b$ is true, line 8), we must check if we have not reached another branch of the process LTS at the same level (line 9), which marks the end of our search (line 10), otherwise we append such a tuple to the result and call $branch$ to the rest of the sequence (line 12). If we have not yet found the $key$, we maintain the result unchanged and recursively call $branch$ to the rest of the sequence (line 15).

\vspace*{-.7cm}

\setcounter{algocf}{10}

\begin{center}
\resizebox{1\textwidth}{!}{
\begin{tabular}{ll}
\begin{minipage}{1\textwidth}
\LinesNumbered
\begin{algorithm*}[H]
branch(key,$<>$, l, b) = $<>$\\
branch(key,s, l, b) =  \\
    \eIf {(e1(head(s)) == key and e3(head(s)) == l)} {$<$head(s)$>$ $\cat$ branch(key,tail(s),l, true)}
       { 
        \eIf {b} 
          {  \eIf {(e1(head(s)) != key and e3(head(s)) == l)} {$<>$}
                {$<$head(s)$>$ $\cat$ branch(key,tail(s),l, b)}}
          {branch(key,tail(s),l, b)}
       }             
\caption{branch function}
\label{lst:branch}
\end{algorithm*}
\end{minipage}
&
\begin{minipage}{1\textwidth}
\begin{algorithm*}[H]
CVG$\_$BUILDER(src,crt) = \\
\eIf {e1(head(crt))==end} {CVG$\_$BUILDER(src,src)}
{
  \eIf {subset(set(e2(head(crt))), inputs)}  
    {EXEC$\_$Q(inputs, set(e2(head(crt))), GAP); CVG$\_$BUILDER(src,crt))\\
    $\intchoice$\\
    $\Extchoice$ x:set(e2(head(crt))) $\at$ x $\then$\\
    \Indp CVG$\_$BUILDER(src, branch(x,tail(crt), e3(head(crt))+1, false)
    } 
    {EXEC(inputs,GAP);\\
     \Indp $\Intchoice$ x:set(e2(head(crt))) $\at$ x $\then$\\
     \Indp \Indp CVG$\_$BUILDER(src, branch(x,tail(crt),   e3(head(crt))+1, false))}
}
\BlankLine
GLB$\_$CVG(P) = CVG$\_$BUILDER(P$\_$serial,P$\_$serial) $\hide$ $\{$end$\}$
\caption{CVG$\_$BUILDER and GLB$\_$cvg}
\label{lst:CVG_BUILDER}
\end{algorithm*}
\end{minipage}
\end{tabular}
}
\end{center}

The process  $CVG\_BUILDER$ (Listing \ref{lst:CVG_BUILDER}) coordinates the auxiliary processes we have seen in the procedure of building $GLB\_CVG(P)$ for an I/O process $P$. Let $src$ be the serialised representation of $P$ and $crt$ the serialisation of a $P$'s current execution branch. Both are parameters of the $CVG\_BUILDER$ process and are equal initially. If we have found the end of a branch (line 2),  $CVG\_BUILDER$ returns to $P$'s initial state by making $crt$ equal to $src$ (line 3) again. Otherwise it must offer an external choice between inputs (line 5) or an internal choice between outputs (line 10). In the first case, it has the chance to execute up to $n$ ($n = GAP ~|~ GAP \in \mathbf{N}$) new-in-context inputs (line 6) before converging to $P$ (lines 8 and 9). Note that only new in-context-inputs ($inputs \ssub set(e2(head(crt))$) are allowed to be executed non-deterministically, as we cannot violate the external choice between expected inputs $set(e2(head(crt))$; therefore we use the helper process $EXEC\_Q$ to do such a task. In the second case (line 10), before an internal choice between outputs is offered, $CVG\_BUILDER$ can do up to $n$ inputs (line 11) before converging to $P$ (lines 12 and 13).

Therefore, if we consider the CSP processes $T$ and $T'$, whose labelled transition systems are depicted in Figures \ref{fig:original_t} and \ref{fig:cvg_t}, respectively, the following FDR4 assertions hold:

\vspace*{-.3cm}

\begin{enumerate}[i.]
\item \verb"assert GLB_CVG(T_serial) [F= T'" 
\item  \verb"assert" \verb"GLB_CVG(T_serial) [F= T"
\end{enumerate}

According to our strategy, the first assertion implies that $\refcvg{T'}{T}$. The second assertion is particularly not surprising: it comes from Lemma \ref{thm:Hierarchical relations}, by which $\refcvg{T}{T}$ because $T \frefinedby T$.

 
\subsection{Building $GLB\_{ECVG}$} 
 
The construction of $GLB\_ECVG$ follows the same principles used to build $GLB\_CVG$. The differences relate to some additional auxiliary processes, mainly because $GLB\_ECVG$ must be able to do, after a new-in-context input, a sequence of any new-in-context events, before converging (Definition \ref{def:I/O extended convergent behaviour}). Processes $EXEC\_D$ (Listing \ref{lst:EXEC_D}) and $EXEC\_Q$ (Listing \ref{lst:EXEC_Q}) remain as presented before.

As we need to detect an occurrence of a new-in-context \textit{input} before allowing \textit{any} new-in-context event we performed a subtle change in the $EXEC$ process (Listing \ref{lst:newEXEC}): it acknowledges (by communicating the $in\_ack$ event) every time some event happens, making it possible to know when a new-in-context \textit{input} was communicated.              

The process $AFT\_IN$ (Listing \ref{lst:AFTER_IN}) acts like a watcher for $in\_ack$s events, reverberating them by communicating the $in\_rdt$ event. We model the process $EXEC\_AFTER\_IN$ (Listing \ref{lst:AFTER_IN}) to catch the $in\_rdt$ event and turn back to $EXEC$. Note that, playing the role of watchers, both processes cannot force anything to happen, so the successful termination, $SKIP$, is always a possibility.


\begin{center}
\resizebox{1\textwidth}{!}{
\begin{tabular}{ll}
\begin{minipage}{1\textwidth}

\begin{algorithm*}[H]
EXEC(evs, n) =  \\
\eIf {($n > 0$)} 
{$\Intchoice$ x : evs $\at$  (x $\then$ in$\_$ack $\then$ EXEC(evs, n -1)\\
  \Indp \Indp \Indp \Indp \Indp \Indp  $\intchoice$ \\
  EXEC(evs, n -1))}
{SKIP}
\caption{processes EXEC}
\label{lst:newEXEC}
\end{algorithm*}  

\LinesNumberedHidden
\begin{algorithm*}[H]
AFT$\_$IN = in$\_$ack $\then$ in$\_$rdt $\then$ SKIP $\extchoice$ SKIP
\BlankLine
EXEC$\_$AFTER$\_$IN(evs,n) =  \\ 
\Indp \Indp \Indp in$\_$rdt $\then$ EXEC(evs, n) $\extchoice$  SKIP
\caption{AFT$\_$IN and EXEC$\_$AFTER$\_$IN}
\label{lst:AFTER_IN}
\end{algorithm*}

\LinesNumbered
\begin{algorithm*}[H]
EXEC$\_$Q$\_$AFT(evs1, evs2, evs3, n) = \\
\Indp (EXEC$\_$Q(evs1, evs2, 1) $\parallel[\eset{in\_ack}]$ AFT$\_$IN ) \\
 $\parallel[\eset{in\_rdt}]$ \\
 EXEC$\_$AFTER$\_$IN(diff(evs3, evs2),n)\\
\BlankLine
\Indm EXEC$\_$AFT(evs1, evs2, evs3, n) = \\
\Indp (EXEC(evs1,1) $\parallel[\eset{in\_ack}]$ AFT$\_$IN) \\
$\parallel[\eset{in\_rdt}]$ \\
EXEC$\_$AFTER$\_$IN(diff(evs3, evs2),n) \\
\caption{EXEC$\_$Q$\_$AFT and EXEC$\_$AFT}
\label{lst:newEXEC_Q}
\end{algorithm*}

\end{minipage}
&
\begin{minipage}{1\textwidth}

\begin{algorithm*}[H]
ECVG$\_$BUILDER(src,crt) = \\
\eIf {e1(head(crt))==end} {ECVG$\_$BUILDER(src,src)}
{
  \eIf {subset(set(e2(head(crt))), inputs)} 
    {EXEC$\_$Q$\_$AFT(inputs, set(e2(head(crt))), all, GAP-1);\\
    \Indp ECVG$\_$BUILDER(src,crt)\\
    \Indm $\intchoice$ \\
    $\Extchoice$ x:set(e2(head(crt))) $\at$ x $\then$\\
    \Indp ECVG$\_$BUILDER(src, branch(x,tail(crt), e3(head(crt))+1, false))
    }
   { 
    EXEC$\_$AFT(inputs, set(e2(head(crt))), all, GAP-1); \\
    $\Intchoice$ x:set(e2(head(crt))) $\at$ x $\then$ \\
    \Indp ECVG$\_$BUILDER(src, branch(x,tail(crt), e3(head(crt))+1, false))
   }
}
\BlankLine
GLB$\_$ECVG(P) = ECVG$\_$BUILDER(P$\_$serial,P$\_$serial) \\
\Indp \Indp \Indp \Indp \Indp \Indp \Indp \Indp \Indp   $\hide \{$end, in$\_$ack,in$\_$rdt$\}$
\caption{ECVG$\_$BUILDER and GLB$\_$ECVG} 
\label{lst:ECVG_BUILDER}
\end{algorithm*}
\end{minipage}
\end{tabular}
}
\end{center}

The process $EXEC\_Q\_AFT$ (Listing \ref{lst:newEXEC_Q}) acts like a wrapper to $EXEC\_Q$. It is parametrised by three sets of events $evs1$, $evs2$ and $evs3$: the first two have the same intent of their counterparts in $EXEC\_Q$, where the purpose of the latter is to receive the set containing the events that can happen after a new-in-context input of $evs1 \ssub evs2$ (line 1). After an input, $EXEC\_Q$ synchronises on $in\_ack$ with $AFT\_IN$ (line 2), which in turn communicates $in\_rdt$, at this time synchronising with $EXEC\_AFTER\_IN$, which is responsible for offering, iteratively, up to $n$ times any $evs3 \ssub evs2$ (inputs or outputs) new-in-context events (line 4).

Process $EXEC\_AFT$ follows the same reasoning used in $EXEC\_Q\_AFT$, but it uses $EXEC$ instead of $EXEC\_Q$: it is designed to be used before an internal choice among outputs, therefore there is no need to retain the external choice of expected inputs, as $EXEC\_Q$ does. 

The process $ECVG\_BUILDER$ (Listing \ref{lst:ECVG_BUILDER}) coordinates the auxiliary processes we have seen in the task of building $GLB\_ECVG(P)$ for an I/O process $P$, based on its serialised representation $src$ (line 1). It distinguishes from $CVG\_BUILDER$ by the use of the auxiliary processes $EXEC\_Q\_AFT$ and $EXEC\_AFT$ instead of $EXEC\_Q$ and $EXEC$, respectively. The process $EXEC\_Q\_AFT$ (line 6) can execute up to $n$ (where $n = GAP$) new-in-context events ($all = inputs \cup outputs$), but the first must be an input; as mentioned earlier, it differs from $EXEC\_AFT$ (line 12) as it preserves the external choice among expected inputs. 

Now if we consider the CSP processes $T$ and $T''$ in Figures \ref{fig:original_t} and \ref{fig:ecvg_t}, respectively, the following FDR4 assertions hold:

\vspace*{-.1cm}

\begin{enumerate}[i.]
\item \verb"assert GLB_ecvg(T_serial)" \verb"[F= T''"
\item  \verb"assert GLB_ecvg(T_serial) [F= T"
\end{enumerate}

\vspace*{-.1cm}

As before, the first assertion holds, which implies, according to our strategy, that $\refecvg{T''}{T}$ and the second one, as a consequence of Lemma \ref{thm:Hierarchical relations}, from where we know $\refecvg{T}{T}$ because $T \frefinedby T$. 

\vspace*{-.2cm}

\section{Case Study}\label{sec:Case Study}

\setcounter{lstlisting}{16}

We model an autonomous healthcare robot that monitors and medicates patients, being able to contact the relevant individuals or systems in case of emergency. It receives data from a number of sensors and actuates by injecting intravenous drugs and/or by calling the emergency medical services and the patient's relatives or neighbours.  We use the following data types (Listing \ref{lst:cs:types}): \verb"BI" (breath intensity), \verb"BT" (body temperature), \verb"DD" (drug dose), \verb"BGL" (blood glucose level), \verb"CL" (call list, the relevant individuals to be called in the case of emergency), \verb"DRUG" (the drugs in the robot's actuators), \verb"QUEST" (the robot's question list, to ask the patient when its voice recognition module is used).

\noindent\begin{minipage}{.485\textwidth}
\begin{lstlisting}[frame=single, basicstyle=\tiny\ttfamily , caption={Types} , label={lst:cs:types}]
nametype BI = {1..5}
nametype BT  = {34..41}
nametype DD = {0..5}
datatype BGL = low | normal | threshold | high
datatype CL = c911 | cFamily | cNeighbor | ack
datatype DRUG = insulin | painkiller | antipyretic
datatype QUEST = chest | head | vision | lst

datatype EVENTS = 
 breath.BI | bodyTemp.BT | bloodGlucose.BGL |numbnessFace.Bool | 
 fainting.Bool | cough.Bool | troubleSpeaking.Bool |
 visionTrouble.Bool | chestDiscomfort.Bool | headache.Bool |
 ask.QUEST| call.CL | administer.DRUG.DD
datatype IO = out.EVENTS | in.EVENTS

subtype BS = breath.BI | bodyTemp.BT | bloodGlucose.BGL
subtype I_BS = in.BS | out.BS

subtype IS = numbnessFace.Bool |  fainting.Bool
subtype I_IS = in.IS | out.IS
\end{lstlisting} 
\end{minipage}\hfill
\begin{minipage}{.49\textwidth}
\begin{lstlisting}[frame=single, basicstyle=\tiny\ttfamily, caption={Types and channels}, label={lst:cs:typesandchannels}]
subtype VS = cough.Bool | troubleSpeaking.Bool
subtype I_VS = in.VS | out.VS

subtype TK = visionTrouble.Bool  |  chestDiscomfort.Bool | 
             headache.Bool | ask.QUEST
subtype I_TK = in.TK | out.TK

subtype PH = call.CL
subtype I_PH = in.PH | out.PH

subtype IVN = administer.DRUG.DD
subtype I_IVN = in.IVN | out.IVN

channel bodySen : I_BS
channel imageRec : I_IS
channel voiceRec : I_VS
channel talk : I_TK
channel phone : I_PH
channel intravenousNeedle : I_IVN
\end{lstlisting}
\end{minipage}

These types are composed into more elaborated ones whose data will be communicated through the channels used to connect sensors, actuators and phones to the robot. The set \verb"EVENTS" encompasses the data sent in and out of: the body attached sensors (\verb"breath.BI", \verb"bodyTemp.BT" and \verb"bloodGlucose.BGL"), the vision recognition devices (\verb"numbnessFace.Bool"  and  \verb"fainting.Bool"), the noise recognition device (\verb"cough.Bool" and \verb"troubleSpeaking.Bool"), the voice interaction devices (\verb"visionTrouble.Bool",  \verb"chestDiscomfort.Bool", \verb"headache"\verb".Bool" and \verb"ask.QUEST"), the phone interface (\verb"call.CL") and the intravenous injection actuator (\verb"administer.DRUG.DD").

An event in \verb"EVENTS" can be communicated as an output to a component and become an input to the other to which it connects by one of the $\BRIC$ composition rules. We define \verb"IO" as the set \verb"EVENTS" where each value is tagged with \verb"in" and \verb"out", which differentiates inputs from outputs.

Each sensor/device communicates with the robot component via a specific channel, according to this schema  (Listing \ref{lst:cs:typesandchannels}): \verb"bodySen", the body attached sensors; \verb"imageRec", the vision recognition devices/sensors; \verb"voiceRec", the noise recognition devices/sensors; \verb"talk", the voice interaction devices/sensors; \verb"phone", the phone's interface and \verb"intravenousNeedle", the intravenous injection actuator. Each channel has its own type (a subset of \verb"IO") that involves only  functionality related events. As an example, consider a channel with the \verb"I_BS" type (\verb"I_BS" $\subset$ \verb"IO" and \verb"BS" $\subset$ \verb"EVENTS"), then it can communicate any event registered by the body attached sensors: breath, body temperature and blood glucose level.

The behaviour of our healthcare robot is defined in terms of the I/O process \verb"HC_BOT" (Listing \ref{lst:cs:HC_BOT}). It waits for the breath level indicator; if this level is critical ($< 3$), then it behaves as \verb"MOD_CALL_P1" (module phone call priority one), which contacts a patient's neighbour, the registered emergency service and relatives, in this order; then it waits for at least two of them to acknowledge before coming back to its initial state. Otherwise, the patient is breathing normally, and the robot reads the noise sensor to check whether he or she is coughing (\verb"voiceRec.in.cough?b"). If so, it reads the body temperature  (\verb"bodySen.in.bodyTemp?t") and blood glucose (\verb"bodySen.in.bloodGluco" \verb"se?g") sensors. If the body temperature exceeds 38$^{\circ}{\rm C}$, then it administers a dose of antipyretic (\verb"intravenousNeedle.out.administer.antipyretic.d_ap"). If the blood glucose level is in the threshold or high, it administers the hormone insulin (\verb"intra"\verb"venousNeedle.out.administer.insulin.d_in"), otherwise it just comes back to its initial state. After administrating any drug, and before coming back to its initial state, the robot must contact the patient's neighbour and relatives by behaving as \verb"MOD_CALL_P2" (module phone call priority two, Listing \ref{lst:cs:HC_BOT}), in which case at least one of them must acknowledge.

If the patient is breathing normally but in silence, the robot asks the image recognition module to inform about: any unusual sign in his face (\verb"imageRec.in." \verb"numbnessFace?nf") or if he fainted  (\verb"imageRec.in.fainting?f"). If at least one condition holds, the robot administers a painkiller (\verb"intravenousNeedle.out." \verb"administer.painkill" \verb"er.d_pk"), calls the relevant individuals by behaving as \verb"MOD_CALL_P1" (Listing \ref{lst:cs:HC_BOT}). In any case, it goes to its initial state.

\noindent\begin{minipage}{.485\textwidth}
\begin{lstlisting}[frame=single, basicstyle=\tiny\ttfamily,caption={\texttt{HC\_BOT}} , label={lst:cs:HC_BOT}]
HC_BOT = bodySen.in.breath?x ->
if (x < 3) then bodySen.out.breath.x ->  MOD_CALL_P1; HC_BOT
else voiceRec.in.cough?b ->  
 if (b) then bodySen.in.bodyTemp?t-> bodySen.in.bloodGlucose?g-> 
  if(t > 38) 
  then |~| d_ap : DD @
   intravenousNeedle.out.administer.antipyretic.d_ap -> 
   MOD_CALL_P2 ; HC_BOT
  else 
   if (g == high or g ==threshold) 
   then |~| d_in : DD @
    intravenousNeedle.out.administer.insulin.d_in -> 
    MOD_CALL_P2 ; HC_BOT
   else HC_BOT
 else 
  imageRec.in.numbnessFace?nf -> 
  imageRec.in.fainting?f -> 
   if (nf or f) then 
    |~| d_pk : DD @
     intravenousNeedle.out.administer.painkiller.d_pk -> 
     MOD_CALL_P1; HC_BOT  
   else HC_BOT
  
MOD_CALL_P1 = phone.out.call.cNeighbor -> phone.out.call.c911 -> 
phone.out.call.cFamily -> phone.in.call.ack -> 
phone.in.call.ack -> SKIP

MOD_CALL_P2 = phone.out.call.cNeighbor -> 
phone.out.call.cFamily -> phone.in.call.ack -> SKIP
\end{lstlisting} 
\end{minipage}\hfill
\begin{minipage}{.49\textwidth}
\begin{lstlisting}[frame=single, basicstyle=\tiny\ttfamily, caption={\texttt{HC\_BOT\_ACC}}, label={lst:cs:HC_BOT_ACC}]
HC_BOT_ACC = bodySen.in.breath?x ->
if (x < 3) then bodySen.out.breath.x ->  MOD_CALL_P1; HC_BOT_ACC
else voiceRec.in.cough?b -> 
 if (b) 
 then bodySen.in.bodyTemp?t -> bodySen.in.bloodGlucose?g -> 
  if(t > 38) 
  then intravenousNeedle.out.administer.antipyretic.t%37 ->
   MOD_CALL_P2 ; HC_BOT_ACC
  else 
   if (g == high) 
   then |~| d_in_h : {3,4,5} @
    intravenousNeedle.out.administer.insulin.d_in_h ->
    MOD_CALL_P2 ; HC_BOT_ACC
   else 
    if (g == threshold) 
    then |~| d_in_t : {1,2} @
     intravenousNeedle.out.administer.insulin.d_in_t ->
     MOD_CALL_P2 ; HC_BOT_ACC
    else HC_BOT_ACC
 else imageRec.in.numbnessFace?nf -> imageRec.in.fainting?f -> 
  if (nf or f) then |~| d_pk : DD @
   intravenousNeedle.out.administer.painkiller.d_pk -> 
   MOD_CALL_P1; HC_BOT_ACC  
  else HC_BOT_ACC

\end{lstlisting}
\end{minipage}
%

In $\BRIC$, the healthcare robot is defined in terms of the $Ctr_{HC\_BOT}$ contract (Figure \ref{fig:CtrHCBOT}). It behaves as  \verb"HC_BOT" and can interact with its environment by one of its visible communication channels: \verb"bodySen",  \verb"imageRec", \verb"voiceRec", \verb"phone" and \verb"intravenousNeedle".

\captionsetup[sub]{labelfont=subfontSmall}

\begin{figure*}
\resizebox{.6\textwidth}{!}{
\begin{tabular}{l}
\begin{minipage}{1\textwidth}
 \begin{subfigure}{\textwidth}
  \begin{equation*}
 Ctr_{HC\_BOT} \defs \\
  \quad \left\langle
        \begin{array}{l}
             \texttt{HC\_BOT}, 
            \left\{
            \begin{array}{l}
                  \texttt{bodySen} \mapsto  \texttt{I\_BS}, \texttt{imageRec} \mapsto  \texttt{I\_IS},\\
                  \texttt{voiceRec} \mapsto  \texttt{I\_VS}, \texttt{phone} \mapsto  \texttt{I\_PH},\\
                  \texttt{intravenousNeedle} \mapsto  \texttt{I\_IVN}\\
            \end{array}
            \right\},
            \left\{
            \begin{array}{l}
            \texttt{I\_BS},  \texttt{I\_IS},\\ \texttt{I\_VS}, \texttt{I\_PH},\\ \texttt{I\_IVN}
            \end{array}
            \right\},\\
            \{  \texttt{bodySen},  \texttt{imageRec}, \texttt{voiceRec}, \texttt{phone}, \texttt{intravenousNeedle}\}
        \end{array}
        \right\rangle
\end{equation*}
  \caption{$Ctr_{HC\_BOT}$}
\label{fig:CtrHCBOT}
  \end{subfigure}
\end{minipage}
\\
  \begin{minipage}{1\textwidth}
  \begin{subfigure}{\textwidth}
 \begin{equation*}
 Ctr_{HC\_BOT\_ACC} \defs \\
  \quad \left\langle
        \begin{array}{l}
             \textbf{\texttt{HC\_BOT\_ACC}}, 
            \left\{
            \begin{array}{l}
                  \texttt{bodySen} \mapsto  \texttt{I\_BS}, \texttt{imageRec} \mapsto  \texttt{I\_IS},\\
                  \texttt{voiceRec} \mapsto  \texttt{I\_VS}, \texttt{phone} \mapsto  \texttt{I\_PH},\\
                  \texttt{intravenousNeedle} \mapsto  \texttt{I\_IVN}\\
            \end{array}
            \right\},
            \left\{
            \begin{array}{l}
            \texttt{I\_BS},  \texttt{I\_IS},\\ \texttt{I\_VS}, \texttt{I\_PH},\\ \texttt{I\_IVN}
            \end{array}
            \right\},\\
            \{  \texttt{bodySen},  \texttt{imageRec}, \texttt{voiceRec}, \texttt{phone}, \texttt{intravenousNeedle}\}
        \end{array}
        \right\rangle
\end{equation*}
 \caption{$Ctr_{HC\_BOT\_ACC}$}
\label{fig:CtrHCBOTACC}
 \end{subfigure}
\end{minipage}
\\
\begin{minipage}{1\textwidth}
  \begin{subfigure}{\textwidth}
\begin{equation*}
 Ctr_{HC\_BOT\_TK} \defs \\
    \left\langle 
        \begin{array}{l}
             \texttt{HC\_BOT\_TK}, 
            \left\{
            \begin{array}{l}
                  \texttt{bodySen} \mapsto  \texttt{I\_BS}, \texttt{imageRec} \mapsto  \texttt{I\_IS},\\
                  \texttt{voiceRec} \mapsto  \texttt{I\_VS}, \texttt{phone} \mapsto  \texttt{I\_PH},\\
                  \texttt{intravenousNeedle} \mapsto  \texttt{I\_IVN}, \\
                  \textbf{\texttt{talk}} \mapsto \textbf{\texttt{I\_TK}}\\
            \end{array}
            \right\},
            \left\{
            \begin{array}{l}
            \texttt{I\_BS},  \texttt{I\_IS},\\ \texttt{I\_VS}, \texttt{I\_PH},\\ \texttt{I\_IVN}, \textbf{\texttt{I\_TK}}
            \end{array}
            \right\},\\
            \{  \texttt{bodySen},  \texttt{imageRec}, \texttt{voiceRec}, \texttt{phone}, \texttt{intravenousNeedle}, \textbf{\texttt{talk}}\}
        \end{array}
        \right\rangle
\end{equation*}
 \caption{$Ctr_{HC\_BOT\_TK} $}
\label{fig:CtrHCBOTTK}
 \end{subfigure}
\end{minipage}
\end{tabular}
}
\caption{Autonomous healthcare robots components}
\end{figure*}

\captionsetup[sub]{labelfont=subfontHuge}

The robot $Ctr_{HC\_BOT}$ is able to diagnose and select the appropriate drug to be administered. Nevertheless, this process abstracts from establishing an appropriate drug dose given the seriousness of the patient condition; in fact it is a nondeterministic decision. For example, consider the indexed nondeterministic choice \verb"|~| ds:DD @ intravenous" \verb"Needle.out.administer.antipyretic.ds", which offers the events 
\verb"intravenousNeedle.out.administer.antipyretic.ds" for all values of \verb"ds" in \verb"DD". No matter the seriousness of the fever, one might know what will be the dose \verb"ds" to be administered to the patient. The I/O process \verb"HC_BOT_ACC" (Listing \ref{lst:cs:HC_BOT_ACC}) addresses the dose issue by using two criteria: each degree above 38$^{\circ}{\rm C}$ corresponds to a unit of the prescribed antipyretic (\verb"intravenousNeedle.out.administer.antipyretic.t%37", where \verb"%" is the CSP operator for modulo) and the insulin dose will be one or two units if the blood glucose level is on the threshold, or greater than two, otherwise.
 
This behaviour extension is defined by the $\BRIC$ contract $Ctr_{HC\_BOT\_ACC}$ (Figure \ref{fig:CtrHCBOTACC}). This healthcare robot version has a better (more deterministic) decision-making mechanism  on the drug dose to be administered to the patient it monitors. By Definition \ref{def:BRIC refinement based on failures}, we have that $Ctr_{HC\_BOT} \frefbric Ctr_{HC\_BOT\_ACC}$: both components share the same channels with equivalent types (interfaces) and have their behaviours related by failures refinement \verb"HC_BOT" $\frefinedby$ \verb"HC_BOT_ACC"; it can be verified by the FDR4 assertion \verb"assert HC_BOT [F= HC_BOT_ACC".

The $Ctr_{HC\_BOT\_ACC}$ brings some improvements to $Ctr_{HC\_BOT}$. Nevertheless, the addition of new functionalities (or the enhancement of the existing ones) cannot be always addressed by refinement, even if we hide the implementation details before trying to establish such a relation, as already discussed. The component we present next, $Ctr_{HC\_BOT\_TK}$ (Figure \ref{fig:CtrHCBOTTK}), extends (inherits from) $Ctr_{HC\_BOT\_ACC}$ (Figure \ref{fig:CtrHCBOTACC}) with the addition of a talk module, which allows this robot to ask patients about their symptoms and thus can possibly better help them. 

The I/O process \verb"HC_BOT_TK" ($Ctr_{HC\_BOT\_TK}$ behaviour, Listing \ref{lst:cs:HCBOTTK}) improves \verb"HC_BOT_ACC" by being able to interact with patients via the voice simulation/recognition device through the new channel \verb"talk". Together with the events \verb"bodySen.in.breath?x", it offers, initially, the possibility of behaving as \verb"MOD_TALK": it receives a chat request (\verb"talk.in.ask.lst"), then collects information about chest discomfort (\verb"talk.in.chestDiscomfort?cd"), headache (\verb"talk.in.headache?" \verb"hd") and vision problems (\verb"talk.in.visionTrouble?vt"). If the patient reports chest discomfort associated with headache or vision problems, the robot understands that a serious situation is under way and calls all the relevant individuals by behaving as \verb"MOD_CALL_P1". In any case, it goes to its initial state.

%
%
\noindent\begin{minipage}{.485\textwidth}
\begin{lstlisting}[frame=single, caption={HC\_BOT\_TK} , basicstyle=\tiny\ttfamily, label={lst:cs:HCBOTTK}]
HC_BOT_TK = 
bodySen.in.breath?x ->
 if (x < 3) 
 then bodySen.out.breath.x ->  MOD_CALL_P1; HC_BOT_TK
 else voiceRec.in.cough?b -> 
  if (b) then bodySen.in.bodyTemp?t -> 
   bodySen.in.bloodGlucose?g -> 
   if(t > 38) 
   then  intravenousNeedle.out.administer.antipyretic.t%37 ->
    MOD_CALL_P2 ; HC_BOT_TK
   else 
    if (g == high) 
    then |~| d_in_h : {3,4,5} @
     intravenousNeedle.out.administer.insulin.d_in_h ->
     MOD_CALL_P2 ; HC_BOT_TK
 \end{lstlisting} 
\end{minipage}\hfill
\begin{minipage}{.49\textwidth}
\begin{lstlisting}[frame=single , basicstyle=\tiny\ttfamily]
    else 
     if (g == threshold) 
     then |~| d_in_t : {1,2} @
      intravenousNeedle.out.administer.insulin.d_in_t ->
      MOD_CALL_P2 ; HC_BOT_TK
     else HC_BOT_TK
  else 
   imageRec.in.numbnessFace?nf -> imageRec.in.fainting?f -> 
   if (nf or f) 
   then |~| d_pk : DD @
    intravenousNeedle.out.administer.painkiller.d_pk -> 
    MOD_CALL_P1; HC_BOT_TK  
   else HC_BOT_TK
[]
MOD_TALK ; HC_BOT_TK

MOD_TALK = talk.in.ask.lst -> 
 	talk.out.ask.chest -> talk.in.chestDiscomfort?cd -> 
 	talk.out.ask.head -> talk.in.headache?hd -> 
 	talk.out.ask.vision -> talk.in.visionTrouble?vt ->
 	if (cd and (hd or vt)) then MOD_CALL_P1 else SKIP
\end{lstlisting}
\end{minipage}

\begin{figure*}[t]  
\resizebox{1\textwidth}{!}{
\begin{tabular}{ll}
\begin{minipage}{1.2\textwidth}
 \begin{subfigure}{\textwidth}
  \includegraphics[scale=1]{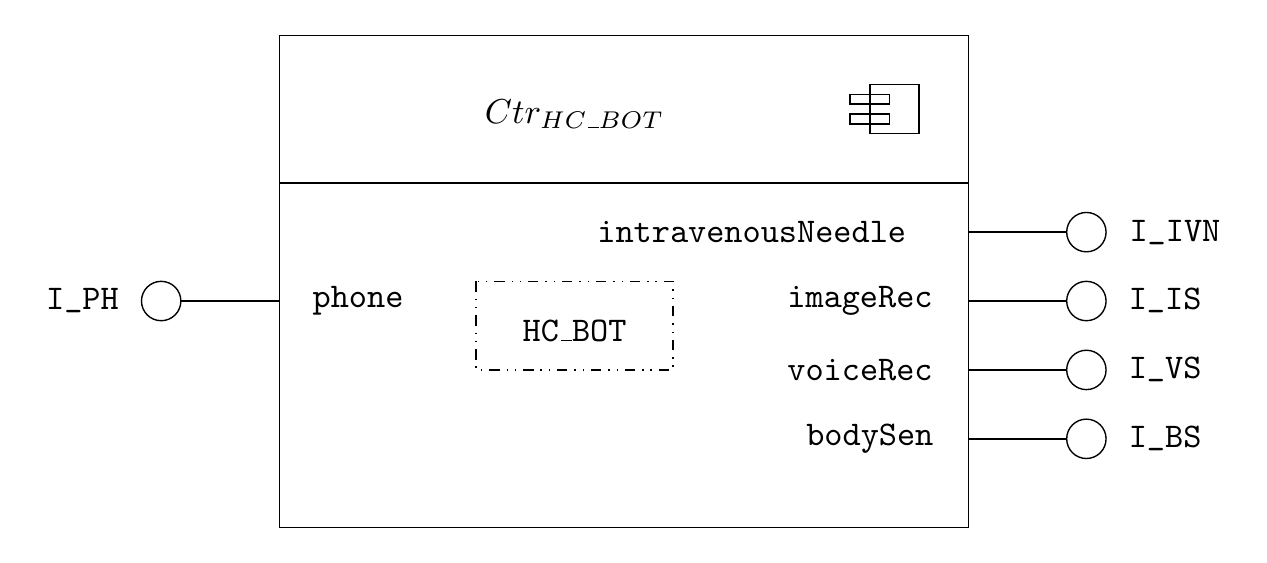}
  \end{subfigure}
\end{minipage} \scalebox{2}{$\frefbric$}
& 
  \begin{minipage}{1.2\textwidth}
  \begin{subfigure}{\textwidth}
 \includegraphics[scale=1]{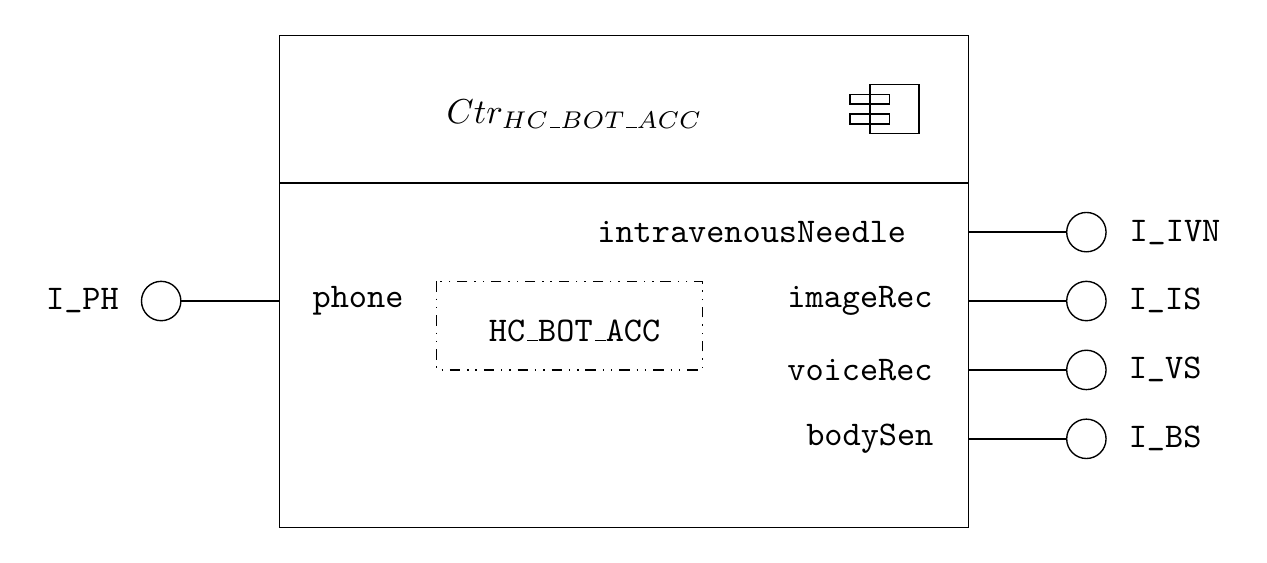}
 \end{subfigure}
\end{minipage}
\\
\begin{minipage}{1.2\textwidth}
\end{minipage} 
& 
\begin{minipage}{1.2\textwidth}
\hspace{5cm}{\rotatebox[origin=c]{-90} { \scalebox{2}{$\subecvg$}}}\\
  \begin{subfigure}{\textwidth}
 \includegraphics[scale=1]{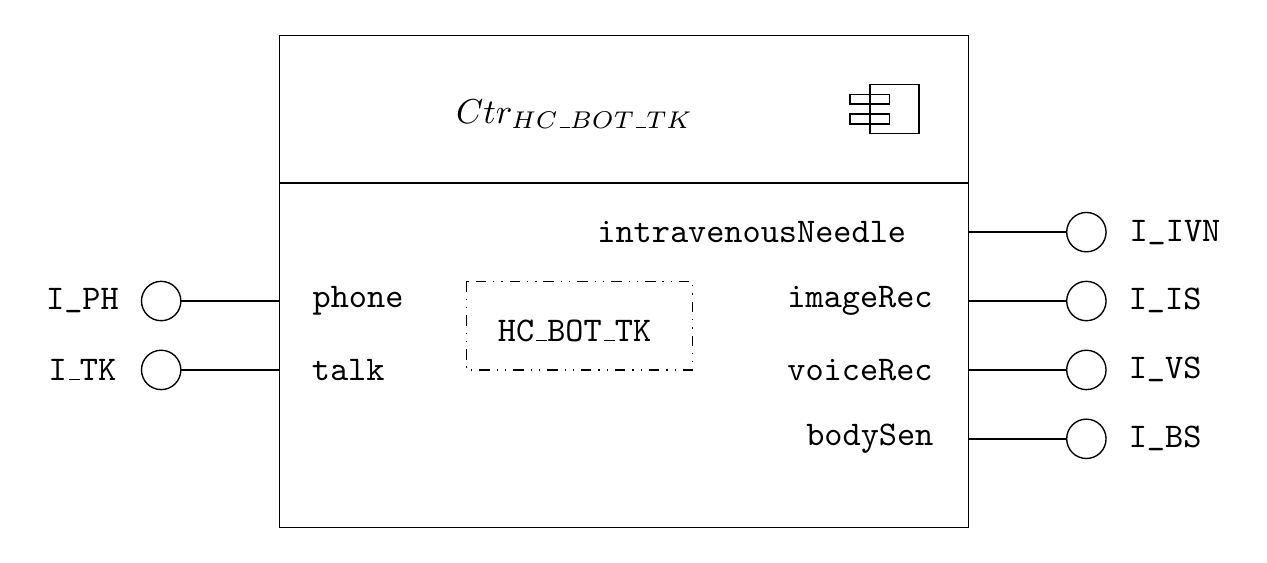}
 \end{subfigure}
\end{minipage}
\end{tabular}
}
\caption{Autonomous healthcare robot hierarchy}
\label{fig:healthcarecasestudy}
\end{figure*}

By Definition \ref{def:BRIC inheritance}, we have that $Ctr_{HC\_BOT\_ACC} \subecvg Ctr_{HC\_BOT\_TK}$. Note that the attempt to establish a failures relation between \verb"HC_BOT_ACC" and \verb"HC_BOT_TK", provided the events communicated through \verb"talk" are hidden on the latter, fails: as the FDR4 assertion \verb"HC_BOT_ACC [F= HC_BOT_TK \ {|talk|}" proves. This shows that convergence and inheritance, in the behavioural and component level perspectives, provide an entire new approach to evolve component based specifications, whilst preserving deadlock freedom. The resulting component hierarchy is depicted in Figure \ref{fig:healthcarecasestudy}.

This hierarchy guarantees important results when composing the healthcare robot. Suppose we have a drug storage component $Ctr_{DRUG\_STR}$ that dispenses drugs, as requested, and informs, afterwards, stock level status; also, consider a communicator hub component $Ctr_{HUB\_COM}$ that handles communications through different mediums: phone calls, messages, audio stream and e-mails (their I/O processes are omitted here for the sake of brevity). Considering the following three compositions:
\vspace*{-.3cm}

\resizebox{.9\textwidth}{!}{
\centering
\begin{minipage}{.5\textwidth}
\begin{align*}
& Ctr_{SYS} = Ctr_{HC\_BOT} \commcomp{ \texttt{intravenousNeedle}}{ \texttt{drugDispenser}} Ctr_{DRUG\_STR}\\
& Ctr_{SYS2} = Ctr_{HC\_BOT\_ACC}  \commcomp{ \texttt{intravenousNeedle}}{ \texttt{drugDispenser}} Ctr_{DRUG\_STR}\\
& Ctr_{SYS3} = (Ctr_{HC\_BOT\_TK}  \commcomp{ \texttt{intravenousNeedle}}{ \texttt{drugDispenser}}  Ctr_{DRUG\_STR})  \\
& \hspace{5cm}\commcomp{ \texttt{talk}}{ \texttt{audioStream}} Ctr_{HUB\_COM} \\  
 \end{align*}
\end{minipage}
}

\vspace*{-.3cm}

We have that (a) since $Ctr_{HC\_BOT} \frefbric Ctr_{HC\_BOT\_ACC}$, we know, by monotonicity of $\BRIC$ component refinement, that $Ctr_{SYS} \frefbric Ctr_{SYS2}$ also holds (both being deadlock free) and, for $Ctr_{DRUG\_STR}$, it is impossible to distinguish between the different healthcare robots; (b) as $Ctr_{HC\_BOT\_ACC}$ $\subecvg Ctr_{HC\_BOT\_TK}$,  $Ctr_{SYS3}$ is deadlock free and, again for $Ctr_{DRUG\_STR}$, it is impossible to distinguish between robots, which is true not only for the component $Ctr_{DRUG\_STR}$ but for any component able to connect to  $Ctr_{HC\_BOT}$, no matter whether new channels on $Ctr_{HC\_BOT\_TK}$ are exercised (as the \verb"talk" channel, which is hooked to \verb"audioStream" from $Ctr_{HUB\_COM}$).

It is important to consider non-convergent component extensions, which can introduce deadlock. Suppose we define a contract $Ctr_{HC\_BOT\_TK\_ECHO}$, which echoes the patient possible responses (output events) asking for confirmation. As these events are new-in-context outputs (not inputs), the contract $Ctr_{HC\_BOT\_TK\_ECHO}$ does not inherit by convergence from $Ctr_{HC\_BOT\_TK}$, as the former can output and, without response lock, where the latter does not. In FDR4 the \verb"assert GLB_CVG(HC_BOT_TK_serial)[F= HC_BOT_TK_ECHO" fails. Although the traces:

\begin{itemize}
\item $\mathtt{\trace{bodySen.in.breath.2,bodySen.out.breath.2, phone.out.call.cNeighbor}}$, from \verb"HC_BOT_TK" 

\item $\langle \mathtt{ bodySen.in.breath.2, bodySen.out.breath.2, phone.out.call.cNeighbor},\\ \mathtt{echo.in.response.yes } \rangle$, from \verb"HC_BOT_TK_ECHO"
\end{itemize}

are convergent, the following traces:

\begin{itemize}
\item $\mathtt{\trace{bodySen.in.breath.2, bodySen.out.breath.2, phone.out.call.cNeighbor}}$, from \verb"HC_BOT_TK") 

\item $\langle  \mathtt{bodySen.in.breath.2, bodySen.out.breath.2, phone.out.call.cNeighbor,}\\ \mathtt{{\bf{echo.out.timeout}}, echo.out.response.yes}\rangle $, from \verb"HC_BOT_TK_ECHO"
\end{itemize}
are not convergent, causing the nonconformance. It happens because of the \verb"echo.out.timeout", which signals a timeout in the patient response, but it is a new-in-context output, not a new-in-context input.

Finally, it is important to notice that this work does not provide a guideline to evolve or correct the models with respect to convergence: it is beyond the scope of this work and is a topic to be addressed as future work.


\section{Related work}\label{sec:Related work}

Apart from the precise definition of component behaviour and interface, a formal approach to CB-MDD must specify how components can be assembled into more complex ones, how they can be refined and, ultimately, how to evolve them into more specialised or functional ones. Several works have proposed a formal foundation for CB-MDD: \cite{Ramos11,Liu09,Jifeng05,Wang2009,Zhenbang09,Chen2007,Meng2006,Hennicker2008,Bauer2010}. Specially, the work reported in \cite{Ramos11} focuses on the development of deadlock free component-based systems by construction. Nevertheless, it does not propose a refinement or an inheritance relation for components, which is the main purpose of the current work. 


In rCOS \cite{Liu09} a component has a provided and a required interface and code associated to each of their method signatures. An interface is a syntactic notion that encompasses typed variable declarations, called fields, and method signatures with input and output parameters alongside with their types. This approach is distinguished from $\BRIC$ by not treating inputs as an environment decision and outputs as an internal decision, by not analysing behavioural properties in its composition rules and by stating the results on traces instead of the failures model, which compromises substitutability, since, for example, the traces model does not allow one to reason about deadlock freedom. 

In rCOS, a contract refines another when there is a corresponding refinement between their required and provided interfaces (given in terms of the Unifying Theories of Programming (UTP) semantics \cite{Hoare1998Unifying}) and between their traces (traces semantics). The works in \cite{Liu09,Zhenbang09,Chen2007} require that both components have the same provided and required interface, as we assume in our refinement relation, although the approaches in \cite{Jifeng05} and \cite{Wang2009} allow interface extension. 

Other works, notably \cite{Hennicker2008,Bauer2010}, have established their roots in the transition systems theory: in \cite{Hennicker2008}, the authors define an I/O labelling as a 3-tuple of outputs, inputs and internal labels used by an I/O-transition system, which encompasses a set of states and transitions. A component is formed of a set of ports and an observable behaviour given in terms of an I/O-transition system. A component refines another if both have the same ports and there is a correspondence between the states of their I/O-transition systems. There is no possibility of functionality extension, since both components must have the same ports and any observable behaviour of one must be possible by the other. The approach in \cite{Bauer2010} uses similar I/O transition systems to define component behaviour. It allows functionality extension by expressing component refinement in terms of a mapping function between states and transitions of two I/O transition systems. It adopts a similar approach to \cite{Meng2006}, whereas ours stands with that of \cite{Hennicker2008}, since our understanding is towards distinguishing refinement from inheritance, considering refinement as a way to achieve non-determinism reduction and inheritance as a way to embed on new system functionalities.

Some works have proposed inheritance relations for behavioural specifications \cite{Liskov94, Wehrheim03, America90,Bowman99,Puntigam96, DihegoPedroAugusto2013}. In \cite{America90,Liskov94} inheritance relations are defined in terms of invariants over state components and by pre and postconditions over defined methods. The remaining works define subtype relations based on models like failures and failures-divergences (denotational models of CSP), relating refinement with inheritance \cite{Wehrheim03}. None of them differentiates inputs from outputs nor considers structural elements besides behavioural specifications, although \cite{Wehrheim03} analyses substitutability and relates it with behavioural properties, as deadlock freedom.

Aligned to the mentioned works on behavioural inheritance, we also state our relations in the failures model, but we distinguish inputs from outputs, not only by putting them in different sets, but restricting the way they can be communicated. The work in \cite{Bertrand2011} presents a relation named I/O abstraction that has some connection with convergence. It allows an implementation to input more but restricts it to output less than its abstractions; however, it does not consider what happens after the implementation communicates a new input, which clearly weakens substitutability and thus the behavioural properties preservation, as deadlock freedom, in the composition rules. In the same way, the conformance relation used for testing, \textit{ioco} \cite{Tretmans96}, allows new inputs to be communicated by an implementation, but as the I/O abstraction relation, it differs from our work by not considering how the implementation behaves after engaging in a new input. We also highlight an important design decision we have taken in this work: we allow functionality extension to be implemented not only in terms of new events but also by existing ones; therefore we allow \textit{new-in-context} (not only \textit{new-in-alphabet}) events to be communicated by an inherited component. It gives more flexibility, but presents a challenge to the inheritance verification as we have demonstrated in the construction of an automated strategy for verifying convergence using the FDR4 model checker.

The treatment of inputs as an environment decision and outputs as being internally resolved by the component itself (I/O process definition) is also considered by the approach proposed in \cite{Cavalcanti2013}, but it differs from ours by developing a new semantic model named \textit{IOFailures}, which is not compositional, in general. Also, such a relation is proposed for testing, whereas we focus on a design that preserves behavioural properties by construction, supported by the $\BRIC$ composition rules.

The concept of evolution by retrenchment is presented in \cite{Duque2009}: a retrenchment  represents a contradiction of the current specification followed by an evolution of it in a different direction. We understand it is useful in the early stages of a specification, but as it evolves, we generally need to define evolutions that conform to the previous ones.
 
Following the lines of \cite{Liskov94, Wehrheim03}, although focusing on data refinement, the work reported in \cite{back1996superposition} presents the concept of evolution of reactive action systems by superposition. Extensions must not change the original behaviour and are restricted to be defined in terms of new events. In our work, we allow extensions to reuse existing events, additionally we develop an automated strategy to ensure such extensions are valid.

It is also important to mention that \cite{Aalst2002} defines an inheritance relation between components, based on the notion of projection, whose semantics is given in terms of Petri nets and equivalence is checked by bisimulation. Projections define loose relationships between behaviour specifications and in general they do not allow reuse of existing events. We differ by defining inheritance for a component model (based on CSP) where behavioural properties are guaranteed by construction and are locally verified by the FDR4 model checker.

Recent works have addressed how inheritance affects substitutability \cite{Maddox2018} and how evolution can benefit from it \cite{Sofiane2016}. In a large-scale study conducted in thousands of Java projects, the work reported in \cite{Maddox2018} has shown that a major portion of inheritance implementations violate the substitutability principle, specially in the contexts where threads were used. It highlights the importance of having theories and tools that guarantee inheritance does not break substitutability, as we propose here. In  \cite{Sofiane2016}, Petri nets are used to model the behaviour of web services and, as $\BRIC$ does, it also distinguishes inputs from outputs and substitutability is given in terms of structural and behavioural aspects. We differ from the latter aspect bacause our inheritance relations ensure behaviours eventually will converge, whereas, in  \cite{Sofiane2016}, a candidate extension is only obligated to contain the original behaviour but it is free do communicate anything else, which clearly does not guarantee deadlock-freedom.

More recently, the work presented in \cite{Lange2019} has discussed the rules for inheritance in multi-level modelling, assuming every abstraction has at least one realization. It provides rules for substitutability in each level of modelling, focusing on the structural and data aspects of inheritance relations, whereas we focus on both structural and behavioural aspects of extensions as a means to guarantee behavioural properties.

From what we have seen and, as far as we are aware, this is the first time component inheritance relations are developed for a formal and sound CB-MDD approach, with a formal semantics, a refinement relation, an analysis on the impact of substitutability and an automated strategy to check conformance.

\section{Conclusions}\label{sec:Conclusions}

\vspace*{-.3cm}

We have developed in this work a novel concept called behavioural convergence for specifications that distinguish inputs from outputs. Based on that we have developed refinement and inheritance relations for an approach to CB-MDD, where we consider structural and behavioural aspects. We incorporated these relations in a set of composition rules that guarantee deadlock freedom by construction.

First, we defined a congruent semantics for $\BRIC$ that considers component structure and behaviour. This makes it possible to understand the precise meaning of a component, and is the basis to define component refinement and inheritance. Our refinement notion guarantees the relevant properties required by the $\BRIC$ component compositions, fulfilling the substitutability principle (a refinement should be usable wherever its abstraction is expected, without a client being able to tell the difference)

As a major contribution of this work, we defined two inheritance relations for $\BRIC$, both based on a novel concept called \textit{behavioural convergence}. It captures the idea that components can evolve by accepting new inputs or establishing a communication session after these inputs, but then they must \textit{converge} to the predicted behaviour exhibited by its abstraction. Our definition of inheritance deals with component structural and behavioural aspects and guarantees substitutability, in the composition rules, preserving deadlock freedom. Therefore a component of a model can evolve by the reduction of non-determinism (refinement) or by the extension of functionality (inheritance)  and still preserving, in the entire model, deadlock freedom and protocol compatibility.

We have two forms of convergence and, consequently, two inheritance relations upon them. We have proved that one is a generalisation of the other. Indeed, we have proved our relations form a hierarchy, where component refinement is the strongest relation between $\BRIC$ components.

We have developed an automated strategy for verifying convergence as refinement assertions using FDR4. We have systematised the construction of a \textit{greatest lower bound} process (under the failures refinement relation), of which all convergent processes are stable-failure refinements. Based on this result, we have converted an assertion about convergence (and extended convergence) into a refinement verification, which can be carried out by FDR4. The overall approach was validated using a case study that involved the modelling and verification of an autonomous healthcare robot.

Quality attributes of programming and modelling, like high cohesion and low coupling, as well as some good practices of object-oriented design, are also useful when adopting the approach we propose for component design evolution. However, the main reuse aspect in our context is control flow behaviour, in contrast to data aspects. Therefore, some patterns, such as {\it decorator\/}, {\it command\/}, {\it template method\/} and {\it chain of responsibility\/}, and concurrent communication patterns~\cite{Roscoe1998Theory} as, for instance, {\it client-server\/}, {\it resource sharing\/} and {\it routing\/}, are likely to potentialise an extensible component design model. 

A major topic for future work is to devise a detailed process, tailored to support the approach proposed in this paper. The main benefit of this process is to allow the user to define, as a separate concern, model extensions that ensure convergence by construction. In this context, the developer would work, for instance, with a more appealing notation like UML or SysML. Then, for verification, the graphical models would be translated into CSP and the verification could be carried out in background, completely hidden from the developer, with proper and transparent traceability to the model.


Also as future work, we aim to develop industrial case studies such as traffic aviation control systems and extend a $\BRIC$ modelling tool (BTS--$\BRIC$ Tool Support \cite{POFS17}) to verify refinement and inheritance. We also plan to mechanise the proofs of our theorems in CSP-Prover \cite{CSPProver08}, an interactive theorem prover for CSP based on Isabelle/HOL \cite{Nipkow2002}. The cost of verifying convergence needs to be further investigated. Finally, it is in our agenda to consider the preservation of other classical concurrency properties like (non)determinism and livelock freedom, based on the approaches in \cite{Filho2016, Madiel2018,Otoni2017}, as well as domain specific properties. 

\vspace*{-.3cm}

\section*{References}

\vspace*{-.2cm}

\bibliography{mybibfile}

\begin{thebibliography}{10}
\expandafter\ifx\csname url\endcsname\relax
  \def\url#1{\texttt{#1}}\fi
\expandafter\ifx\csname urlprefix\endcsname\relax\def\urlprefix{URL }\fi
\expandafter\ifx\csname href\endcsname\relax
  \def\href#1#2{#2} \def\path#1{#1}\fi

\bibitem{Szyperski98}
C.~Szyperski, Component Software: Beyond Object-oriented Programming, ACM
  Press/Addison-Wesley Publishing Co., New York, NY, USA, 1998.

\bibitem{Papazoglou2003}
M.~P. Papazoglou, D.~Georgakopoulos,
  \href{http://doi.acm.org/10.1145/944217.944233}{Introduction:
  Service-oriented computing}, Commun. ACM 46~(10) (2003) 24--28.
\newblock \href {http://dx.doi.org/10.1145/944217.944233}
  {\path{doi:10.1145/944217.944233}}.
\newline\urlprefix\url{http://doi.acm.org/10.1145/944217.944233}

\bibitem{Russell71}
R.~L. Ackoff, Towards a system of systems concepts, Management Science 17~(11)
  (1971) 661--671.
\newblock \href {http://dx.doi.org/10.1287/mnsc.17.11.661}
  {\path{doi:10.1287/mnsc.17.11.661}}.

\bibitem{Jifeng2006}
H.~Jifeng, X.~Li, Z.~Liu, rcos: A refinement calculus of object systems,
  Theoretical Computer Science 365 (2006) 109 -- 142, formal Methods for
  Components and Objects Formal Methods for Components and Objects.
\newblock \href {http://dx.doi.org/http://dx.doi.org/10.1016/j.tcs.2006.07.034}
  {\path{doi:http://dx.doi.org/10.1016/j.tcs.2006.07.034}}.

\bibitem{Meng2006}
S.~Meng, L.~S. Barbosa,
  \href{http://dx.doi.org/10.1016/j.tcs.2005.09.072}{Components as coalgebras:
  The refinement dimension}, Theor. Comput. Sci. 351~(2) (2006) 276--294.
\newblock \href {http://dx.doi.org/10.1016/j.tcs.2005.09.072}
  {\path{doi:10.1016/j.tcs.2005.09.072}}.
\newline\urlprefix\url{http://dx.doi.org/10.1016/j.tcs.2005.09.072}

\bibitem{Hennicker2008}
R.~Hennicker, S.~Janisch, A.~Knapp, Refinement of components in connection-safe
  assemblies with synchronous and asynchronous communication, in: Proceedings
  of the 15th Monterey Conference on Foundations of Computer Software: Future
  Trends and Techniques for Development, Monterey'08, Springer-Verlag, Berlin,
  Heidelberg, 2010, pp. 154--180.
\newblock \href {http://dx.doi.org/10.1007/978-3-642-12566-9_9}
  {\path{doi:10.1007/978-3-642-12566-9_9}}.

\bibitem{Arbab2004}
F.~Arbab, \href{http://dx.doi.org/10.1017/S0960129504004153}{Reo: A
  channel-based coordination model for component composition}, Mathematical.
  Structures in Comp. Sci. 14~(3) (2004) 329--366.
\newblock \href {http://dx.doi.org/10.1017/S0960129504004153}
  {\path{doi:10.1017/S0960129504004153}}.
\newline\urlprefix\url{http://dx.doi.org/10.1017/S0960129504004153}

\bibitem{Chen2009}
Z.~Chen, Z.~Liu, A.~P. Ravn, V.~Stolz, N.~Zhan, Refinement and verification in
  component-based model-driven design, Sci. Comput. Program. 74~(4) (2009)
  168--196.
\newblock \href {http://dx.doi.org/10.1016/j.scico.2008.08.003}
  {\path{doi:10.1016/j.scico.2008.08.003}}.

\bibitem{Buchi1998}
M.~B{\"u}chi, E.~Sekerinski,
  \href{http://dx.doi.org/10.1007/3-540-69687-3_68}{Formal Methods for
  Component Software: The Refinement Calculus Perspective}, Springer Berlin
  Heidelberg, Berlin, Heidelberg, 1998, pp. 332--337.
\newblock \href {http://dx.doi.org/10.1007/3-540-69687-3_68}
  {\path{doi:10.1007/3-540-69687-3_68}}.
\newline\urlprefix\url{http://dx.doi.org/10.1007/3-540-69687-3_68}

\bibitem{KurkiSuonio1999}
R.~Kurki-Suonio, \href{http://dx.doi.org/10.1007/3-540-48119-2_10}{Component
  and Interface Refinement in Closed-System Specifications}, Springer Berlin
  Heidelberg, Berlin, Heidelberg, 1999, pp. 134--154.
\newblock \href {http://dx.doi.org/10.1007/3-540-48119-2_10}
  {\path{doi:10.1007/3-540-48119-2_10}}.
\newline\urlprefix\url{http://dx.doi.org/10.1007/3-540-48119-2_10}

\bibitem{Liskov94}
B.~H. Liskov, J.~M. Wing, A behavioral notion of subtyping, ACM Trans. Program.
  Lang. Syst. 16~(6) (1994) 1811--1841.
\newblock \href {http://dx.doi.org/http://doi.acm.org/10.1145/197320.197383}
  {\path{doi:http://doi.acm.org/10.1145/197320.197383}}.

\bibitem{Wegner88}
P.~Wegner, S.~B. Zdonik,
  \href{http://portal.acm.org/citation.cfm?id=646148.679054}{Inheritance as an
  incremental modification mechanism or what like is and isn't like}, in:
  Proceedings of the European Conference on Object-Oriented Programming, ECOOP
  '88, Springer-Verlag, London, UK, 1988, pp. 55--77.
\newline\urlprefix\url{http://portal.acm.org/citation.cfm?id=646148.679054}

\bibitem{America90}
P.~America, Designing an object-oriented programming language with behavioural
  subtyping, in: Proceedings of the REX School/Workshop on Foundations of
  Object-Oriented Languages, Springer-Verlag, London, UK, 1991, pp. 60--90.

\bibitem{Liskov1987}
B.~Liskov, Keynote address - data abstraction and hierarchy, SIGPLAN Not.
  23~(5) (1987) 17--34.
\newblock \href {http://dx.doi.org/10.1145/62139.62141}
  {\path{doi:10.1145/62139.62141}}.

\bibitem{Bowman99}
H.~Bowman, J.~Derrick,
  \href{http://portal.acm.org/citation.cfm?id=646816.708754}{{A Junction
  between State Based and Behavioural Specification (Invited Talk)}}, Kluwer,
  B.V., Deventer, The Netherlands, 1999, pp. 213--239.
\newline\urlprefix\url{http://portal.acm.org/citation.cfm?id=646816.708754}

\bibitem{Puntigam96}
F.~Puntigam, {Types for Active Objects Based on Trace Semantics}, in:
  Proceedings FMOODS 96, Chapman and Hall, 1996, pp. 4--19.

\bibitem{Wehrheim03}
H.~Wehrheim, Behavioral subtyping relations for active objects, Form. Methods
  Syst. Des. 23~(2) (2003) 143--170.
\newblock \href {http://dx.doi.org/http://dx.doi.org/10.1023/A:1024764232069}
  {\path{doi:http://dx.doi.org/10.1023/A:1024764232069}}.

\bibitem{DihegoPedroAugusto2013}
J.~Dihego, P.~Antonino, A.~Sampaio, Algebraic laws for process subtyping, in:
  L.~Groves, J.~Sun (Eds.), Formal Methods and Software Engineering, Vol. 8144
  of Lecture Notes in Computer Science, Springer Berlin Heidelberg, 2013, pp.
  4--19.
\newblock \href {http://dx.doi.org/10.1007/978-3-642-41202-8_2}
  {\path{doi:10.1007/978-3-642-41202-8_2}}.

\bibitem{Rubinger2010}
A.~L. Rubinger, B.~Burke, Enterprise JavaBeans 3.1, O'Reilly Media, Inc., 2010.

\bibitem{Krasner1988}
G.~E. Krasner, S.~T. Pope,
  \href{http://dl.acm.org/citation.cfm?id=50757.50759}{A cookbook for using the
  model-view controller user interface paradigm in smalltalk-80}, J. Object
  Oriented Program. 1~(3) (1988) 26--49.
\newline\urlprefix\url{http://dl.acm.org/citation.cfm?id=50757.50759}

\bibitem{Ramos2009Systematic}
R.~Ramos, A.~Sampaio, A.~Mota, Systematic development of trustworthy component
  systems, in: 2nd World Congress on Formal Methods, Vol. 5850 of Lecture Notes
  in Computer Science, Springer, 2009, pp. 140--156.

\bibitem{DihegoAugustoMarcel2015}
J.~Dihego, A.~Sampaio, M.~Oliveira,
  \href{http://doi.acm.org/10.1145/2695664.2695916}{Constructive extensibility
  of trustworthy component-based systems}, in: Proceedings of the 30th Annual
  ACM Symposium on Applied Computing, SAC '15, ACM, New York, NY, USA, 2015,
  pp. 1808--1814.
\newblock \href {http://dx.doi.org/10.1145/2695664.2695916}
  {\path{doi:10.1145/2695664.2695916}}.
\newline\urlprefix\url{http://doi.acm.org/10.1145/2695664.2695916}

\bibitem{RobinsonABR14}
T.~Gibson{-}Robinson, P.~J. Armstrong, A.~Boulgakov, A.~W. Roscoe, {FDR3} - {A}
  modern refinement checker for {CSP}, in: Tools and Algorithms for the
  Construction and Analysis of Systems - 20th International Conference, TACAS
  2014, 2014, Grenoble, France, April 5-13, 2014. Proceedings, 2014, pp.
  187--201.
\newblock \href {http://dx.doi.org/10.1007/978-3-642-54862-8_13}
  {\path{doi:10.1007/978-3-642-54862-8_13}}.

\bibitem{Roscoe1998Theory}
A.~W. Roscoe, Theory and Practice of Concurrency, {Prentice-Hall} Series in
  Computer Science, {Prentice-Hall}, 1998.

\bibitem{Ramos11}
R.~T. Ramos, {Systematic Development of Trustworthy Component-based Systems},
  Ph.D. thesis, {Center of Informatics - Federal University of Pernambuco},
  Brazil (2011).

\bibitem{Pedro14}
P.~R.~G. Antonino, A.~Sampaio, J.~Woodcock, A refinement based strategy for
  local deadlock analysis of networks of {CSP} processes, in: {FM} 2014: Formal
  Methods, Singapore, May 12-16, 2014. Proceedings, 2014, pp. 62--77.
\newblock \href {http://dx.doi.org/10.1007/978-3-319-06410-9_5}
  {\path{doi:10.1007/978-3-319-06410-9_5}}.

\bibitem{Filho2016}
M.~S.~C. Filho, M.~V.~M. Oliveira, A.~Sampaio, A.~Cavalcanti,
  \href{http://dx.doi.org/10.1007/978-3-319-47846-3_18}{{Local Livelock
  Analysis of Component-Based Models}}, in: K.~Ogata, M.~Lawford, S.~Liu
  (Eds.), Formal Methods and Software Engineering - 18th International
  Conference on Formal Engineering Methods, ICFEM 2016, Vol. \textbf{10009} of
  Lecture Notes in Computer Science, Springer International Publishing, 2016,
  pp. 279--295.
\newblock \href {http://dx.doi.org/10.1007/978-3-319-47846-3_18}
  {\path{doi:10.1007/978-3-319-47846-3_18}}.
\newline\urlprefix\url{http://dx.doi.org/10.1007/978-3-319-47846-3_18}

\bibitem{Madiel2018}
M.~C. Filho, M.~Oliveira, A.~Sampaio, A.~Cavalcanti,
  \href{http://www.sciencedirect.com/science/article/pii/S0020019018300036}{Compositional
  and local livelock analysis for csp}, Information Processing Letters 133
  (2018) 21 -- 25.
\newblock \href {http://dx.doi.org/https://doi.org/10.1016/j.ipl.2017.12.011}
  {\path{doi:https://doi.org/10.1016/j.ipl.2017.12.011}}.
\newline\urlprefix\url{http://www.sciencedirect.com/science/article/pii/S0020019018300036}

\bibitem{Otoni2017}
R.~Otoni, A.~Cavalcanti, A.~Sampaio, Local analysis of determinism for csp, in:
  S.~Cavalheiro, J.~Fiadeiro (Eds.), Formal Methods: Foundations and
  Applications, Springer International Publishing, Cham, 2017, pp. 107--124.

\bibitem{Roscoe1987pursuit}
B.~Roscoe, N.~Dathi, The pursuit of deadlock freedom, Information and
  Computation 75~(3) (1987) 289--327.

\bibitem{Roscoe2006Confluence}
A.~W. Roscoe, Confluence thanks to extensional determinism, Electronic Notes in
  Theoretical Computer Science 162 (2006) 305--309.

\bibitem{Petrenko04}
M.~van~der Bijl, A.~Rensink, J.~Tretmans,
  \href{http://dx.doi.org/10.1007/978-3-540-24617-6_7}{Compositional testing
  with ioco}, in: A.~Petrenko, A.~Ulrich (Eds.), Formal Approaches to Software
  Testing, Vol. 2931 of Lecture Notes in Computer Science, Springer Berlin
  Heidelberg, 2004, pp. 86--100.
\newblock \href {http://dx.doi.org/10.1007/978-3-540-24617-6_7}
  {\path{doi:10.1007/978-3-540-24617-6_7}}.
\newline\urlprefix\url{http://dx.doi.org/10.1007/978-3-540-24617-6_7}

\bibitem{COMPASS}
{Oliveira, M. and Sampaio, A. and Antonino,P. and Dihego, J. and Filho, M. C.
  and Bryans, J.},
  \href{http://www.compass-research.eu/Project/Deliverables/D24.4.pdf}{{Compositional
  Analysis and Design of CML Models}}, Tech. rep., Comprehensive Modelling for
  Advanced Systems of Systems (2014).
\newline\urlprefix\url{http://www.compass-research.eu/Project/Deliverables/D24.4.pdf}

\bibitem{Enderton77}
H.~B. Enderton, \href{http://opac.inria.fr/record=b1084859}{Elements of set
  theory}, Academic Press, New York, 1977.
\newline\urlprefix\url{http://opac.inria.fr/record=b1084859}

\bibitem{Wehrheim02checkFDR}
H.~Wehrheim, Checking behavioural subtypes via refinement, in: B.~Jacobs,
  A.~Rensink (Eds.), Formal Methods for Open Object-Based Distributed Systems
  V, Springer US, Boston, MA, 2002, pp. 79--93.

\bibitem{Liu09}
Z.~Liu, C.~Morisset, V.~Stolz, {rCOS}: theory and tool for component-based
  model driven development, in: Proceedings of the Third IPM international
  conference on Fundamentals of Software Engineering, FSEN'09, Springer-Verlag,
  Berlin, Heidelberg, 2009, pp. 62--80.
\newblock \href {http://dx.doi.org/10.1007/978-3-642-11623-0_3}
  {\path{doi:10.1007/978-3-642-11623-0_3}}.

\bibitem{Jifeng05}
H.~Jifeng, X.~Li, Z.~Liu,
  \href{http://dx.doi.org/10.1007/11560647_5}{Component-based software
  engineering}, in: Proceedings of the Second International Conference on
  Theoretical Aspects of Computing, ICTAC'05, Springer-Verlag, Berlin,
  Heidelberg, 2005, pp. 70--95.
\newblock \href {http://dx.doi.org/10.1007/11560647_5}
  {\path{doi:10.1007/11560647_5}}.
\newline\urlprefix\url{http://dx.doi.org/10.1007/11560647_5}

\bibitem{Wang2009}
Z.~Wang, H.~Wang, N.~Zhan, Towards a theory of refinement of component-based
  systems, Report 427, UNU-IIST (October 2009).

\bibitem{Zhenbang09}
Z.~Chen, Z.~Liu, A.~P. Ravn, V.~Stolz, N.~Zhan, Refinement and verification in
  component-based model-driven design, Sci. Comput. Program. 74~(4) (2009)
  168--196.
\newblock \href {http://dx.doi.org/10.1016/j.scico.2008.08.003}
  {\path{doi:10.1016/j.scico.2008.08.003}}.

\bibitem{Chen2007}
X.~Chen, J.~He, Z.~Liu, N.~Zhan,
  \href{http://dl.acm.org/citation.cfm?id=1775223.1775236}{A model of
  component-based programming}, in: Proceedings of the 2007 International
  Conference on Fundamentals of Software Engineering, FSEN'07, Springer-Verlag,
  Berlin, Heidelberg, 2007, pp. 191--206.
\newline\urlprefix\url{http://dl.acm.org/citation.cfm?id=1775223.1775236}

\bibitem{Bauer2010}
S.~Bauer, P.~Mayer, A.~Schroeder, R.~Hennicker, On weak modal compatibility,
  refinement, and the mio workbench, in: J.~Esparza, R.~Majumdar (Eds.), Tools
  and Algorithms for the Construction and Analysis of Systems, Vol. 6015 of
  Lecture Notes in Computer Science, Springer Berlin Heidelberg, 2010, pp.
  175--189.
\newblock \href {http://dx.doi.org/10.1007/978-3-642-12002-2_15}
  {\path{doi:10.1007/978-3-642-12002-2_15}}.

\bibitem{Hoare1998Unifying}
C.~Hoare, J.~He, Unifying Theories of Programming, {Prentice-Hall}, 1998.

\bibitem{Bertrand2011}
N.~Bertrand, T.~J{\'e}ron, A.~Stainer, M.~Krichen,
  \href{http://dl.acm.org/citation.cfm?id=1987389.1987402}{Off-line test
  selection with test purposes for non-deterministic timed automata}, in:
  Proceedings of the 17th international conference on Tools and algorithms for
  the construction and analysis of systems, TACAS'11/ETAPS'11, Springer-Verlag,
  Berlin, Heidelberg, 2011, pp. 96--111.
\newline\urlprefix\url{http://dl.acm.org/citation.cfm?id=1987389.1987402}

\bibitem{Tretmans96}
J.~Tretmans, Test generation with inputs, outputs and repetitive quiescence,
  Software - Concepts and Tools 17~(3) (1996) 103--120.

\bibitem{Cavalcanti2013}
A.~Cavalcanti, R.~M. Hierons, Testing with inputs and outputs in csp, in:
  Proceedings of the 16th international conference on Fundamental Approaches to
  Software Engineering, FASE'13, Springer-Verlag, Berlin, Heidelberg, 2013, pp.
  359--374.
\newblock \href {http://dx.doi.org/10.1007/978-3-642-37057-1_26}
  {\path{doi:10.1007/978-3-642-37057-1_26}}.

\bibitem{Duque2009}
J.~Garc\'{\i}a-Duque, J.~J. Pazos-Arias, M.~L\'{o}pez-Nores,
  Y.~Blanco-Fern\'{a}ndez, A.~Fern\'{a}ndez-Vilas, R.~P. D\'{\i}az-Redondo,
  M.~Ramos-Cabrer, A.~Gil-Solla,
  \href{https://doi.org/10.1007/s00766-009-0074-z}{Methodologies to evolve
  formal specifications through refinement and retrenchment in an
  analysis–revision cycle}, Requir. Eng. 14~(3) (2009) 129–153.
\newblock \href {http://dx.doi.org/10.1007/s00766-009-0074-z}
  {\path{doi:10.1007/s00766-009-0074-z}}.
\newline\urlprefix\url{https://doi.org/10.1007/s00766-009-0074-z}

\bibitem{back1996superposition}
R.~J. Back, K.~Sere, Superposition refinement of reactive systems, Formal
  Aspects of Computing 8~(3) (1996) 324--346.

\bibitem{Aalst2002}
W.~M.~P. van~der Aalst, K.~M. van Hee, R.~A. van~der Toorn,
  \href{https://doi.org/10.1016/S0167-6423(01)00005-3}{Component-based software
  architectures: A framework based on inheritance of behavior}, Sci. Comput.
  Program. 42~(2–3) (2002) 129–171.
\newblock \href {http://dx.doi.org/10.1016/S0167-6423(01)00005-3}
  {\path{doi:10.1016/S0167-6423(01)00005-3}}.
\newline\urlprefix\url{https://doi.org/10.1016/S0167-6423(01)00005-3}

\bibitem{Maddox2018}
J.~Maddox, Y.~Long, H.~Rajan,
  \href{https://doi.org/10.1145/3236024.3236075}{Large-scale study of
  substitutability in the presence of effects}, in: Proceedings of the 2018
  26th ACM Joint Meeting on European Software Engineering Conference and
  Symposium on the Foundations of Software Engineering, ESEC/FSE 2018,
  Association for Computing Machinery, New York, NY, USA, 2018, p. 528–538.
\newblock \href {http://dx.doi.org/10.1145/3236024.3236075}
  {\path{doi:10.1145/3236024.3236075}}.
\newline\urlprefix\url{https://doi.org/10.1145/3236024.3236075}

\bibitem{Sofiane2016}
S.~{Bourouz}, N.~{Zeghib}, Towards formal checking of web services
  substitutability, in: 2016 International Conference on Advanced Aspects of
  Software Engineering (ICAASE), Vol. 2016, 2016, pp. 1--8.

\bibitem{Lange2019}
A.~{Lange}, C.~{Atkinson}, On the rules for inheritance in lml, in: 2019
  ACM/IEEE 22nd International Conference on Model Driven Engineering Languages
  and Systems Companion (MODELS-C), 2019, pp. 113--118.

\bibitem{POFS17}
D.~I. de~A.~Pereira, M.~V.~M. Oliveira, M.~S.~C. Filho, S.~R. D.~R. Silva,
  \href{https://doi.org/10.1007/978-3-319-66845-1}{{BTS: A Tool for Formal
  Component-based Development}}, in: N.~Polikarpova, S.~Schneider (Eds.),
  Proceedings of the 13th International Conference on Integrated Formal Methods
  - IFM 2017, Vol. \textbf{10510} of Lecture Notes in Computer Science,
  Springer, 2017, pp. 211--226.
\newline\urlprefix\url{https://doi.org/10.1007/978-3-319-66845-1}

\bibitem{CSPProver08}
Y.~ISOBE, M.~ROGGENBACH, Csp-prover: a proof tool for the verification of
  scalable concurrent systems, Computer Software 25~(4).

\bibitem{Nipkow2002}
T.~Nipkow, M.~Wenzel, L.~C. Paulson, Isabelle/HOL: A Proof Assistant for
  Higher-order Logic, Springer-Verlag, Berlin, Heidelberg, 2002.

\end{thebibliography}

\clearpage

\appendix

\section{CSP}\label{appendix:CSP}

This appendix presents the relevant definitions, the syntax and the failures semantics of CSP \cite{Roscoe1998Theory}.

\begin{table}[ht]
\centering 
\begin{tabular}{|c|c|c|}
 \hline
  \bf{CSP} & $\bf{CSP}_{M}$ & \bf{description}  \\
  \hline
  \hline
$STOP$ & \verb"STOP" & termination \\
\hline
$SKIP$ & \verb"SKIP" & successful termination\\
\hline
$c \then P$ & \verb"c-> P" & prefix\\
\hline
$P;Q$ & \verb"P;Q" & sequential composition\\
\hline
$P \hide X$ & \verb"P \ X" & hiding\\
\hline
$P \extchoice Q$ & \verb"P[]Q" & external choice\\
\hline
$P \intchoice Q$ & \verb"P|~|Q" & internal choice\\
\hline
$b ~\& ~P$ & \verb"b & P" & boolean guard\\
\hline
if $b$ then $P$ else $Q$ & \verb"if b then P else Q" & if-then-else\\
\hline
$P ~\rensubs{a}{b}$ & \verb"P[[a <- b]]" & renaming\\
\hline
$P ~\interleave~ Q$ & \verb"P ||| Q" & interleaving\\
\hline
$P ~\parallel[a]~ Q$ & \verb"P [|a|] Q" & alphabetized parallel\\
\hline
$\extchoice~ e :X \at P$ & \verb"[] e :X @ P" &  replicated external choice\\
\hline
$\intchoice~ e :X \at P$ & \verb"|~| e :X @ P" &  replicated internal choice\\
\hline
$\interleave~ e : X \at P$ & \verb"||| e : X @ P" &  replicated interleave\\
\hline
$\parallel~ e : X \at [X'] P$ & \verb"|| e : X @ [X'] P" & replicated alphabetized parallel\\
 \hline
\end{tabular}
\caption{CSP processes}
\end{table}

\begin{table}[ht]
\centering 
\begin{tabular}{|c|c|c|}
  \hline
  \bf{CSP} & $\bf{CSP}_{M}$ & \bf{description}  \\
  \hline
  \hline
 $\Sigma$ & \verb"all" & alphabet of all communications\\
 \hline
 $\eset{c}$ & \verb"{| c |}" & the events communicated through channel $c$\\
 \hline
 $X \ssub Y$ & \verb"diff(X,Y)" & $\set{e ~|~ e \in X \land e \notin Y}$\\ 
 \hline
 $\tick$  & & (tick) termination event\\
 \hline
 $\tau$ &  & (tau) invisible event\\
 \hline
 $\Sigma^{\tick}$ &  & $\Sigma \cup \set{\tick}$\\
 \hline
 $\Sigma^{\tick,\tau}$ &  & $\Sigma \cup \set{\tick, \tau}$\\
 \hline
\end{tabular}
\caption{Events}
\end{table}

\begin{table}[ht]
\centering 
\begin{tabular}{|c|c|c|}
  \hline
  \bf{CSP} & $\bf{CSP}_{M}$ & \bf{description}  \\
  \hline
  \hline
 $\Sigma^{*}$ & & set of all finite traces over $\Sigma$\\
 \hline
 $\trace{}$  & \verb"<>" & the empty trace\\
 \hline
 $t \cat s$ & \verb"t ^ s" & concatenation of traces\\
 \hline
 $s \leq t$ & \verb"s <= t" & $\equiv \exists u. s \cat u = t$ (prefix order)\\
 \hline
 $\#s$ & \verb"#s" & length of $s$\\
  \hline
 $t - s$ & \verb"t - s" & $\begin{array}{c} 
                          t - \trace{} = t \\
                          \trace{} - s = \trace{}\\
                          \trace{e} \cat t - \trace{e} \cat s = t - s\\
                          (\trace{e_1} \cat t) - (\trace{e_2} \cat s) = \trace{e_1} \cat (t -  (\trace{e_2} \cat s)) ~|~ e_1 \neq e_2
 						 \end{array}$\\
 \hline
 $R^* t$ &  & $\begin{array}{c} 
 			R : \Sigma \rightarrow \Sigma\\
 			R^* \trace{} = \trace{}\\
 			R^* (\trace{e} \cat s) = \trace{R~e~} \cat R^* s ~|~ e \in \Sigma \land s \in \Sigma^*\\
              \end{array}$\\
  \hline
\end{tabular}
\caption{Traces}
\end{table}

\begin{table}[ht]
\begin{tabular}{|@{\hskip1pt}c|@{\hskip1pt}c|}
 \hline
  \bf{CSP} & \bf{failures semantics}  \\
  \hline
  \hline
$STOP$ & $\fmodel(STOP) = \set{(\emptyseq, X) \ | \ X \subseteq \Sigmatick}$ \\
\hline
$SKIP$ & $\fmodel(SKIP) = \begin{array}{l}
\set{(\emptyseq, X) \ | \ X \subseteq \Sigma} \cup \\
\set{(\trace{\tick}, X) \ | \ X \subseteq \Sigmatick}
\end{array} $\\
\hline
$ev \then P$ & $\fmodel(ev \then P) = \begin{array}{l} \set{(\trace{},X) \ | \ ev \notin X} \cup \\ \set{(\trace{ev} \cat s, X) \ | \ (s, X) \in \fmodel(P)} \end{array}$\\
\hline
$P;Q$ &  $  \fmodel(P;Q)  =   \begin{array}{l}
\set{(s, X) \ | \ s \in \Sigma^{*} \land (s, X \cup \set{\tick}) \in \fmodel(P)}  \\
\cup 
\set{(s \cat t, X) \ | \ s \cat \trace{\tick} \in \tmodel(P) \land (t, X) \in \fmodel(Q)}
\end{array}$\\
\hline
$P \hide X$ & $\fmodel(P \hide X) =\set{ (s \hide X, Y) \ | \ (s, X \cup Y) \in \fmodel(P)}$\\
\hline
$P \extchoice Q$ & $\fmodel(P \extchoice Q) = \begin{array}{l}\set{(\trace{},X) \ | \ (\trace{},X) \in \fmodel(P) \cap \fmodel(Q)} \\ 
\cup
\set{(\trace{},X) \ | \ X \subseteq \Sigma \land \trace{\tick} \in \tmodel(P) \cup \tmodel(Q)}\\
\cup
\set{(s,X) \ | \ (s,X) \in \fmodel(P) \cup \fmodel(Q) \land s \neq \trace{}}\end{array}$\\
\hline
$P \intchoice Q$ & $\fmodel(P \intchoice Q) = \fmodel(P) \cup \fmodel(Q)$\\
\hline
if $b$ then $P$ else $Q$ & $\fmodel(if \ \ c \ \ then \ \ P \ \ else \ \  Q) =$ if $c$ then $\fmodel(P)$, else $\fmodel(Q)$\\
\hline
$P \rename{R}$ & $\fmodel(P \rename{R}) = \set{(t, R(X)) \ | \ \exists s. s \ R^{*} \ t \land (s, X) \in \fmodel(P)}$\\
\hline
$P \parallel[X] Q$ & $\fmodel(P \parallel[X] Q) = 
\left\lbrace\begin{array}{r}
(u, Y \cup Z) \ | \exists s, t. (s,Y) \in \fmodel(P) \land\\
 (t,Z) \in \fmodel(Q) \land \\
Y \backprime X^{\tick} = Z \backprime X^{\tick} \land u = s \parallel[X] t,\\
\quad \quad \quad \text{where } X^{\tick}= X \cup \set{\tick}
\end{array} \right\rbrace$\\
 \hline
\end{tabular}
\caption{CSP failures semantics}
\end{table}

\clearpage

\section{$\BRIC$}\label{appendix:BRIC}

This appendix details the $\BRIC$ composition rules \cite{Ramos2009Systematic}, which are represented in terms of binary and unary asynchronous composition. 

When a pair of channels are connected by a $\BRIC$ composition rule, outputs from one will be inputs for the other and vice versa. There are two directional flows of communication. Therefore, two buffers are required to emulate an asynchronous medium between these channels, one for each flow.

\begin{definition}[$\BRIC$ buffer] \label{def:BRIC Buffer} 
A $\BRIC$ buffer maps inputs to outputs and vice versa (by the relations $L$ and $R$), without loss or reordering. 
\begin{align*}
BUFF^{n}_{IO}(L, R) \ = \  B^{n}(L) \ \interleave \ B^{n}(R), \text{where}\\
\end{align*}

\vspace*{-1.5cm}

{\small\begin{align*} 
B^{n}(R) \ = \ & B^{n}_{\emptyseq} (R) \ = \ ?x:\dom R \then B^{n}_{\seq{x}} \\
B^{n}_{s} (R) \ = \ & \# s < n \ \& \ ?x:\dom R \then B^{n}_{s \cat \seq{x}}(R)\\
& \extchoice \  R(head(s)) \then B^{n}_{tail(s)} (R)                    
\end{align*}}
\end{definition}

The asynchronous binary composition (Definition \ref{def:Asynchronous binary composition}) hooks two components, say $P$ and $Q$, with disjoint communication points, by their respective channels $c$ and $z$. Instead of communicating directly, their communications are buffered by $BUFF^{n}_{IO}(R_{IO}^{~c \rightarrow z}, R_{IO}^{~z \rightarrow c})$.

\begin{definition}[Asynchronous binary composition] \label{def:Asynchronous binary composition} 
Let $P$ and $Q$ be two distinct component contracts, and $c \in \cc{C}{P}$ and $z \in \cc{C}{Q}$ two channels, such that $\cc{C}{P}$ and $\cc{C}{Q}$ are disjoint. Then, the asynchronous binary composition of $P$ and $Q$, $P  {}_{\langle c \rangle}\asymp_{\langle z \rangle} Q$, is given by:
\begin{align*}
P  {}_{\langle c \rangle}\asymp_{\langle z\rangle} Q & = 
\langle \cc{B}{P} 
     	\parallel[\eset{c}] BUFF^{n}_{IO}(R_{IO}^{~c \rightarrow z}, R_{IO}^{~z \rightarrow c}) \parallel[\eset{z}] 
         \cc{B}{Q}, 
         \ce{R}', \ce{I}', \ce{C}' \rangle
\end{align*}

\text{where } $\ce{C}' = (\cc{C}{P} \cup \cc{C}{Q}) \backprime \set{c,z}$, $\ce{R}' = \ce{C}' \lhd (\cc{R}{P} \cup \cc{R}{Q})$,\\
\hspace*{1.6cm} $\ce{I}' = \ran \ce{R}'$ and $R_{IO}^{~a \rightarrow b} = \set{a.out.x \mapsto b.in.x}$.
\end{definition}

In this composition, the channels $c$ and $z$ are combined such that output events from one channel are consumed by input events of the other, and vice versa. This correspondence is made by two mapping relations, $R_{IO}^{~c \rightarrow z}$ and $R_{IO}^{~z \rightarrow c}$, which are used to input/output from the buffer  $BUFF^{n}_{IO}$. The resulting component behaviour is that of $P$ synchronised with the buffer $BUFF^{n}_{IO}(R_{IO}^{~c \rightarrow z}, R_{IO}^{~z \rightarrow c})$ on $c$ and with $Q$ on $z$. The interface, $\ce{C}'$, of the resulting component, $P  {}_{\langle c \rangle}\asymp_{\langle z\rangle} Q$, contains channels of both $P$ and $Q$ except for the hooked channels $c$ and $z$ ($(\cc{C}{P} \cup \cc{C}{Q}) \backprime \set{c,z}$). Therefore, only channels in $\ce{C}'$ appear in the resulting relation $\ce{R}'$ ($S \lhd R$ restricts the domain of $R$ to $S$) and in the resulting interface $\ce{I}'$ ($\ran \ce{R}'$).

The asynchronous unary composition (Definition \ref{def:Asynchronous unary composition}) hooks two channels, say $c$ and $z$ of the same component $P$. It allows $P$ to send and receive information to/from itself. It can be very useful if $P$ has inner components, that must be connected ( for example, the dining of philosophers). 

\begin{definition}[Asynchronous unary composition] \label{def:Asynchronous unary composition}
Let $P$ be a component contract, and $\set{c,z} \subseteq \cc{C}{P}$ two of its channels. Then, the asynchronous unary composition of $P$ by hooking $c$ and $z$, $P  \asymp\!\!|_{\langle z \rangle} ^{\langle c \rangle}$, is given by:
\begin{align*}
P  \asymp\!\!|_{\langle z \rangle} ^{\langle c \rangle} & = 
\langle \cc{B}{P} 
     	\parallel[\eset{c,z}] BUFF^{n}_{IO}(R_{IO}^{~c \rightarrow z}, R_{IO}^{~z \rightarrow c}), 
         \ce{R}', \ce{I}', \ce{C}' \rangle
\end{align*}

where $\ce{C}' = \cc{C}{P} \backprime \set{c,z}$, $\ce{R}' = \ce{C}' \lhd \cc{R}{P}$, $\ce{I}' = \ran \ce{R}'$ and\\
\hspace*{1.5cm} $R_{IO}^{~a \rightarrow b} = \set{a.out.x \mapsto b.in.x}$.
\end{definition}

The unary composition, like the binary one, combines two channels $c$ and $z$ such that output events from one channel are consumed by input events of the other, and vice versa; the difference is that both channels are from the same component. The resulting component behaviour is that of $P$ synchronised with the buffer $BUFF^{n}_{IO}(R_{IO}^{~c \rightarrow z}, R_{IO}^{~z \rightarrow c})$ on both channels $c$ and $z$. The resulting component $P  \asymp\!\!|_{\langle z \rangle} ^{\langle c \rangle}$ interface is the same as $P$ except for the hooked channels $c$ and $z$ ($\cc{C}{P} \backprime \set{c,z}$). As the binary composition, only channels in $\ce{C}'$ appear in the resulting relation $\ce{R}'$ and in the resulting interface $\ce{I}'$.

Not all components can be connected, and some connections using binary or unary compositions can lead to deadlock. The $\BRIC$ rules are defined in terms of the unary and binary asynchronous compositions and define the conditions where components can be safe (deadlock free) connected. They differ by the preconditions and the number of components involved, which are detailed in \cite{Ramos2009Systematic}. What follows summarises the $\BRIC$ composition rules.

The interleave composition rule is the simplest form of composition. It aggregates two independent components such that, after composition, these components still do not communicate between themselves. They directly communicate with the environment as before, with no interference from each other. 

\begin{definition}[Interleave composition] \label{def:interleaveComposition}
Let $P$ and $Q$ be two component contracts, such that $\ce{C}_P \cap~\ce{C}_Q = \emptyset$. The interleave composition of $P$ and $Q$  (namely $P \intercomp Q$) is given by:
\begin{align*}
    P \intercomp Q = P {}_{\langle\rangle}\dcomp_{\langle\rangle} Q
\end{align*}%
\end{definition}

This composition requires $P$ and $Q$ to have disjoint sets of channels $(\ce{C}_P \cap~\ce{C}_Q = \emptyset)$. The resulting component $P \intercomp Q$ is given by the binary composition (Definition \ref{def:Asynchronous binary composition}) of $P$ and $Q$ without hooking any of its channels (${}_{\langle\rangle}\dcomp_{\langle\rangle}$), which implies they run independently, in interleaving.

The second rule, the communication composition represents the most common way for linking  channels of two different components. As interleaving, it is given in terms  of asynchronous binary composition, but it connects channels from $P$ and $Q$ components. The assembled channels cannot be used in subsequent compositions, as imposed by Definition \ref{def:Asynchronous binary composition}.

\begin{definition}[Communication composition] \label{def:communicationComposition}
Let $P$ and $Q$ be two component contracts, and $ic$ and $oc$ two communication channels. The communication composition of $P$  and $Q$  (namely $P \commcomp{ic}{oc} Q$) via $ic$ and $oc$ is defined as follows:
\begin{align*}
    P \commcomp{ic}{oc} Q  =  P {}_{\langle ic\rangle}\dcomp {}_{\langle oc \rangle} Q
\end{align*}%
\end{definition}

This rule assumes the components behaviours on channels $ic$ and $oc$ are I/O confluent, strong compatible and satisfy the finite output property (FOP). These properties are detailed in \cite{Roscoe1987pursuit,Roscoe2006Confluence,Ramos11}: I/O confluence means that choosing between inputs (deterministically) or outputs (non-deterministically) does not prevents other inputs/outputs offered alongside from being communicated afterwards; two processes are strong compatible if all outputs produced by one are consumed by the other, and vice versa; finally, FOP guarantees that a process cannot output forever, so eventually it inputs after a finite sequence of outputs. The resulting component $P {}_{\langle ic\rangle}\dcomp {}_{\langle oc \rangle} Q$ is the binary composition of $P$ and $Q$ on channels $ic$ and $oc$ (Definition \ref{def:Asynchronous binary composition}).

The next two compositions allow the link of two  channels of a same component by means of asynchronous unary composition. The feedback composition provides the possibility of creating safe cycles for components with a tree topology. It achieves this by, among others conditions, ensuring that the channels being connected are decoupled: the behaviour projection over them are equivalent to the interleaving of each one's projection. 

\begin{definition}[Decoupled channels] \label{def:independentchannels}
Let $\ce{B}$ be an I/O process and $\set{c,z}$ two I/O channels. Then, $c$ and $z$ are decoupled in $\ce{B}$ if, and only, if: 
\begin{align*}
& \ce{B} \hide \Sigma \backprime \eset{c,z} \fequiv (\ce{B} \hide \Sigma \backprime \eset{c}) \interleave (\ce{B} \hide \Sigma \backprime \eset{z})
\end{align*}
\end{definition}

\begin{definition}[Feedback composition] \label{def:feedbackcomposition}
Let $P$ be a component contract, and $ic$ and $oc$ two communication channels. The feedback composition $P$ ($P\feedcomp{oc}{ic}$) hooking $oc$ to $ic$ is defined as follows:
\begin{align*}
  P \feedcomp{oc}{ic} = P \asymp\!\!\big\vert_{\langle oc \rangle}^{\langle ic \rangle}
\end{align*}%
\end{definition}

As the communication rule, the feedback composition assumes the component behaviour on channels $ic$ and $oc$ are I/O confluent, strong compatible and satisfy the finite output property (FOP). Moreover, it requires that $P$ behaves on $ic$ and $oc$ independently (as it were two distinct components), i.e., $ic$ and $oc$ are decoupled channel \cite{Ramos11}. The resulting component  $P \asymp\!\!\big\vert_{\langle oc \rangle}^{\langle ic \rangle}$ is achieved by synchronous unary composition of $P$ on channels $ic$ and $oc$.

The last composition rule, reflexive, is more general than the feedback composition; it is also more costly regarding verification, since it is able to assemble dependent channels (feedback assembles only independent channels), and so in general it requires a global analysis to ensure deadlock freedom. On the other hand, reflexive composition allows to connect channels in a cyclic topology, whereas feedback is restricted to tree topologies. 

\begin{definition}[Reflexive composition] \label{def:reflexiveComposition}
Let $P$ be a component contract, and $ic$ and $oc$ two communication channels. The reflexive composition $P$ (namely $P\refcomp{oc}{ic}$) hooking $oc$ to $ic$ is defined as follows:
\begin{align*}
  P \refcomp{ic}{oc} = P \asymp\!\!\big\vert_{\langle oc \rangle}^{\langle ic \rangle}
\end{align*}%

\end{definition}

This last rule relaxes feedback composition restrictions by allowing the connection of non decoupled channels but, nevertheless, it requires that outputs produced by $ic$ are consumed in the same rate (differing by at least one) by inputs from $oc$, and vice versa, i.e., the behaviour of $P$ is self-injection compatible on the hooked  channels \cite{Ramos11} .

\clearpage

\section{Proofs}\label{appendix:Proofs}

In this appendix we present proofs of some of our lemmas and theorems. 

\noindent \textbf{Lemma \ref{lem:cvg_implies_ecvg_on_prefixing} ($cvg$ implies $ecvg$ on trace prefixing).} \textit{Consider two I/O processes $T$ and $T'$ such that, $t_1 \cat t_3 \in \tmodel(T)$ and $t' \in \tmodel(T')$. If $\cvg{t'}{t_1 \cat t_3}$, where $t_1 \leq t'$, then $\ecvg{t'}{t_1 \cat t_3}$.}

\allowdisplaybreaks
\begin{proof}
\begin{align*}
& \text{Proof by induction on }t_3\\
& \textbf{Basis step: } t_3 = \trace{}\\
& t_1 \leq t'  ~.~ \cvg{t'}{t_1 \cat \trace{}}  \\
& \equiv[\text{empty trace concatenation}]\\
& t_1 \leq t'  ~.~ \cvg{t'}{t_1}\\
& \equiv[\text{Definition \ref{def:I/O convergent traces}}]\\
& t_1 = t'  ~.~ \cvg{t_1}{t_1}\\
& \equiv [\text{Definition \ref{def:I/O extended convergent traces}}]\\
& t_1 = t'  ~.~ \ecvg{t_1}{t_1}\\
& \textbf{Inductive step:}\\
& \text{Inductive hypothesis: }  t_1 \leq t'  ~.~ \cvg{t'}{t_1 \cat t_3} \implies t_1 \leq t'  ~.~ \ecvg{t'}{t_1 \cat t_3}\\
& \text{Prove that }  t_1 \leq t' \land e \in \Sigma ~.~\cvg{t'}{t_1 \cat (t_3 \cat \trace{e})} \implies  \ecvg{t'}{t_1 \cat (t_3 \cat \trace{e})}\\
& t_1 \leq t' \land e \in \Sigma ~.~ \cvg{t'}{t_1 \cat (t_3 \cat \trace{e})}\\
& \textbf{Cases: } e \text{ is new input } (e \notin in(T,t_1 \cat t_3) \cup out(T,t_1 \cat t_3)) \text{ or } \\
& \quad e \text{ is a possible event after }  t_1 \cat t_3 ~~ ( t_1 \cat (t_3 \cat \trace{e}) \in \tmodel(T))\\
& \text{Case 1: }e \text{ is a new input }\\
& \text{rewrite LHS as } (t_1 \cat t_3) \cat \trace{ne'} \cat t_3' \text{ where } t_3'= \trace{} \text{ and } ne' =  e\\
& \implies[\text{Definition \ref{def:I/O convergent traces}}]\\
& \cvg{t'}{(t_1 \cat t_3)} \\
& \implies[ \text{by inductive hypothesis}]\\
&  \ecvg{t'}{t_1 \cat t_3}\\
&\text{Case 2: } e \text{ is a possible event after }  t_1 \cat t_3\\
& \text{rewrite LHS as } (t_1 \cat t_3) \cat \trace{e'} \cat t_3'\text{ where }t_3'= \trace{} \text{ and } e' = e\\
&\implies [\text{Definition \ref{def:I/O convergent traces}}]\\
&  \cvg{t'}{(t_1 \cat t_3) \cat \trace{e'}} \text{ and } (t_1 \cat t_3) \cat \trace{e'} = t' \implies \ecvg{t'}{(t_1 \cat t_3) \cat \trace{e'}}
\end{align*}
\end{proof}

\noindent \textbf{Lemma \ref{lem:cvg subseteq ecvg} ($\mathtt{cvg} \subseteq \mathtt{ecvg}$).} \textit{Consider two I/O processes $T$ and $T'$, and $t$ and $t'$ two of its traces, respectively ($t \in \tmodel(T)$ and $t' \in \tmodel(T')$). If $\cvg{t'}{t}$ then $\ecvg{t'}{t}$.}

\allowdisplaybreaks
\begin{proof}
{\small\begin{align*}
& \cvg{t'}{t}\\
& \equiv [\text{Definition \ref{def:I/O convergent traces}}]
\\
& \begin{array}{l}
(t' = t) \lor
	\left( 
	\begin{array}{l}
 	(\# t' > \# t) ~~\land~~ \exists t_1, t_3 :  \Sigma^{*}, \exists ne : \Sigma  ~|~ \\
 		\quad \left( 
		\begin{array}{c}
 		t' = t_1 \cat \trace{ne}  \cat t_3 ~~\land~~ t_1 \leq t \land \\
 		ne \in inputs \land ne \notin in(T,t_1 ) \land\\
 		\cvg{t_1 \cat t_3}{t}
 		\end{array}
		\right)
	\end{array}
	\right)
\end{array}	\\
& \implies[\text{predicate calculus}]\\	
& \begin{array}{l}
	(t' = t) \lor \left( 
	\begin{array}{l}
 	(\# t' > \# t) ~~\land~~ \exists t_1,t_2, t_3 :  \Sigma^{*}, \exists ne \in \Sigma  ~|~ t_2 = \emptyseq \\
 		\quad \land \left( 
		\begin{array}{c}
 		t' = t_1 \cat \trace{ne} \cat t_2  \cat t_3 \land t_1 \leq t \land \\
 		ne \in inputs \land ne \notin in(T,t_1 ) \land\\
 		set(t_2) \cap (in(T, t_1 ) \cup out(T, t_1 )) = \emptyset \land\\
 		\cvg{t_1 \cat t_3}{t}
 		\end{array}
		\right)
	\end{array}
	\right)
\end{array}	\\
& \implies [\text{Lemma \ref{lem:cvg_implies_ecvg_on_prefixing}}]\\
& \begin{array}{l}
	(t' = t) \lor \left( 
	\begin{array}{l}
 	(\# t' > \# t) ~~\land~~ \exists t_1,t_2, t_3 :  \Sigma^{*}, \exists ne \in \Sigma  ~|~ t_2 = \emptyseq \\
 		\quad \land \left( 
		\begin{array}{c}
 		t' = t_1 \cat \trace{ne} \cat t_2  \cat t_3 \land t_1 \leq t \land \\
 		ne \in inputs \land ne \notin in(T,t_1 ) \land\\
 		set(t_2) \cap (in(T, t_1 ) \cup out(T, t_1 )) = \emptyset \land\\
 		\ecvg{t_1 \cat t_3}{t}
 		\end{array}
		\right)
	\end{array}
	\right)
\end{array}	\\
& \equiv [\text{Definition \ref{def:I/O extended convergent traces}}] ~~~
\ecvg{t'}{t}\\
\end{align*}}
\end{proof}

\noindent \textbf{Lemma \ref{lem:io_cvg subseteq io_ecvg} ($\mathtt{io \un cvg} \subseteq \mathtt{io \un ecvg}$).} \textit{Consider two I/O processes $T$ and $T'$. If $\refcvg{T'}{T} $ then $\refecvg{T'}{T}$.}
\vspace*{-.3cm}

\allowdisplaybreaks
\begin{proof}
{\small\begin{align*}
&  \refcvg {T'}{T}\\
& \equiv [\text{Definition \ref{def:I/O convergent behaviour}}]\\
& \forall (t',X) \in \fmodel(T'), \exists(t,Y) \in \fmodel(T)  \spot 
	   \left( 
		\begin{array}{c}
 		\cvg{t'}{t} \land \\
 		Y \cap inputs \supseteq X \cap inputs \land \\
 		Y \cap outputs \subseteq X \cap outputs\\
 		\end{array}
		\right)\\
& \equiv [\text{Lemma \ref{lem:cvg subseteq ecvg}}] \\
& \forall (t',X) \in \fmodel(T'), \exists(t,Y) \in \fmodel(T)  \spot 
	   \left( 
		\begin{array}{c}
 		\ecvg{t'}{t} \land \\
 		Y \cap inputs \supseteq X \cap inputs \land \\
 		Y \cap outputs \subseteq X \cap outputs\\
 		\end{array}
		\right)\\
& \implies  [a \implies a \lor b]\\	
& \forall (t',X) \in \fmodel(T'),   \exists(t,Y) \in \fmodel(T)  \spot
 \left( 
	 \begin{array}{c}  
	  \left( 
		\begin{array}{c}
		\ecvg{t'}{t} \land\\
 		Y \cap inputs \supseteq X \cap inputs \land \\
 		Y \cap outputs \subseteq X \cap outputs\\
 		\end{array}
 	   \right) \\
 	   	\begin{array}{l}
        \lor ( \Sigma \backprime Y \subseteq X )
         \end{array}
   \end{array}
 	   \right) \\	
& \implies [\text{Definition \ref{def:I/O extended convergent behaviour}}]~~~
 \refecvg{T'}{T} \\
\end{align*}}
\end{proof}

\noindent \textbf{Theorem \ref{thm:Hierarchical relations}  (Hierarchy).} \textit{The relations $\frefbric$, $\subcvg$ and $\subecvg$ form a hierarchy: $\frefbric \subseteq \subcvg \subseteq \subecvg$. In this proof, assume $T$ and $T'$ are components.}

\vspace*{-.1cm}

\allowdisplaybreaks
\begin{proof}
\begin{align*}
& \text{Hypothesis: }\frefbric \subseteq \subcvg\\
& T \frefbric T' \\
& \equiv [\text{Definition \ref{def:BRIC refinement based on failures}}]\\
& (\cc{B}{T} \frefinedby  \cc{B}{T'}) \land (\cc{C}{T} = \cc{C}{T'}) \land (\forall c: \cc{C}{T} \spot \cc{R}{T}(c) \subseteq \cc{R}{T'}(c))\\
& \implies [\text{failures semantics, hiding semantics}]\\
& (\cc{B}{T} \frefinedby  \cc{B}{T'}) \land (\cc{C}{T} = \cc{C}{T'}) \land (\forall c: \cc{C}{T} \spot \cc{B}{T} \project c \frefinedby  \cc{B}{T'} \project c)\\
& \implies [\text{Definition \ref{def:Default channel congruence}}]\\
& (\cc{B}{T} \frefinedby  \cc{B}{T'}) \land (\cc{C}{T} = \cc{C}{T'}) \land 
(\forall c : \cc{C}{T} \spot \defcong{\cc{B}{T'}}{\cc{B}{T}}{c})\\
& \implies [\text{Definition \ref{def:I/O convergent behaviour}}]\\
&  (\refcvg{\cc{B}{T'}}{\cc{B}{T}}) \land (\cc{C}{T} = \cc{C}{T'}) \land 
(\forall c : \cc{C}{T} \spot \defcong{\cc{B}{T'}}{\cc{B}{T}}{c})\\
& \implies [\text{Definitions \ref{def:BRIC inheritance}}]~~~T \subcvg T'\\
& \text{Hypothesis: } \subcvg \subseteq \subecvg\\
& T \subcvg T'\\
& \equiv [\text{Definition \ref{def:BRIC inheritance}}]\\
& \refcvg{\cc{B}{T'}}{\cc{B}{T}} \land \cc{R}{T} \subseteq \cc{R}{T'} \land  \forall c : \cc{C}{T} \spot 
\left(\begin{array}{l}
\defcong{\cc{B}{T'}}{\cc{B}{T}}{c} \quad \lor \\ \inpcong{\cc{B}{T'}}{\cc{B}{T}}{c}
\end{array}\right)
\\
& \implies [\text{Lemma \ref{lem:io_cvg subseteq io_ecvg}}]\\
&  \refecvg{\cc{B}{T'}}{\cc{B}{T}} \land \cc{R}{T} \subseteq \cc{R}{T'} \land  \forall c : \cc{C}{T} \spot\left(\begin{array}{l}
\defcong{\cc{B}{T'}}{\cc{B}{T}}{c} \quad \lor \\ \inpcong{\cc{B}{T'}}{\cc{B}{T}}{c}
\end{array}\right)\\
& \implies [\text{Definition \ref{def:BRIC inheritance}}]~~~ T \subecvg T'\\
\end{align*}
\end{proof}

\noindent \textbf{Lemma \ref{lem:BRIC inheritance preserves deadlock freedom} (Inheritance preserves deadlock freedom).} \textit{Consider $T$ and $T'$ two $\BRIC$ components, such that $T$ is deadlock free. If $T \subecvg T'$ then $T'$ is deadlock free.}

\allowdisplaybreaks
\begin{proof}
\begin{align*}
& T \subecvg T' \land T \text{ is deadlock free}\\
& \equiv [\text{Definition \ref{def:BRIC inheritance}},\text{deadlock free process}]\\
& \refecvg{\cc{B}{T'}}{\cc{B}{T}} \ \ \land \ \ \forall s \in \Sigma^{*} \spot (s, \Sigma^{\tick}) \notin \failures(\cc{B}{T})\\
& \equiv [\text{Definition \ref{def:I/O extended convergent behaviour}}]\\
& \forall (t',X) \in \fmodel(\cc{B}{T'}),   \exists(t,Y) \in \fmodel(\cc{B}{T})  \spot \begin{array}{c}
 \ecvg{t'}{t} \land\\
  \left( 
	 \begin{array}{c}  
	  \left( 
		\begin{array}{c}
 		Y \cap inputs \supseteq X \cap inputs \land \\
 		Y \cap outputs \subseteq X \cap outputs\\
 		\end{array}
 	   \right) \\
 	   	\begin{array}{l}
        \lor ( \Sigma \backprime Y \subseteq X )
         \end{array}
   \end{array}
 	   \right)
\end{array}\\ 	   
& \land \forall s \in \Sigma^{*} \spot (s, \Sigma^{\tick}) \notin \failures(\cc{B}{T})\\
& \implies [\text{failures semantics, predicate calculus}]\\
& \forall (t',X) \in \fmodel(\cc{B}{T'}),   \exists(t,Y) \in \fmodel(\cc{B}{T})  \spot \begin{array}{c}
 (\ecvg{t'}{t}) \land (\Sigma^{\tick} \not\subseteq Y) \land \\
  \left( 
	 \begin{array}{c}  
	  \left( 
		\begin{array}{c}
 		Y \cap inputs \supseteq X \cap inputs \land \\
 		Y \cap outputs \subseteq X \cap outputs\\
 		\end{array}
 	   \right) \\
 	   	\begin{array}{l}
        \lor ( \Sigma \backprime Y \subseteq X )
         \end{array}
   \end{array}
 	   \right)
\end{array}\\
& \implies [\text{predicate calculus}]\\
& \forall (t',X) \in \fmodel(\cc{B}{T'}) \spot (\Sigma^{\tick} \not\subseteq X)\\   
& \implies [\text{deadlock freedom}]\\
& \cc{B}{T'} \text{ is deadlock free process} \implies T' \text{ is a deadlock free component}
\end{align*}
\end{proof}

\noindent \textbf{Theorem \ref{thm:Substitutability for BRIC inheritance} (Substitutability).} \textit{Let $T$, $T'$  be two components such that $T \subecvg T'$. Consider $S[T]$ a deadlock free component contract, where $T$ is a deadlock free component contract that appears within the context $S$, then $S[T']$ is deadlock free.}

The proof follows by structural induction on the composition operators of $\BRIC$. Assuming it holds for $T$, we prove that it holds for the following cases: $T \intercomp Q$,  $T \commcomp{c}{z} Q$,   $T \feedcomp{c}{z}$ and $T \refcomp{c}{z}$, where $Q$ is a $\BRIC$ component.

\allowdisplaybreaks
\begin{proof}
\begin{align*}
& \textbf{Base case} \\
& S[T] = T\\
& \implies[T \subecvg T', S[T] \text{ is deadlock free}, \text{Lemma \ref{lem:BRIC inheritance preserves deadlock freedom}}] \\
& S[T'] = T' \text{ is deadlock free}\\
& \textbf{Interleaving composition}\\
&  S[T] = T \intercomp Q\\
& \implies [T \subecvg T', S[T] \text{ is deadlock free}, \text{Lemma \ref{lem:BRIC inheritance preserves deadlock freedom}}]\\
&  \cc{C}{T'} \cap \cc{C}{Q} = \emptyset  \land T' \text{ is deadlock free}\\
& \implies [Q \text{ is deadlock free}, \text{Definition \ref{def:interleaveComposition}}, \text{Theorem \ref{thm:Deadlock-free Component Systems}}]\\
& S[T'] = T' \intercomp Q \text{ is deadlock free}\\
& \textbf{Communication composition }\\
& T \commcomp{c}{z} Q\\
& \implies [T \subecvg T', S[T] \text{ is deadlock free}, \text{Lemma \ref{lem:BRIC inheritance preserves deadlock freedom}}]\\
&  \cc{C}{T'} \cap \cc{C}{Q} = \emptyset  \land T' \text{ is deadlock free}\\
& \implies [\text{Definition \ref{def:communicationComposition}}]\\
&  \cc{C}{T'} \cap \cc{C}{Q} = \emptyset  \land T' \text{ is deadlock free} \land\\
& \cc{B}{T \commcomp{c}{z} Q} = \cc{B}{T}  \parallel[\eset{c}] BUFF^{n}_{IO}(R_{IO}^{~c \rightarrow z}, R_{IO}^{~z \rightarrow c}) \parallel[\eset{z}] \cc{B}{Q}\\
& \implies [\cc{B}{T \commcomp{c}{z} Q} \text{ is deadlock free}, \  \text{ hiding semantics}, {R_{IO}^{~c \rightarrow z}}^{*} t \text{ \ref{appendix:CSP}}~~]\\
&  \cc{C}{T'} \cap \cc{C}{Q} = \emptyset  \land T' \text{ is deadlock free } \land\\
& \forall t \in \tmodel(\cc{B}{T} \project c) \spot  {R_{IO}^{~c \rightarrow z}}^{*} t \in \tmodel(\cc{B}{Q} \project z) \implies\\
& \hspace*{3cm} \left( 
 \begin{array}{c}
R_{IO}^{~c \rightarrow z} out(\cc{B}{T} \project c, t) \subseteq in(\cc{B}{Q} \project z, {R_{IO}^{~c \rightarrow z}}^{*} t),   \\
R_{IO}^{~z \rightarrow c} out(\cc{B}{Q} \project z, {R_{IO}^{~c \rightarrow z}}^{*} t) \subseteq in(\cc{B}{T} \project c, t)
\end{array}
\right)\\
& \implies [\defcong{\cc{B}{T'}}{\cc{B}{T}}{c} \lor \inpcong{\cc{B}{T'}}{\cc{B}{T}}{c}]\\
&  \cc{C}{T'} \cap \cc{C}{Q} = \emptyset  \land T' \text{ is deadlock free } \land\\
& \forall t : \Sigma^{*} \ | \ t \in \tmodel(T \project c) \land t \in \tmodel(T' \project c) \spot \\
& \quad out(T', t) \subseteq out(T, t) \land in(T,t) \subseteq in(T', t) \ \land \\
& \quad \tmodel(\cc{B}{T'} \project c) \cap \tmodel(\cc{B}{Q} \project z \rename{R_{IO}^{~z \rightarrow c}}) \subseteq \tmodel(\cc{B}{T} \project c) \cap \tmodel(\cc{B}{Q} \project z \rename{R_{IO}^{~z \rightarrow c}})\\
& \implies [\text{Definition \ref{def:communicationComposition}}, \text{ Theorem \ref{thm:Deadlock-free Component Systems}}] \\
& S[T'] = T' \commcomp{c}{z} Q \text{ is deadlock free}\\
& \textbf{Feedback composition} \\
& T \feedcomp{c}{z}\\
& \implies [T \subecvg T', S[T] \text{ is deadlock free}, \text{Lemma \ref{lem:BRIC inheritance preserves deadlock freedom}}]\\
&  \cc{C}{T'} \cap \cc{C}{Q} = \emptyset  \land T' \text{ is deadlock free}\\
& \implies [ \text{Definition \ref{def:feedbackcomposition}}]\\
&   \cc{C}{T} \subseteq \cc{C}{T'}  \land T' \text{ is deadlock free} \land \\
& \cc{B}{T \feedcomp{c}{z}} = \cc{B}{T} \parallel[\eset{c,z}] BUFF^{n}_{IO}(R_{IO}^{~c \rightarrow z}, R_{IO}^{~z \rightarrow c})\\
& \implies [\cc{B}{T \feedcomp{c}{z}} \text{ is deadlock free, hiding semantics}]\\
&   \cc{C}{T} \subseteq \cc{C}{T'}  \land T' \text{ is deadlock free} \land \\
& \forall t \in \tmodel(\cc{B}{T} \project c) \spot  {R_{IO}^{~c \rightarrow z}}^{*} t \in \tmodel(\cc{B}{T} \project z) \implies 
\\
& \hspace*{3cm}\left( 
 \begin{array}{c}
R_{IO}^{~c \rightarrow z} out(\cc{B}{T} \project c, t) \subseteq in(\cc{B}{T} \project z, {R_{IO}^{~c \rightarrow z}}^{*} t)
\\ \land  \\
R_{IO}^{~z \rightarrow c} out(\cc{B}{T} \project z, {R_{IO}^{~c \rightarrow z}}^{*} t) \subseteq in(\cc{B}{T} \project c, t)
\end{array}
\right)\\
& \implies [\defcong{\cc{B}{T'}}{\cc{B}{T}}{c} \lor \inpcong{\cc{B}{T'}}{\cc{B}{T}}{c}]\\
&   \cc{C}{T} \subseteq \cc{C}{T'}  \land T' \text{ is deadlock free} \land \\
& \forall t : \Sigma^{*} \ | \ t \in \tmodel(T \project c) \land t \in \tmodel(T' \project c) \spot \\
& \quad out(T', t) \subseteq out(T, t) \land in(T,t) \subseteq in(T', t) \land\\
& \quad \tmodel(\cc{B}{T'} \project c) \cap \tmodel(\cc{B}{T} \project z \rename{R_{IO}^{~z \rightarrow c}}) \subseteq \tmodel(\cc{B}{T} \project c) \cap \tmodel(\cc{B}{T} \project z \rename{R_{IO}^{~z \rightarrow c}})\\
& \implies [\text{Definition \ref{def:feedbackcomposition},  Theorem \ref{thm:Deadlock-free Component Systems}}]] \\
& S[T'] =  T' \feedcomp{c}{z} \text{ is deadlock free}\\
& \textbf{Reflexive composition}. \text{ It is almost identical to the feedback composition.}
\end{align*}
\end{proof}

\end{document}